\documentclass[12pt]{iopart}
\usepackage{hyperref}
\usepackage{iopams}
\usepackage[english]{babel}
\usepackage{graphicx}
\usepackage{epsfig}
\usepackage{color}
\usepackage{latexsym}
\usepackage{bm}
\makeatletter

\newcommand{\Leftarrowfill}[0]{$\m@th  \mathord\Leftarrow  \mkern-6mu
\cleaders\hbox{$\mkern-2mu \mathord= \mkern-2mu$}\hfill
\mkern -6mu \mathord=$}

\newcommand{\StemPullBack}[2]{
  \vtop{\mathsurround=0pt
  \ialign{##\crcr$\textstyle{#1}\strut$\crcr
    \noalign{\kern-0.4ex\nointerlineskip}{\tiny#2}\crcr}}}

\newcommand{\IndxPullBack}[2]{
  \vtop{\mathsurround=0pt
  \ialign{##\crcr\hfil$\scriptstyle{#1}$\hfil\crcr
    \noalign{\kern+0.4ex\nointerlineskip}{\tiny#2}\crcr}}}

\newcommand{\sct}{\mbox{\textsc{t}}}
\newcommand{\tilbox}{\tilde{\mbox{\rule{0pt}{2.5mm}$\Box$}}} 
\newcommand{\Real}{\mathbb{R}}

\newcommand{\cM}{{\cal M}}

\newcommand{\cN}{{\cal N}}
\newcommand{\cA}{{\cal A}}
\newcommand{\cB}{{\cal B}}
\newcommand{\cC}{{\cal C}}

\newcommand{\ag}{\alpha} 
\newcommand{\bg}{\beta} 
\newcommand{\Dg}{\Delta}
\newcommand{\dg}{\delta} 
\newcommand{\cg}{\gamma}

\newcommand{\lam}{\lambda}

\newcommand{\veg}{\varepsilon} 
\newcommand{\vareg}{\varepsilon}
\newcommand{\Sg}{\Sigma} 
\newcommand{\sg}{\sigma} 

\newcommand{\di}{\partial} 
\newcommand{\be}{\begin{equation}} 
\newcommand{\ee}{\end{equation}} 
\newcommand{\bearr}{\begin{eqnarray}}
\newcommand{\eearr}{\end{eqnarray}} 

\newcommand{\bftheta}{\bm{\theta}} 
\newcommand{\QED}{\rule{1.5mm}{3mm}} 
\newcommand{\Cyc}[1]{
  \vtop{\mathsurround=0pt
  \ialign{##\crcr$\textstyle{\rm Cyc}\strut$\crcr
    \noalign{\kern-0.4ex\nointerlineskip}{\tiny#1}\crcr}}\ }

\newtheorem{definition}{Definition}  

\newtheorem{proposition}{Proposition}

\begin{document} 

\title[The Poisson brackets of free null initial data for vacuum general relativity]{The Poisson brackets of free null initial data for vacuum general relativity}
\author{Michael P Reisenberger}
\address{Instituto de F\a'{\i}sica, Facultad de Ciencias,\\
        Universidad de la Rep\a'ublica Oriental del Uruguay,\\
        Igu\a'a 4225, esq. Mataojo, Montevideo, Uruguay}
\ead{miguel@fisica.edu.uy}
%\maketitle

\begin{abstract}
A hypersurface composed of two null sheets, or "light fronts", swept out by 
the two congruences of future null normal geodesics emerging from a spacelike 2-disk 
can serve as a Cauchy surface for a region of spacetime. Already in the 1960s free
(unconstrained) initial data for vacuum general relativity were found for
hypersurfaces of this type. Here the Poisson brackets of such free initial data are 
calculated from the Hilbert action. The brackets obtained can form the starting 
point for a constraint free canonical quantization of general relativity and may be 
relevant to holographic entropy bounds for vacuum gravity. Several of the results 
of the present work have been presented in abreviated form in the letter \cite{PRL}.

\end{abstract}

\pacs{04.20.Fy}% PACS, the Physics and Astronomy

\maketitle

\section{Introduction}

Initial data for general relativity (GR) is usually subject to constraints. That is,
not all valuations of the initial data correspond to solutions of the field 
equations, only those that satisfy the constraint equations do. This complicates 
canonical formulations of GR in terms of initial data, which is reflected, for instance, 
in canonical approaches to quantizing gravity. It seems fair to say that at present 
the handling of constraints absorbs most of the effort invested in canonical quantum 
gravity.

\begin{figure}
\begin{center}

\includegraphics[height=5cm]{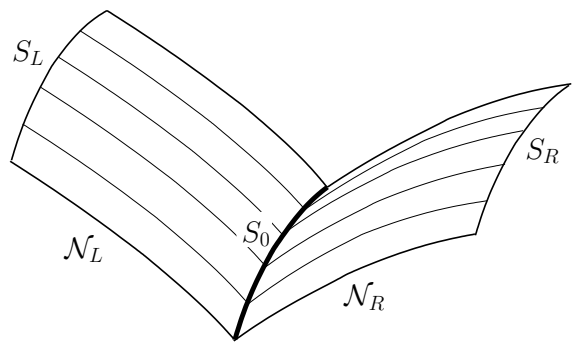}
\caption{A double null sheet in 2+1 dimensional spacetime. In 3+1 dimensions $S_0$ is a disk instead of a line segment, and $\cN_L$ and $\cN_R$ are solid cylinders
instead of 2-surfaces.}
\label{Nfigure}
\end{center}
\end{figure}

Instead of working with constrained data one can look for free data. Such data are not subject
to constraints: all valuations correspond to solutions of the field equations.
In the 1960s complete free initial data were identified on certain piecewise null hypersurfaces
\cite{Sachs, Dautcourt, Penrose}. In particular Sachs \cite{Sachs} and Dautcourt \cite{Dautcourt}
found free initial data for vacuum GR on what we call a {\em double null sheet}, illustrated in Fig.\ \ref{Nfigure}.
This is a compact hypersurface $\cN$, consisting of two null branches, $\cN_L$ and $\cN_R$, that 
meet on a spacelike 2-disk $S_0$. The branches are swept out by the two congruences of future directed normal null geodesics 
from $S_0$ (called {\em generators}), and are truncated on disks $S_L$ and $S_R$ respectively
before these geodesics form caustics or cross. These data on $\cN$ determine a solution to the Einstein field equations
in a portion of spacetime to the future of $\cN$ \cite{Rendall}.

In the present work a Poisson bracket corresponding to the Einstein-Hilbert action is obtained 
for a complete set of free initial equivalent to that of Sachs and Dautcourt. An abreviated presentation
of these brackets has appeared in the letter \cite{PRL}, and a preliminary form of their calculation can 
be found in the e-print \cite{MR07}. The first half of the calculation, expressing the symplectic form
in terms of the free data on $\cN$, is reported in \cite{MR13}. Here we complete the calculation, obtaining 
the brackets from the symplectic form.

A canonical formulation of GR in terms of free null data, that is ``physical degrees of freedom'', should 
be useful in the analytical and numerical study of classical solutions, and of course it constitutes a new 
avenue for the canonical quantization of GR, free of the difficulties associated with constraints.
(See \cite{FR17} for steps in the latter direction.)
A particularly interesting issue that such a theory seems well suited to clarify 
is the conjectured holographic entropy bound \cite{Bekenstein,tHooft,Susskind,Bousso,BCFM14}
in the case of full, non-linear, vacuum gravity. In Bousso's formulation this bound 
limits entropy on ``light sheets'', and the null branches $\cN_A$ ($A = L\ \mbox{or}\ R$) 
of $\cN$ are light sheets in Bousso's sense, provided their generators do not diverge at $S_0$. 

Given that free null initial data for GR has been available for such a long time the question arises as 
to why a canonical framework based on such data was not developed sooner. In fact canonical formulations 
of GR in terms of {\em constrained data} on null hypersurfaces has been developed by several researchers. 
See for example \cite{Torre, Goldberg1, Goldberg2, Vickers, Alexandrov_Speziale, Hopfmueller_Freidel}. Several partial results on 
the Poisson brackets of free null initial have also been published \cite{GR, Goldberg2}.
In \cite{GR} Gambini and Restuccia express the bracket of the so called conformal 2-metric on $\cN$ 
with itself in terms of a perturbation series in Newton's constant. The conformal 2-metric, which encodes the 
conformal geometry of $\cN$ induced by the spacetime metric, is one of the free inital data in our formalism,
and the only one among these that is set on all of $\cN$. Gambini and Restuccia do not consider the remaining
free data, which are set on the intersection 2-surface $S_0$. However, their result was essential for
the genesis of the present work. The bracket of the conformal 2-metric presented in \cite{MR07}
(and with a slight modification here and in \cite{PRL}) was first obtained by summing the series of \cite{GR} 
in closed form, before being derived more systematically from the symplectic form. 
Finally, in \cite{Goldberg2} Goldberg, Robinson and Soteriou present distinct free data
in place of the conformal 2-metric, which are claimed to form a canonically conjugate pair
on the basis of a machine calculation of their Dirac brackets. It would be interesting to see if 
these data are also conjugate according to the Poisson structure obtained here.

Nevertheless there are conceptual problems which must be overcome in order to develop a null canonical formalism
for gravity, and which may have slowed this development.
First and foremost is the problem of caustics and generator crossings: Generically the generators of 
a null hypersurface will cross and enter the chronological future of the hypersurface (see \cite{Wald} 
theorem 9.3.8). To have an achronal initial data surface it is necessary to truncate the generators at 
or before the crossings. One must then deal either with a non-smooth initial data surface, or - as is 
done here - with one having boundaries. Moreover, the truncation of the generators at or before crossings 
must somehow be expressed as a limitation on the allowed initial data.

A second problem arises because a null hypersurface for one solution to the field equations is in general not null
for an even slightly different solution. Thus, for instance only a special class of solutions can be represented 
by our null initial data on a hypersurface $\cN$ fixed in spacetime, namely those that make $\cN$ a double null sheet.
To represent other solutions by these data one has to move $\cN$ to coincide with one of their double null sheets, and 
provide additional data that specifies this displacement. This complication is of course greatly reduced by 
the diffeomorphism gauge invariance of general relativity, but, as is well known in the case of asymptotically flat
canonical gravity, not all diffeomorphisms are gauge when the initial data hypersurface has boundaries, and this issue 
is not trivial here. 

\begin{figure}
%\begin{center}
\centering
%\includegraphics[width=\textwidth]{\input{fig3ext2.pstex_t}}
%\scalebox{0.77}{\input{fig3ext2(copy).pstex_t}}
%\input{fig3ext.pstex_t}
\includegraphics[width=0.75\textwidth]{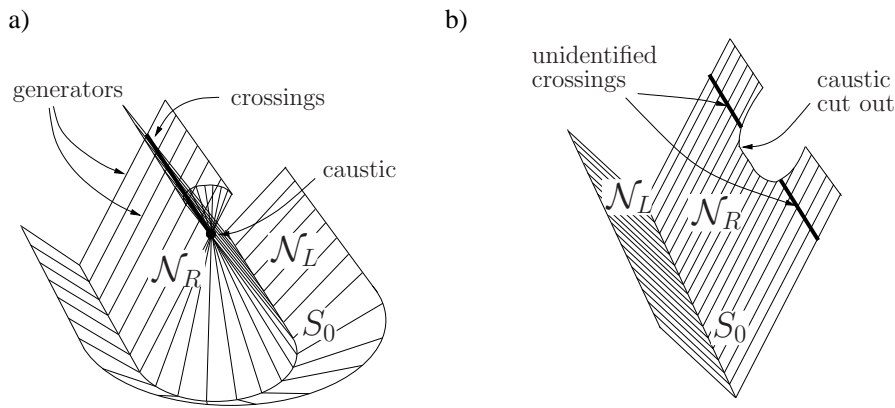}%[width=0.5\textwidth]
\caption{Panel a) shows a simple example of a caustic and intersections of
generators in $2 + 1$ Minkowski space: $S_0$ is a spacelike curve having the
shape of a half racetrack - a semicircle extended at each end by a tangent
straight line. The congruence of null geodesics normal to $S_0$ and directed
inward and to the future sweep out $\cN_R$, which takes the form of a ridge roof,
terminated by a half cone over the semicircle. The generators from the
semicircle form a caustic at the vertex of the cone. There neighbouring
generators intersect. On the other hand generators from the two straight
segments of $S_0$ cross on a line (the ridge of the roof) starting at the
caustic, but the generators that cross there are not neighbours at $S_0$.
Clearly the generator segments beyond the crossing points enter the interior
of the domain of dependence of $\cN$.
In Panel b) the double null sheet defined by $S_0$ in the covering space
%defined in Appendix \ref{causal_appendix}
is shown, with the points that are
identified in the original spacetime indicated.}
\label{crossing}
%\end{center}
\end{figure}

The remainder of this introduction outlines how these issues are resolved, and how the Poisson bracket on initial data 
will be defined here. First let us consider caustics and generator crossings, two related but distinct fenomena. At a 
caustic point neighboring generators focus, so that the congruence of generators is singular there. This can be detected
in the initial data: If the functions $\theta^1$ and $\theta^2$ on $\cN$ form a chart on the disk $S_0$ and are constant along the 
generators then the matrix of components $h_{pq}$ in the $\theta$ chart of the metric induced on 2-surfaces $S \subset \cN$ transverse 
to the generators is singular at caustic points. Caustics will be excluded from $\cN$ by allowing only initial data such that $h_{pq}$
is regular everywhere on $\cN$.  

The problem of non-neighboring generators crossing is resolved in a completely different manner. Non-neighboring 
generators can cross even if there are no caustic points on $\cN$. See Figure \ref{crossing}. But when there are no
caustics on $\cN$ there exists an isometric covering of a spacetime neighborhood of $\cN$ by a spacetime in which no 
generators cross. This covering spacetime is (a subset of) the normal bundle of $S_0$ equipped with the metric obtained
by pulling back the metric of the original spacetime using the exponential map. See \cite{MR07} appendix B for details.
The covering spacetime of course induces the same data on $\cN$ as does the original spacetime.
Thus we can (and will) always suppose that the generators segment within $\cN$ do not cross in the 
spacetimes matching the data given on $\cN$, even though for some data there exist alternative matching spacetimes, 
obtained by identifying isometric spacetime regions, in which the generators do cross.
In fact, we will impose the slightly stronger requierments that any point $p \in \cN$ is causally connected only to points on the 
generators through $p$ and also that $\cN$ is achronal, because $\cN$ has these properties in the covering spacetime 
\cite{MR07} appendix B.\footnote{
It is worth noting that the same issue arises in the spacelike Cauchy
problem, and is resolved in the same way. Spacelike hypersurfaces that enter
the interior of their own domains of dependence are easily constructed in
any solution spacetime $M$. But the unique maximal Cauchy development of the
initial data induced from $M$ on such a hypersurface is a covering manifold
of the original domain of dependence, in which the hypersurface is
achronal.}

Now let us turn to the issue of representing variations of the solution spacetime metric in terms of 
the variations of null initial data on $\cN$ when these variations of the spacetime metric do
not preserve the nullness of $\cN$. This problem is largely resolved by exploiting the diffeomorphism gauge 
invariance of GR: one describes a given variation of the spacetime metric by the corresponding variations of null 
initial data in a gauge equivalent variation of the solution metric which does preserve the nullness of
the branches of $\cN$. 

This is relevant to the definition of the Poisson brackets of null initial data.
In mechanics the Poisson bracket is the inverse of the symplectic form. That is,
\be\label{Poisson_def_0}
\dg \phi = \omega[\{\phi,\cdot\},\dg]
\ee
for all variations $\dg$ tangent to the space of solutions and all functions $\phi$ on this space. Here 
$\{\cdot,\cdot\}$ is the Poisson bracket and $\omega$ is the symplectic form, a 2-form on the space of 
solutions defined in Section \ref{definition_bracket_on_data}. (See \cite{Lee_Wald}.)
The definition (\ref{Poisson_def_0}) works also when the system has gauge symmetries, that is when $\omega$ 
is degenerate \cite{Lee_Wald}, provided $\phi$ is restricted to gauge invariant functions on the space of solutions. It 
then defines the Poisson bracket between such gauge invariant functions.   

Proceeding analogously in the case of vacuum GR one fixes the diffeomorphism gauge freedom of solutions
by requiering that a given hypersurface $\cN$ is always a double null sheet, and imposing further conditions
until no diffeomorphism gauge freedom remains, that is, each gauge equivalence class is represented by 
precisely one allowed solution. The solutions will then always induce null data on $\cN$ and these are, moreover, 
gauge invariant functions on the space of solutions. (\ref{Poisson_def_0}) defines the Poisson bracket 
between these gauge fixed data.

In fact not all diffeomorphisms are gauge in the sense of being generated by null vectors of the symplectic 
form (see (\ref{boundary_diff})). And in particular it is not clear that all variations of the metric are gauge 
equivalent to variations that preserve the double null sheet character of $\cN$. In other words, the gauge fixing 
imagined above may be impossible.

To proceed we adopt an apparently completely different approach, based on Peierls' definition of the Poisson bracket, 
which in the end leads us back to an equation for the Poisson bracket which is simply a slightly weaker 
version of the condition (\ref{Poisson_def_0}). The Peierls bracket $\{A,B\}$ between two functions, $A$ and $B$ 
on the space of solutions is the retarded perturbation in $B$ due to the addition of $\cg A$ to the action, 
calculated to first order in the parameter $\cg$, minus the analogous first order advanced perturbation \cite{Peierls}. 
It provides an expression for the Poisson bracket in terms of advanced and retarded Green's functions which does not 
involve the symplectic form or Cauchy surfaces. It's simplicity gives it a good claim to being a more fundamental definition than 
(\ref{Poisson_def_0}) in terms of the symplectic form. Furthermore it agrees with the latter definition when 
both are defined \cite{Peierls, DeWitt}.

The Peierls bracket does not provide the Poisson bracket between initial data on $\cN$ directly.
It is ambiguous on these data because the advanced and retarded Green's functions between two points 
are discontinuous when these points are lightlike separated. However the Peierls bracket {\em is} 
well defined on what we call "observables", diffeomorphism invariant functionals $F[g]$ of the metric, 
with smooth functional derivatives $\dg F/\dg g_{ab}$ of compact support (called the {\em domain of 
sensitivity} of the observable and denoted $\sg_F$).  
Our approach \cite{PRL, MR07} is to look for a Poisson bracket $\{\cdot,\cdot\}_\bullet$ on initial 
data that reproduces the Peierls brackets between observables having domains of sensitivity in the 
interior of the the causal domain of dependence of $\cN$.\footnote{
The causal domain of dependence $D[S]$ of a set $S$ in a Lorentzian signature
spacetime is the set of all points $p$ such that every inextendible causal curve
through $p$ intersects $S$. If $S$ is a closed achronal hypersurface one expects
in physical theories that initial data on $S$ fixes the solution in $D[S]$.
See \cite{Wald}.}

The construction of such observables in vacuum GR is not easy in general, but can be done in principle in most vacuum spacetimes.
On generic spacetimes charts can be formed from scalar contractions of the Weyl tensor. If $\cN$ is chosen so
that $D[\cN]$ is small enough that it can be covered by an atlas of such charts then it is clear 
that observables in our sense can be defined in terms of the metric components in this chart, and 
that these characterize the geometry of the interior, $D[\cN]^\circ$, of $D[\cN]$ completely.  
See \cite{Komar58}\cite{Bergmann_Komar60} for an approach to observables along these lines. 
\footnote{
Cartan's solution to the equivalence problem shows that in fact all Lorentzian spacetime geometries
can be completely characterized (i.e. distinguished from each other) by means of diffeomorphism invariant
quantities constructed from the curvature tensor and a finite set of its derivatives \cite{Cartan}\cite{Brans}
\cite{Karlhede}\cite{Stephani} (even though polynomials in the curvature tensor and its derivatives will not 
always suffice \cite{Page}). It has not yet been shown, to the best of my knowledge, that this can 
always be turned into a description in terms of observables as defined here.} 
Of course each such observable will work, that is satisfy the definition of observable, only on a limited 
subset of the space of solutions (see \cite{Khavkine} for a discussion), but this is the case for almost all
physical observables we know, so it does not disqualify them. Note that the role of observables 

Note that observables will ultimately play no direct role in the calculation of the Poisson brackets between initial 
data presented here. Their role is to motivate the definition adopted for this bracket. The existence of a rich set 
of observables in generic vacuum spacetimes suffices for this purpose.

Another aspect of the motivating framework that is ultimately not used in obtaining our results is the existence and 
uniqueness of solutions of the Einstein field equations corresponding to the Sachs/Dautcourt initial data. The expression for 
the symplectic form in terms of initial data was obtained in \cite{MR13} without invoking this result, and in the present work 
it is again unnecessary. In fact, both the symplectic form and the Peierls bracket at a given solution metric $g$ are features
of the theory linearized about $g$. Thus only properties of GR linearized about $g$ are needed.

We will see that a Poisson bracket $\{\cdot,\cdot\}_\bullet$ on the free null initial data on $\cN$ reproduces the 
Peierls bracket on observables of the interior of $D[\cN]$ if for any such observable $A$ expressed in terms of the 
initial data
\be                     \label{auxbracketdef}
        \dg A = \omega_\cN[\{A,\cdot\}_\bullet,\dg],
\ee
for any $\dg$ in the space $L_g^0$ of smooth variations which satisfy the field equations linearized about $g$ and vanish 
in a spacetime neighbourhood of $\di\cN$.

This is just a weakened version of the condition (\ref{Poisson_def_0}) that the Poisson bracket be 
inverse to the symplectic form. But the restriction that the variations $\dg g_{ab}$ vanish in a 
neighborhood of $\di\cN$ makes it possible to express both sides of this equation purely in terms of 
the free null initial data on $\cN$ and their variations, despite the fact that not all diffeomorphisms are gauge.
Specifically, one may add any diffeomorphism generator that vanishes in a neighborhood of $\di\cN$ to $\dg$ 
and any diffeomorphism generator at all to $\{A,\cdot\}_\bullet$ without changing the value of either side of 
the equation. This allows one to restrict both arguments of $\omega_\cN$ in (\ref{auxbracketdef}) to what are called 
``admissible variations`` in \cite{MR13}; in the case of $\dg$, without weakening the condition it imposes on 
$\{A,\cdot\}_\bullet$, and in the case of $\{A,\cdot\}_\bullet$ without leaving the space of solutions to this condition. 

Admissible variations are smooth variations of the spacetime metric, satisfying the linearized field equations, 
that leave the branches $\cN_A$ swept out by the generators in spacetime invariant (and also satisfy further 
conditions detailed in \ref{a_chart_appendix}). Admissible variations therefore correspond to variations of null 
initial data on a fixed hypersurface $\cN$. In fact within $D[\cN]$ they are determined, modulo diffeomorphism 
generators, by these variations and the Green's functions for the linearized field equations. See \ref{a_chart_appendix} 
or \cite{MR07} Appendix C. It therefore becomes possible to write (\ref{auxbracketdef}) entirely in terms of null 
initial data turning it into a condition on the $\bullet$ Poisson brackets of these data:
\be                     \label{auxbracketdef1}
        \dg A = \omega_\cN[\{A,\cdot\}_\bullet,\dg]\ \ \ \forall\: A\ \mbox{observable},\: \dg\in \cB
\ee
where $\cB$ is the set of variations of the initial data corresponding to admissible variations in $L_g^0$, i.e. 
that vanish in a spacetime neighborhood of $\di\cN$. The observable $A$ enters (\ref{auxbracketdef1}) only through
its linearization about the spacetime metric $g$, $\Delta A = \int_{\sg_A} \dg A/\dg g^{ab} \Delta g^{ab} \vareg$
for any solution $\Dg g^{ab}$ to the linearized field equations. (Here $\vareg$ is the spacetime metric volume form.) 
Linearized observables can in turn be expressed as smeared variations of the initial data, with a particular class of 
smooth smearing functions.

This condition is almost what we need. It is a condition directly on the $\bullet$ bracket of the initial data, but the
set $\cB$ of variations of the initial data is not at all easy to characterize, nor is the set of smearing functions on the 
initial data that correspond to linearized observables. It would be better to have a simpler condition to define $\{\cdot,\cdot\}$. 

Another weakness of (\ref{auxbracketdef1}) is that it does not guarantee that the $\bullet$ bracket satisfies the Jacobi relations.
If the Poisson bracket is defined as the inverse of the symplectic 2-form then the Jacobi relations follow as a consequence of the
fact that the symplectic form is closed, but (\ref{auxbracketdef1} is weaker than demanding that $\{\cdot,\cdot\}_\bullet$ be inverse 
to $\omega_\cN$. If one solves (\ref{auxbracketdef1}) as a linear equation for the $\bullet$ brackets between the initial data, 
assuming that $\{\cdot,\cdot\}_\bullet$ is a derivation in each of its arguments but not that it satisfies the Jacobi relations, 
then the result is not unique, and in general does not satisfy the Jacobi relations. 
For example, the ``pre-Poisson bracket'' $\{\cdot,\cdot\}_\circ$ of \cite{MR07} satisfies (\ref{auxbracketdef1}) but violates the Jacobi relations. 

For these reasons (\ref{auxbracketdef}) will be replaced by a simpler and stronger set of conditions which 
implies the original condition as a corollary, and thus that the $\bullet$ bracket reproduces the Peierls 
bracket on observables, and also yields a unique solution which satisfies the Jacobi relation. 

% Are there boundary conditions on smearing functions? Few. Stated later.

In these new conditions determining the $\bullet$ bracket the set of linearized observables is replaced by a 
larger set $\Phi$ of smeared initial data, defined by simple smoothness and boundary conditions on the weighting 
functions; The set of variations $\cB$ is replaced by a larger set $\cC$ of variations of the data, also satisfying 
simple smoothness and boundary conditions; And finally it will be required that also the variations 
$\{\varphi, \cdot\}_\bullet$ generated by smeared initial data $\varphi$ via the bracket, lie in $\cC$, so that 
the $\bullet$ bracket in fact is inverse to $\omega_\cN$ restricted to $\cC$. Thus we require that for all $\varphi \in \Phi$
\be	\label{final_bracket_condition1}
        \dg \varphi = \omega_\cN[\{\varphi,\cdot\}_\bullet,\dg]\ \ \ \forall\:\dg \in \cC, %\:\varphi \in \Phi,
\ee
and
\be     \label{final_bracket_condition2}
\{\varphi,\cdot\}_\bullet \in \cC.  %\ \ \ \forall\:\varphi \in \Phi.
\ee

The fact that the bracket thus defined on smeared initial data in $\Phi$ is unique will be demonstrated in Section \ref{bracket_calc},
and that it satisfies the Jacobi relation will proved in Section \ref{definition_bracket_on_data}. There is, however, one apparent problem 
with the system (\ref{final_bracket_condition1}, \ref{final_bracket_condition2}): (\ref{auxbracketdef1} guarantees that the $\bullet$ 
bracket reproduces the Peierls bracket on observables {\em if} $\{A,\cdot\}_\bullet$ is admissible for all observables $A$. But all that 
is obvious from (\ref{final_bracket_condition1}, \ref{final_bracket_condition2}) is that $\{A,\cdot\}_\bullet \in \cC$, and $\cC$ is not 
contained in the set $\cA$ of admissible variations of the initial data.

In fact $\{A,\cdot\}_\bullet$ is admissible for all observables $A$. To prove this it is enough to show that (\ref{final_bracket_condition2})
and the slightly weakened form of (\ref{final_bracket_condition1}),
\be	\label{final_bracket_condition1w}
        \dg \varphi = \omega_\cN[\{\varphi,\cdot\}_\bullet,\dg]\ \ \ \forall\:\dg \in \cC \cap \cA,
\ee
suffice to determine $\varphi,\cdot\}_\bullet$. Then, we shall see, that there exists for each observable $A$ a manifestly admissible variation 
of the initial data which satisfies the same conditions as $\{A,\cdot\}_\bullet$ and is hence equal to it.
One could therefore say that the $\bullet$ bracket is defined by (\ref{final_bracket_condition1w}) and (\ref{final_bracket_condition2}), but 
it also satisfies (\ref{final_bracket_condition1}), which permits an easy proof of the Jacobi relations.

The brackets obtained from these conditions are stated in Section \ref{structure_relations}. They are identical to 
the ones anounced in \cite{PRL}.

Notice that while the requierment that the $\bullet$ bracket reproduces the Peierls bracket on observables was used to motivate
the conditions (\ref{final_bracket_condition1}), (\ref{final_bracket_condition2}), and (\ref{final_bracket_condition1w}), the conditions themselves
do not involve the observables. They are simply modified versions of the standard requierment that Poisson bracket be inverse to the symplectic 
form. 

Of course the question arises: To what extent is the $\bullet$ bracket determined by the requierment that it matches the Peierls 
bracket on observables, and to what extent is it determined by the additional conditions placed on it to arrive at 
(\ref{final_bracket_condition1}, \ref{final_bracket_condition2})? We will see that the brackets of certain data, the so called 
''diffeomorphism data`` (see Section \ref{data}), are not determined by the matching to the Peierls brackets of observables 
of the interior of the domain of dependence at all because these observables not depend on the diffeomorphism data. On the other hand, experience 
solving weaker forms of the condition (\ref{final_bracket_condition1}) suggests that the brackets of the remaining data, the 
''geometrical data`` are essentially uniquely determined by the matching. That is, if one assumes {\em a priori} that the $\bullet$ 
bracket has the properties of a Poisson bracket, including the Jacobi relations, then (\ref{auxbracketdef1}) suffices to determine 
it uniquely on these data. Note, however, that this is not proved.

The $\bullet$ bracket that we obtain (see Section \ref{structure_relations}) has one strange property. 
It does not preserve the reality of the metric. That is, the Hamiltonian flow generated by smeared data that
is real on real solutions generates an imaginary component of the metric. This is, however, much less serious 
than it seems at first because the imaginary mode generated in the initial data does not affect observables. It is 
a shock wave that skims along a branch of $\cN$ without entering the interior of the domain of dependence. See 
Subsection \ref{complex_modes}. Such a shock wave mode is not seen in the analyses of the initial value problem 
of \cite{Sachs} and \cite{Dautcourt} because it is excluded by continuity conditions on the data at $S_0$ which 
are, however, not natural for the Poisson bracket on initial data. It is thus interesting that in \cite{FR17} 
alternative data is found that captures all the information of the data used here, {\em except} this mode.

The remainder of the paper is organized as follows: The next section provides the technical underpinning of
part of the preceding discussion. The Peierls bracket and the symplectic form are defined and related to each other,
and the Jacobi relation is established for brackets satisfying (\ref{final_bracket_condition1}, \ref{final_bracket_condition2}).
In Section \ref{data} free initial data that will be used are presented, and the spaces of variations $\cC$ and smeared data $\Phi$ 
are defined. In Section \ref{omega_free_data} the symplectic form on admissible variations is expressed in terms of the initial 
data and their variations. This completes the preparation for the calculation of the $\bullet$ brackets of the data.
In Section \ref{structure_relations} the result of this calculation is presented and discussed. The calculation that
leads to these brackets is presented in detail in Section \ref{bracket_calc}. The article closes with 
reflections on the results obtained.

\section{The Peierls bracket on observables and the definition of the Poisson bracket on initial data}\label{definition_bracket_on_data}

Unless otherwise indicated the solution spacetimes we consider will always be smooth ($C^\infty$). 
The double null sheet $\cN$ will also be smooth in the sense that $S_0$, $\cN_L$, and $\cN_R$ are smoothly embedded.\footnote{
A smooth function on an arbitrary domain is defined to be one that posseses a
smooth extension to an open domain. See the appendix of \cite{Lee_smooth} or \cite{AMR} chapter 7. 
Consequently a smooth manifold with boundary necessarily has an extension to a smooth manifold without boundary, 
and an embedding of a manifold with boundary is smooth iff there exists a smooth extension of the 
embedding to a manifold without boundary.}
(The smoothness of $\cN_L$ and $\cN_R$ actually follows from the smoothness $S_0$ and the spacetime metric, 
together with the assumptions that $\cN$ is compact and contains neither caustics nor intersections of generators.)  

The solution spacetime will also be assumed to be globally hyperbolic. This is not a very strong restriction
because we may take as our spacetime a small neighborhood about $\cN$.

At a solution $g$ the Peierls bracket \cite{Peierls} between two observables $A$ and $B$ is 
\be
\{A,B\} = \Dg_A B
\ee
where 
$\Dg_A = \dg^+_A - \dg^-_A$, and $\dg^\pm_A$ is the retarded ($+$)/advanced ($-$) first order perturbation 
of $g$ due to the addition of a source term $\gamma A$ to the action, with $\gamma$ the perturbation parameter.
That $\dg^+_A$ is retarded means that the support of $\dg^+_A g_{ab}$ is contained in the causal 
future $J^+[\sg_A]$ of the domain of sensitivity $\sg_A$ of $A$, and similarly the advanced perturbation is 
supported in the causal past $J^-[\sg_A]$ of $\sg_A$.

$\dg^\pm_A g_{ab}$ are solutions to the linearized field equation
with source $\dg A/\dg g_{cd}$, and may be obtained from the retarded/advanced
Green's function. The definition of $\dg^\pm_A g_{ab}$ requiers the choice of a gauge, which we will take to be
de Donder gauge,\footnote{
The de Donder gauge fixes most of the diffeomorphism freedom in solutions $\dg g_{ab}$ of the linearized 
field equations by imposing the condition $\nabla^a \dg g_{ab} - \frac{1}{2} \nabla_b \dg g^a{}_a = 0$.}  
but the Peierls brackets of observables are independent of this choice. 
The Peierls bracket is well defined between all observables of $D[\cN]$ and has all the properties of a 
Poisson bracket \cite{DeWitt}. 

Note that since spacetime is assumed to be globally hyperbolic with a smooth metric, and the linearized vacuum 
Einstein equations in de Donder gauge are normally hyperbolic $\dg^\pm_A g_{ab}$ amd $\Dg_A g_{ab}$ are smooth 
\cite{Bar_Ginoux_Pfaffle}.

Note also that these perturbations, and thus the Peierls bracket, depends only on the linearized field equations
and the linearized observables, or equivalently the functional derivatives $\dg A/\dg g_{cd}$ of the observables. 
The definition of the Peierls bracket and the general properties we have mentioned require only that these functional 
derivatives be smooth, compactly supported, symmetric tensor fields with vanishing divergence, $\nabla_a \dg F/\dg g_{ab} = 0$ 
(the latter being a consequence of the diffeomorphism invariance of $F$). They are therefore valid not only for observables, but for all
{\em linear observables}, functionals of the variations $\dg g^{ab}$ of the metric about $g$ of the form 
$\int \theta^{ab}\dg g^{ab} \vareg$, with $\theta^{ab}$ a smooth, compactly supported, symmetric tensor 
field with vanishing divergence.

This will apply quite generally to all our results involving observables: they hold just as well if we replace the set 
of observables with the set of linear observables. Linear observables, which include the linearizations of the full observables as a subset, 
may be thought of as the observables of the linearized theory.

The symplectic form is determined by the action. We shall adopt the Hilbert action,
	\be		\label{EH}
                I = \frac{1}{16\pi G}\int_Q R \vareg,
        \ee
where $Q$ is the chosen domain of integration and $\vareg$ is the metric spacetime volume form.
The sign conventions for the curvature tensor and scalar are those of \cite{Wald}, that is,
$R = R_{ab}{}^{ab}$ with $[\nabla_a,\nabla_b]\bg_c = R_{abc}{}^d \bg_d$ for any 1-form $\bg$.

The variation of the action due to a variation $\dg$ of the metric consists of a bulk term,
which vanishes on solutions, and a boundary term
        \be
    \phi[\dg] =  -\frac{1}{8\pi G}\int_{\di Q} \dg\Gamma^{[c}_{cb} g^{a]b}\vareg_{a\cdot\cdot\cdot}, \label{boundaryEH}
        \ee
where the dots represent uncontracted abstract indices, that is, the integrand is a 3-form. 
Restricting this boundary integral to a portion $\Sg$ of $\di Q$ one obtains the {\em symplectic potential}, 
$\Theta_\Sg [\dg]$, of $\Sg$. We will be interested in the case in which $\Sg$ is the double null sheet $\cN$.

The {\em presymplectic form} on field histories, evaluated on a pair of variations $\dg_1$ and $\dg_2$, is
\bearr
\Omega_{\Sg}[\dg_1,\dg_2] & \equiv & \dg_1\Theta_{\Sg}[\dg_2] - \dg_2\Theta_{\Sg}[\dg_1]
                                - \Theta_{\Sg}[[\dg_1,\dg_2]] \label{Omega_def}\\
			  & = & -\frac{1}{8\pi G} \int_\Sg \delta_2
\Gamma^{[c}_{cb} \delta_1 (g^{a]b}\vareg_{a\cdot\cdot\cdot}) - (1 \leftrightarrow 2)\nonumber \label{presymp0}	
\eearr
\cite{Crnkovic_Witten}. 
$\Omega_{\Sg}$ can be interpreted as the curl (exterior derivative) of $\Theta_\Sg$ in the 
space of metric fields, contracted with two tangent vectors, $\dg_1$ and $\dg_2$, to this space. (See \cite{AMR} 
for the definition of exterior differentiation in the infinite dimensional context.) 

The field histories on which $\Omega_\Sg$ is defined need not satisfy the field equations, nor need the variations 
preserve these field equations.
Restricting $\Omega_\Sg$ to metrics that satisfy the field equation and variations $\dg_1$, $\dg_2$ that satisfy 
the linearized field equations one obtains $\omega_\Sg$, the presymplectic form on the space of solutions. 
Here the space $L_g$ of smooth solutions to the field equations linearized about the metric $g$ plays the role 
played by the tangent space to the phase space in the definition of the presymplectic form in finite dimensional mechanics. 

The degeneracy vectors of the presymplectic form on the space of solutions are variations $\cg$ such that 
$\omega_\Sg[\cdot,\cg] = 0$. These generate the gauge transformations of the system in the domain of dependence of $\Sg$ 
\cite{Lee_Wald}. Lie derivatives, which generate diffeomorphisms, are degeneracy vectors of the presymplectic form 
on the space of solutions of vacuum gravity defined by \ref{presymp0}. Suppose $\dg$ is an arbitrary variation in 
$L_g$ and $\xi$ is a smooth vector field then $\pounds_\xi \in L_g$, because the space of solutions is diffeomorphism 
invariant, and \cite{MR13} 
\be
\omega_\cN[\pounds_\xi,\dg] = \frac{1}{16\pi G} \int_{\di\cN} 3 \xi^{[a}\dg\Gamma^c_{cd} g^{b]d}\veg_{ab\cdot\cdot}
      + \dg[g^{ca}\veg_{ab\cdot\cdot}]\nabla_c \xi^b.    \label{boundary_diff}
\ee
It follows that any $\xi$ that vanishes on $\di\cN$ and has vanishing derivatives there generates gauge diffeomorphisms,
but some other diffeomorphisms, generated by $\xi$ which have non-zero values or derivatives on $\cN$ are not gauge.

The preceding results apply to more or less arbitrary compact hypersurfaces with boundary, $\Sg$, in spacetime, and in particular
to the case that $\Sg$ is a double null sheet $\cN$. 
Equation (\ref{boundary_diff}) therefore demonstrates the equivalence of (\ref{auxbracketdef}) and (\ref{auxbracketdef1}) because 
$\omega_\cN[\pounds_\xi,\dg] = 0$ for any $\dg \in L_g$ if $\xi = 0$ in a neighborhood of $\di\cN$, and for any $\xi$
if $\dg \in L_g^0$.

In standard terminology a {\em symplectic form} is a presymplectic form that has no non-zero degeneracy vectors. But 
of course whether or not it has non-zero degeneracy vectors depends on the variables used to describe the system.  
Since it is always in a sense ``the same'' object the author prefers to 
refer to it from here on as the symplectic form of the system, independently of the variables used and whether or not it has 
degeneracy vectors.

Note that in the calculation of the symplectic form no boundary term has been added to the Hilbert action because 
such a boundary term would not affect the Peierls bracket, since it does not affect the advanced and retarded Green's functions. 
At first sight it also seems that it would not affect the symplectic form either because it contributes only a total 
variation to the boundary term $\phi[\dg]$ in the variation of the action, and thus, apparently, to the symplectic potential. 
However, $\phi[\dg]$ determines its integrand only up to an exact form, and thus the symplectic potential $\Theta_\cN$ only up 
to a $\di\cN$ boundary term. Depending on the prescription chosen for determining the integrand of $\phi[\dg]$ a boundary term in the action 
may or may not contribute a boundary term to $\omega_\cN$ \cite{Lee_Wald}. A change in $\omega_\cN$ could in turn 
affect the $\bullet$ bracket on initial data since this bracket will ultimately be defined by the symplectic form, 
via conditions (\ref{final_bracket_condition1}) and (\ref{final_bracket_condition2}). However, whatever the boundary 
term in the action and whatever the prescription used to determine the integrand of $\phi[\dg]$, the $\bullet$ bracket 
we obtain will reproduce the same Peierls bracket. Since this is all that is required of the $\bullet$ bracket the 
simplest option is adopted: no boundary term is added to the Hilbert action, and the integrand of $\phi[\dg]$
is as indicated in (\ref{boundaryEH}).

The symplectic form is closely related to the Peierls bracket. %, $\{A,B\} \equiv \Dg_A B = [\dg^+_A - \dg^-_A] B$: 
Let $A$ and $B$ be two (full or linear) observables of the interior of the domain of dependence of $\cN$, that is, having 
domains of sensitivty in $D[\cN]^\circ$, and let us choose the domain of integration $Q$ of the action so that 
it is bounded to the past by $\cN$ and to the future by another achronal hypersurface, and contains the domains 
of sensitivity of $A$ and $B$ in its interior. At stationary points of the action $\dg I - \phi[\dg] = 0$ for any variation $\dg$.
(In vacuum GR this is $G_{ab}\dg g^{ab} = 0$, with $G$ the Einstein curvature.) Thus, if a variation $\dg_0$ preserves the 
field equations then 
\be  \label{lin_field}
\dg_0 \dg I = \dg_0 \phi[\dg],
\ee 
which is equivalent to the linearized field equation $\dg_0 G_{ab} = 0$.

Now consider the perturbed action $I + \gamma A$. If there were an exact power series solution $g + \cg \dg_A g + ...$ of the
corresponding field equations, then at the first order aproximate solution $g + \cg\dg_A g$   
\be \label{perturbed_stationarity}
\dg I + \gamma \dg A - \phi[\dg] = o(\gamma)
\ee
for any variation $\dg$, where $o(\gamma)$ vanishes more rapidly than $\gamma$ as $\gamma \rightarrow 0$. 
Note that the boundary term $\phi[\dg]$ in the variation is the same as for the unperturbed action since 
$\dg A/\dg g_{ab}$ vanishes in a neighborhood of $\di Q$.

This equation is taken as the definition of first order solutions, even when no corresponding exact power series solution exists.
It is equivalent to an order zero equation obtained by setting $\cg = 0$, which requires $g$ to be a stationary 
point of the unperturbed action, and an order one equation obtained by taking the derivative in $\cg$ at $\cg = 0$: 
\be  \label{perturbation}
\dg_A \dg I + \dg A - \dg_A \phi[\dg] = 0.
\ee
This is just the linearized field equation with a source term, $[16\pi G]^{-1}\dg_A G_{ab} + \dg A/\dg g^{ab} = 0$.

% Here there is a hypothesys that the derivative of the o(\cg) term vanishes at $\cg = 0$. This is true: [o(\cg) - 0]/\cg -> 0 as \cg -> 0, so o'(0) = 0.

Putting $\dg = \dg_0$ in (\ref{perturbation}) and $\dg = \dg_A$ in (\ref{lin_field}), and subtracting
the two one obtains
\be      \label{inverse0}
\dg_0 A = \dg_A \phi[\dg_0] - \dg_0 \phi[\dg_A] - \phi[[\dg_A,\dg_0]].
\ee 

% Here use achronality of $\cN$.

This equation applies in particular to the retarded and advanced parturbations $\dg^\pm_A$ that appear in the 
definition of the Peierls bracket. But $\dg^+_A$, being retarded, vanishes on $\cN$, while the support
of $\dg^-_A$ on $\di Q$ is contained in $J^-[\sg_A]\cap \di Q$ which lies entirely in the interior of $\cN$
\cite{MR07} (prop. B.21). Consequently $\Omega_{\cN}[\dg^+_A,\dg_0] = 0$ and 
\begin{equation}
\fl  \dg^-_A \phi[\dg_0] - \dg_0 \phi[\dg^-_A] - \phi[[\dg^-_A,\dg_0]]
  = -(\dg^-_A \Theta_{\cN}[\dg_0] - \dg_0 \Theta_{\cN}[\dg^-_A] - \Theta_{\cN}[[\dg^-_A,\dg_0]]) = - \Omega_{\cN}[\dg^-_A,\dg_0],
\end{equation}
where the minus arises because we take $\cN$ to be future oriented, while $\di Q$ is past oriented where it coincides with $\cN$.
Thus (\ref{inverse0}) implies
\be      \label{inverse}
     \dg_0 A = \Omega_{\cN}[\dg^+_A,\dg_0] - \Omega_{\cN}[\dg^-_A,\dg_0] = \omega_\cN[\Dg_A,\dg_0],
\ee
an equation valid for $\dg_0$ any solution to the linearized field equations. Note that by 
(\ref{perturbation}) $\Dg_A = \dg^+_A - \dg^-_A$ also satisfies the linearized field equation (\ref{lin_field}).

We are now in a position to demonstrate that (\ref{auxbracketdef}) is sufficient 
to ensure that the $\bullet$ bracket reproduces the Peierls bracket on observables. 
Suppose $\{\cdot,\cdot\}_\bullet$ is a Poisson bracket on the free null initial data such that for any observable $B$, expressed in terms of
initial data, the variation of initial data $\{B,\cdot\}$ corresponds to a smooth solution of the linearized field equations, which will also 
be denoted $\{B,\cdot\}$. Then (\ref{inverse}), and the antisymmetry of the $\bullet$ Poisson bracket and the symplectic form, imply that
\be\label{bb_omega_dA}
 \{A,B\}_\bullet = - \{B, A\}_\bullet = -\omega_\cN[\Dg_A, \{B,\cdot\}_\bullet] = \omega_\cN[\{B,\cdot\}_\bullet,\Dg_A].
\ee
for all observables $A$ and $B$. 

On the other hand, by definition the Peierls bracket is $\{A,B\} = \Dg_A B$, so the $\bullet$ bracket reproduces the 
Peierls bracket on observables if and only if
\be \label{auxbracketdef0}
\Dg_A B = \omega_\cN[\{B,\cdot\}_\bullet,\Dg_A].
\ee
But $\Dg_A \in L_g^0$, since the support of $\dg^\pm_A$, contained in $J^\pm[\sg_A]$, is disjoint from a neighborhood of 
$\di\cN$ \cite{MR07} (prop. B.21). It follows that (\ref{auxbracketdef}) implies (\ref{auxbracketdef0}).

Note that the preceding argument only involves the linearizations of the observables. The result applies to all linear 
observables which can be expressed in terms of the variations of the initial data about those of the reference solution $g$: 
For any pair of such linear observables (\ref{auxbracketdef}) guarantees that $\{A,B\}_\bullet = \{A,B\}$. 

In fact, all linear observables of $D[\cN]^\circ$ may be expressed in terms of the variations of the initial data.
For $A$ any such linear observable $\Dg_A$ lies in $L_g^0$, so it can be made admissible by a suitable adjustment of gauge, specifically, 
by addition of a diffeomorphism generator that vanishes in a neighborhood of $\di\cN$. (See \ref{a_chart_appendix}.)\footnote{
It is easy to show that the diffeomorphism generator may be chosen to have support only in $J^-[\sg_A]$, so this new, admissible $\Dg_A$
still has support only in the causal domain of influence of $\sg_A$.} 
Let us suppose from here on that this has been done. This of course leaves (\ref{inverse}) unchanged. If $\dg_0$ is also required to be admissible, 
then the right side of (\ref{inverse}) is expressible in terms of the free initial data and their variations. Indeed, (\ref{inverse}) becomes the 
sought for expression for the linear observable $A[\dg] \equiv \dg A$ in terms of the variations of the initial data. It is valid on the set $\cA$ of 
variations of the initial data that correspond to admissible variations, described in detail in \ref{a_chart_appendix}. 

This also allows us to verify that the linear observable $A$ is contained in the space $\Phi$ of 
smeared data. Since $\Dg_A$ is admissible and vanishes in a neighborhood of $\di\cN$ the variation of the initial data it defines belongs 
to $\cB \subset \cA$. It thus satisfies all the smoothness conditions of variations in $\cA$ described in \ref{a_chart_appendix} 
and also vanishes in a neighborhood of $\di\cN$ in $\cN$. It is then straightfoward to check that $\omega_\cN[\Dg_A,\dg]$, for $\dg$ admissible, is 
a smearing of the variations under $\dg$ of the initial data with weighting functions that satisfy the smoothness and boundary conditions that 
define $\Phi$, given in Definition \ref{Phi_def}.

Now let us consider the conditions (\ref{final_bracket_condition1w}) and (\ref{final_bracket_condition2}). 
In the Introduction these were presented as the last in a chain of sufficient conditions for the agreement 
on observables between the $\bullet$ bracket and the Peierls bracket, starting with (\ref{auxbracketdef}),
in order to introduce the ideas one at a time.
% the symplectic form, and also they key difference.
It is not difficult to show that (\ref{final_bracket_condition1w}) and (\ref{final_bracket_condition2}) imply the
other sufficient conditions in the chain, but it is easier still to show directly that they imply the matching of 
$\bullet$ and Peierls brackets.

The key result is that if $\{\cdot,\cdot\}_\bullet$ satisfies (\ref{final_bracket_condition1w}) and (\ref{final_bracket_condition2})
for all $\varphi \in \Phi$ then for all linear observables $A$ 
\be\label{Aflow_DeltaA}
\{A,\cdot\}_\bullet = \Dg_A
\ee
on initial data. In Section \ref{bracket_calc} it is shown that (\ref{final_bracket_condition1w}) and (\ref{final_bracket_condition2}) define
$\{\varphi,\cdot\}_\bullet$ uniquely for all $\varphi \in \Phi$ and thus in particular they define $\{A,\cdot\}_\bullet$ uniquely. On the other hand $\Dg_A$
satisfies the same conditions as $\{A,\cdot\}_\bullet$: $\cB$ is a subset of $\cC$ as can be seen from the definition of $\cC$ in Subsection \ref{ABCPhi}, 
so $\Dg_A$ satisfies(\ref{final_bracket_condition1w}) and (\ref{final_bracket_condition2}) (\ref{final_bracket_condition2}), and by (\ref{inverse}) 
it also satisfies (\ref{final_bracket_condition1w}). This establishes (\ref{Aflow_DeltaA}).

The result (\ref{Aflow_DeltaA}) shows that $\{A,\cdot\}_\bullet \in \cA$, as claimed in the Introduction. It also shows immediately that the $\bullet$ bracket 
reproduces the Peierls bracket on observables: $\{A,B\} \equiv \Dg_A B = \{A,B\}_\bullet$ for all linear observables $A$, $B$ (and thus {\em a fortiori} for all 
full observables).

In Section \ref{bracket_calc} it is also demonstrated that the bracket that satisfies (\ref{final_bracket_condition1w}) and (\ref{final_bracket_condition2})
for all $\varphi \in \Phi$ satisfies (\ref{final_bracket_condition1}) as well for all these $\varphi$. This makes it easy to prove that the $\bullet$
bracket satisfies the Jacobi relations
\begin{equation}\label{Jacobi}
 \{\ag,\{ \bg, \cg \}_\bullet\}_\bullet  + \{\cg,\{ \ag, \bg \}_\bullet\}_\bullet  + \{\bg,\{ \cg, \ag \}_\bullet\}_\bullet = 0\ \ \ \forall\:\ag,\bg,\cg \in \Phi
\end{equation}
In finite dimensional mechanics the Poisson bracket $\{\phi,\theta\}$ is a bilinear on the differentials $d\phi$, $d\theta$ of its arguments
(or equivalently a derivation on each argument) and inverse to the symplectic form, which is closed, and this implies that it satisfies the 
Jacobi relation. The following proposition provides an analogous result for a bracket $\{\cdot,\cdot\}_\bullet$ satisfying 
(\ref{final_bracket_condition1}) and (\ref{final_bracket_condition2}).

\begin{proposition}\label{Jacobi_proof}
If $\{\cdot,\cdot\}_\bullet$ is a derivation on each of its arguments defined on a domain that includes $\Phi$ and also $\{\varphi_1,\varphi_2\}_\bullet$ 
for all $\varphi_1,\varphi_2 \in \Phi$, and
\be                     \label{auxbracketdef2}
  \dg \varphi = \omega_\cN[\{\varphi,\cdot\}_\bullet,\dg]\ \ \ \ \mbox{and}\ \ \ \{\varphi,\cdot\}_\bullet \in \cC\ \ \forall\:\dg \in \cC, \varphi \in \Phi.
\ee
then $\{\cdot,\cdot\}_\bullet$ satisfies the Jacobi relations.
\end{proposition}

{\em Proof}: By (\ref{auxbracketdef2}) $\{\varphi_1,\varphi_2\}_\bullet 
= - \omega_\cN[\{\varphi_1,\cdot\}_\bullet,\{\varphi_2,\cdot\}_\bullet\}]$. On the other hand
(\ref{Omega_def}) restricted to variations in $\cC$ implies that  
$\Cyc{i,j,k} \dg_i\omega_\cN[\dg_j,\dg_k] = \Cyc{i,j,k} \omega_\cN[[\dg_i,\dg_j],\dg_k]$ for 
all $\dg_1,\dg_2,\dg_3 \in \cC$, where $\mbox{Cyc}$ indicates a cyclic sum.
(In differential terms, $\omega_\cN$ is a closed form because it is the curl of $\Theta_\cN$.)
Thus, letting $X = \Cyc{i,j,k} \{\varphi_i,\{ \varphi_j, \varphi_k \}_\bullet\}_\bullet$ and 
$\Dg_i = \{\varphi_i,\cdot\}_\bullet$ for $i = 1,2,3$,
\bearr
X  & = & - \Cyc{i,j,k} \Dg_i \omega_\cN[\Dg_j,\Dg_k]
= - \Cyc{i,j,k} \omega_\cN[[\Dg_i,\Dg_j],\Dg_k]\nonumber \\
& &  =  \Cyc{i,j,k} [\Dg_i,\Dg_j] \varphi_k.
\eearr
But then $X = \Cyc{i,j,k} \{\varphi_i,\{ \varphi_j, \varphi_k \}_\bullet\}_\bullet 
- \{\varphi_j,\{ \varphi_i, \varphi_k \}_\bullet\}_\bullet = 2X$, so $X = 0$, which is 
precisely the Jacobi relation.\QED

Notice that the first equation in the proof, $\{\chi, \varphi\}_\bullet = \omega_\cN[\{\varphi,\cdot\}_\bullet, \{\varphi,\cdot\}_\bullet]$
shows that the $\bullet$ bracket is antisymmetric as a consequence of (\ref{final_bracket_condition1}, \ref{final_bracket_condition2}), so
a biderivation satisfying (\ref{final_bracket_condition1}) and (\ref{final_bracket_condition2}) automatically has all the properties of a 
Poisson bracket. 

\section{The free data}\label{data}

The data will be referred to charts $(v_A, \theta^1, \theta^2)$ constructed on each branch $\cN_A$ of $\cN$
from a common smooth chart $(\theta^1,\theta^2)$ on $S_0$: The coordinates $\theta^1$ and $\theta^2$ 
are extended to $\cN_A$ by holding them constant along the generators, and the coordinate $v_A$ parametrizes 
each generator. Since $\di_{v_A}$ is then tangent to the generators and hence null, and also normal to $\cN_A$
(see \cite{MR13} for a proof), the line element on $\cN_A$ takes the form
\be   \label{line_element}
ds^2 = h_{pq} d\theta^p d\theta^q
\ee
with no $dv$ terms. $v_A$ is taken to be proportional to the square root of $\rho \equiv \sqrt{\det h}$,
the area density in $\theta$ coordinates on two dimensional cross sections of $\cN_A$, and normalized to $1$ at
$S_0$. Thus $\rho(v_A, \theta^1, \theta^2) = \rho_0(\theta^1, \theta^2) v_A^2$, with $\rho_0$ the area
density on $S_0$. The coordinate $v$ will be called the {\em area parameter}.
%\footnote{
%
(The index $A$ specifying the branch $\cN_A$ that a quantity pertains to will often be dropped
when there is little risk of confusion.)

\subsection{Geometrical data}

The free initial data we will use consists of geometrical data and diffeomorphism data. The geometrical data are:
\begin{enumerate}
 \item \label{bulk_data} The {\em conformal 2-metric} $e_{pq} = h_{pq}/\rho$ given as a function of the $v\theta$ chart on 
 $\cN_L$ and $\cN_R$. $e$ is a unit determinant, symmetric, $2\times 2$ matrix, which captures the equivalence class with 
 respect to local rescalings of the induced metric $h_{pq}$ on the branches of $\cN$. It is the only datum which is given 
 on all of $\cN$.

 \item \label{geom_surface_data} Three fields specified on $S_0$ only, as functions of $(\theta^1, \theta^2)$: 
 \begin{itemize}
  \item $\rho_0$,
  \item $\lam = - \ln|n_L\cdot n_R|$,
 \end{itemize}
 and
 \begin{itemize}
  \item the {\em twist}
 \be	\label{twist}
 \tau_p = \frac{n_L\cdot\nabla_p n_R - n_R\cdot\nabla_p n_L}{n_L\cdot n_R}.
 \ee
 \end{itemize}
 Here $n_A = \di_{v_A}$ is the tangent to the generators of $\cN_A$, and inner products ($\cdot$) are taken 
 with respect to the spacetime metric. 
\end{enumerate}

An advantage of using $v \equiv \sqrt{\rho/\rho_0}$ as the parameter along generators is that it makes excluding 
caustics easy. It suffices to require that $e_{pq}$ be non-singular and $v > 0$ on $\cN$. For then 
$h_{pq} = \rho_0 v^2 e_{pq}$ is non-singular on $\cN$. ($\rho_0 > 0$ is a consequence of $S_0$ being spacelike and
$(\theta^1,\theta^2)$ being a good chart.)

But $v$ is not always a good parameter on generators. For instance, it fails in the important special case in which 
$\cN_A$ is a null hyperplane in Minkowski space, because the generators neither converge nor diverge, resulting in 
a $v$ that is constant on each generator. On the other hand, it is a good parameter in several important cases: In 
the case of greatest interest from the point of view of the holographic entropy bound, in which the generators of 
$\cN_A$ are converging everywhere on $S_0$ ($dv_A$ negative along the generators in direction toward $S_A$), the 
vacuum field equations guarantee that $v_A$ continues to decrease right up to the trunction surface $S_A$.
(See \cite{Wald} Section 9.2.) 

The area parameter $v$ is also a good parameter if the generators are all diverging at $S_0$, provided the generators 
are truncated before they stop diverging. Finally, even if the expansion of the generators has a definite sign only
on a patch of $S_0$ one can define a new, smaller double null sheet formed by the generators through this patch, and 
thus define the $\bullet$ bracket on these generators. This is more significant than it might seem because all the  
brackets between data living on different generators are expected to vanish. This expectation comes from the micro-causality 
principle, which requires the bracket between fields at causally unconnected points to commute, and the fact that on 
$\cN$ only points lying on the same generator are causally connected, points on different generators being spacelike 
to each other. (see \cite{MR07} prop. B.7.)
Micro-causality is an important property of commutators in quantum field theory, and the Peierls bracket also satisifies 
it in the sense that only observables with causally connected domains of sensitivity have non-zero Peierls bracket.\footnote{
One can even argue hueristically that the Peierls bracket between our data fields ought to satisfy micro-causality: 
One may use the conformal metric $e$ in the $a$ charts defined as in \ref{a_chart_appendix} as data instead
of $e$ in the $v\theta$ charts, without loss or gain of information, because the transforation between the charts is determined
by the other data. Regarding the data, including $e$ in the $a$ charts, at fixed values of the coordinates they are referred to
as functionals of the spacetime metric one finds that their domains of sensitivity are subsets of the generators that pass through
the point $p \in \cN$ corresponding to the coordinate values. And in the case of the datum $e$ it is contained in the generator segment
from $p$ to the corresponding truncation surface $S_A$. This suffices to show that only those data nominally living a 
causally connected points have causally connected domains of sensitivity. This in turn implies that only data living on the same generator
can have non-zero Peierls brackets. Transforming back to $e_{pq}(v,\theta^1,\theta^2)$ one sees that this datum also satisfies this rule.
But the argument is only hueristic because the data are not observables according to our 
definition, both because they do not have smooth functional derivatives and because they are not fully diffeomorphism invariant, 
making the Peierls bracket, as we have defined it, ambiguous. See \cite{MR07} for more discussion.}
And in fact, the $\bullet$ bracket does satisfy micro-causality on those double null sheets for which it has been calculated, 
that is, those on which $v$ is a good parameter.

It is therefore without too great a loss of generality that we limit attention to double null sheets
consisting of branches on which $v_A$ is everywhere increasing, or everywhere decreasing as one moves away from 
$S_0$ along the generators.

Sachs \cite{Sachs} and Dautcourt \cite{Dautcourt} have argued that a set of data quite similar to our geometrical data is free,
and complete in the sense that any valuation of their data determines a matching solution metric uniquely 
up to diffeomorphisms. In fact it has been proved by Rendall that any smooth Sachs Dautcourt (SD) data 
matches a unique smooth solution {\em in some neighbourhood of $S_0$} \cite{Rendall}, and, by the arguments of
\cite{Sachs} and \cite{Dautcourt} it is a reasonable conjecture that it matches a unique smooth solution 
on all of $\cN$, provided $\cN$ is free of caustics. 

In \cite{MR13} it is demonstrated that if our geometrical data are smooth 
on their domains, with $e$ continuous at $S_0$, then they are equivalent to similarly smooth valuations of 
the SD data, in the sense that a solution matches our data if and only if it matches the 
corresponding SD data. Our geometrical data are thus as free and complete as the SD data.

\subsection{Diffeomorphism data}

In addition to the geometrical data there are diffeomorphism data.
As mentioned earlier, not all diffeomorphisms are gauge in canonical GR on $\cN$. Some infinitesinmal 
diffeomorphisms which are non-trivial at the boundary $\di\cN$ are not degeneracy vectors of the symplectic 
form (see (\ref{boundary_diff})). It should thus come as no surprise that some diffeomorphism data, 
that measure these non-gauge diffeomorphisms, are needed to express the symplectic 2-form. These are
\begin{itemize}
 \item[(iii)] $s_A^k$, $A \in \{L,R\}, k\in\{1,2\}$ given as a function of the $\theta$ chart. $s_A^k(\theta)$ is the position of 
 the end point on $S_A$ of the generator labeled by $(\theta^1,\theta^2)$ in {\em fixed}, smooth coordinates 
 $y^k_A$ on $S_A$. That is, it is the transformation from the $\theta$ chart to the $y_A$ chart.
\end{itemize}
A fixed chart is one that is, so to speak, ``painted on the manifold'', unlike
moving charts like Riemann normal coordinates or our $v\theta$ chart, which
change at a given point when the metric field changes. Recall that a
manifold consists of a set of {\em a priori} identifiable points and an atlas of
charts on these. These charts are the fixed charts. For any given metric field
the moving charts are also identified with elements of this atlas, but the element depends 
on the metric. Fixed charts are important for the concept of the variation of a field, 
since they facilitate the comparison of different valuations of a field at the same point, 
and of course they are necessary for the description of the gauge equivalence classes of 
solutions if not all diffeomorphisms are gauge. See the appendix of \cite{MR13} for a more 
detailed discussion of fixed and moving charts.

The $s_A^k$ are independent of of each other and of the geometrical data: Imagine that the spacetime metric is acted on by a 
diffeomorphism that leaves a neighborhood of $S_0$ fixed. This leaves the geometrical data, given as functions of the 
$v\theta$ charts, unchanged. But the congruences of null geodesics normal to $S_0$ are carried along 
by the diffeomorphism of the metric, so the $s_A^k(\theta)$ can be changed to any desired value by means of such a diffeomorphism. 

The data $s_A$ are not enough to specify all the non-gauge diffeomorphism degrees of freedom. 
But recall that in equation \ref{auxbracketdef}, which assures that the $\bullet$ bracket on the data reproduces the Peierls bracket 
on observables, the variations can be restricted to only admissible ones. 
The non-gauge diffeomorphism generators in this space {\em are} determined by $s_A^k$, the geometric data, and their variations,
as can be seen from the fact that the symplectic form on admissible variations can be expressed entirely in terms of the variations
of $s_A^k$ and of the geometric data \cite{MR13}. (See equations (\ref{A} - \ref{C}).)

The diffeomorphism equivalence class of the solution to the field equations corresponding 
to the data is independent of the diffeomorphism data, and so are, therefore, the diffeomorphism
invariant observables. The condition that the $\bullet$ bracket reproduces the Peierls 
brackets of the observables thus does not involve the diffeomorphism data at all. This means
that (\ref{auxbracketdef}) cannot tell us anything about the brackets of $s^k_A$.
The fact that our calculation nevertheless fixes the brackets of $s^k_A$ with all the data is
due to the fact that we strengthen (\ref{auxbracketdef}) to get a definite solution. 
It also means that the conditions that (\ref{auxbracketdef}) places on the brackets of the geometric
data cannot involve the diffeomorphism data. In fact, under the $\bullet$ bracket we obtain 
the geometrical data form a closed Poisson algebra among themselves: the brackets between 
the geometrical data do not depend on $s^k_A$. This closed subalgebra is the main result derived in
the present paper.

Should the diffeomorphism data then be discarded as irrelevant?
If we take the point of view that the observables, or the geometry, of the interior
of the domain of dependence is all that is ``physical'' then the datum $s^k_A$, and its $\bullet$
bracket are indeed superfluous. Only the geometrical data and their brackets matter. But the
diffeomorphism data ought to be important when a wider context than just the interior of the domain of 
dependence is considered, because they encode non-gauge information about the boundary of $\cN$.
Their role may be analogous to that of the center of mass coordinates in an isolated mechanical system,
which are not gauge but are superfluous to a description of the internal dynamics of the system. 
In fact the author expects that the non-gauge diffeomorphism degrees of freedom will be associated to 
quasi-local charges, such as linear and angular momentum, for $\cN$.

Here the data $s_A^k$ will be retained because this is natural in our formalism. They play a central role in
the calculation of the symplectic form in \cite{MR13} and also in the calculation of the $\bullet$ bracket here,
even though they ultimately play no role in the brackets of the geometrical data.

One more pair of data fields is needed to express the symplectic form on admissible variations, namely
\begin{itemize}
 \item[(iv)] $\dot{v}_A(\theta^1,\theta^2)$, $A = L,R$, the value of $v$ on the truncating surfaces of the
two branches. 
\end{itemize}

Under admissible variations $\dot{v}_A$ can vary, but its variation is determined by those of the other data
because admissible variations maintain $\dot{\rho}_A$, the $y_A$ chart area density on $S_A$, invariant 
(see \ref{a_chart_appendix}), and 
\be \label{barv_from_other data}
\dot{v}_A(\theta) = \sqrt{\frac{\dot{\rho}_A(s_A(\theta))|\det[\di s_A/\di\theta]|}{\rho_0}}.
\ee
It is therefore possible, and in fact natural, to write the symplectic form on admissible variations without 
the appearance of variations of $\dot{v}$.

$\dot{v}_A$, like $s_A^k$, does not affect the spacetime geometry, although it does affect the shape of the 
domain of dependence of $\cN$ in spacetime. It is really another diffeomorphism degree of freedom which, like 
the other data, can be specified freely, but since it is equivalent to $\dot{\rho}_A$, which is frozen, it does 
not vary independently of these under admissible variations. $\bullet$ brackets for $\dot{v}_A$ can be calculated 
from (\ref{barv_from_other data}) and the brackets of the other data, but it is not clear what the significance of 
these brackets is.

\subsection{Summary of free data and smoothness conditions}

These are all our free initial data. To summarize, they consist of 
\begin{itemize}

\item 10 real $C^\infty$ functions, $\rho_0$,
$\lam$, $\tau_p$, $\dot{v}_A$, and $s_A^k$, on a domain $D \in \mathbb{R}^2$
having the topology of a closed disk, with $\rho_0>0$ and $\dot{v}_A > 0$ and $\neq 1$

\item two $C^\infty$, real, symmetric, unit determinant $2 \times 2$ matrix
valued functions ($e_{pq}$ on $\cN_L$ and $\cN_R$) on the domains
$\{(v_a,\theta^1,\theta^2) \in \Real^3 | \theta \in D, \min(1,\dot{v}_A(\theta)) \leq v_A \leq \max(1,\dot{v}_A(\theta))\}$,
$A = L,R$ which match at $v_L = v_R = 1$ (i.e. on $S_0$).

\end{itemize}
Our phase space is the space of valuations of these data. 

The smoothness and continuity conditions on the data, and the inequalities on $\rho_0$ and $\dot{v}$, ``regularity'' 
conditions for short, follow from the assumed properties of the spacetime geometry, of $\cN$ and its generators, 
and of the charts used: $\rho_0(\theta)$ is $> 0$ and smooth because $S_0$ is spacelike and smooth in a smooth spacetime 
geometry, and $\theta$ is a smooth chart on $S_0$. 
The fact that in a smooth geometry a point on a geodesic depends smoothly on the starting point and tangent 
of the geodesic, and on the corresponding affine parameter (\cite{Hawking_Ellis} p. 33) ensures that the induced 
2-metric components $h_{pq}$ on $\cN$ is a smooth function of $\theta^1$, $\theta^2$ and an affine parameter on the 
congruence of generators. It follows that $v = \sqrt{\rho/\rho_{0}}$ is also smooth, which together with the assumed 
non-stationarity of $v$ along the generators along the generators ensures that the remaining geometrical data on $S_0$ 
is smooth. As to $e$, the non-stationarity of $v$ implies that the induced metric is smooth in $(v, \theta^1,\theta^2)$. 
The absence of caustics on $\cN$ implies that the matrix $h_{pq}$ is everywhere invertible, so $\rho > 0$, and 
thus that $e_{pq}$ is smooth. Finally, the smoothness of $\dot{v}_A$ and of $s_A$ follows from that of $S_A$ and of the 
fixed chart $y_A$, and $\dot{v} \neq 1$ and $\dot{v} > 0$ follow from the nonstationarity of $v$ and $\rho > 0$ respectively.
 
Conversely, suppose a model $\cN$ of a double null sheet is chosen in $\Real^4$, satisfying all the smoothness conditions, 
and equipped with smooth charts $\theta$, $y_L$ and $y_R$ on $S_0$, $S_L$ and $S_R$ respectively, and suppose smooth charts 
$(v_A, \theta^1, \theta^2)$ are chosen such that the $\theta^p$ constant curves remain in $\cN_A$ from $S_0$ to $S_A$, and 
$v_A = \dot{v}_A(\theta)$ and $y_A^k = s_A^k(\theta)$ on $S_A$. 
If regular geometrical data are set on $\cN$ according to these coordinates then in any $C^1$ Lorentzian metric spacetime geometry 
matching the data $S_0$ is spacelike because $\rho_0 e_{pq}$ is a positive definite metric, and $\cN$ is a double null 
sheet with the predetermined generators, the constant $\theta$ curves, because the tangents to these are null vectors 
orthogonal to $\cN$ and the vanishing of the torsion of the metric connection implies that their integral curves are geodesics: 
If $f$ is a smooth function on an open domain in spacetime which vanishes on $\cN$ and $k \equiv df$ is non-zero there
then $k^a$ is orthogonal to $\cN$ and thus tangent to the generators, and
\be
0 = 2 k^a\nabla_{[a}\nabla_{b]} f = \nabla_k k_b - k^a \nabla_b k_a = \nabla_k k_b - \frac{1}{2}\nabla_b k^2 = \nabla_k k_b,
\ee
establishing the claim.

It remains to see
whether a spacetime metric matching the data that satisfies the vacuum field equations is necessarily smooth.
In \cite{MR13} it is demonstrated that regular valuations of our geometrical data are equivalent to similarly smooth 
Sachs Dautcourt data and Rendall has proved that any smooth SD data matches a unique smooth 
solution {\em in some neighbourhood of $S_0$} \cite{Rendall}, and, as we said earlier, it is a reasonable conjecture 
that it matches a unique smooth solution on all of $\cN$. If the conjecture is valid then the regularity of the data 
implies the smoothness of the spacetime geometry. Thus, modulo this last conjectural element, a complete correspondence 
has been established between our initial data and a the set of spacetime solution metrics, equipped with double null sheets and
$\theta$ and $y$ charts, modulo certain diffeomerphisms. 

Note that we will not actually need all of this correspondence, in particular not the conjectured exitence of and uniqueness 
of smooth solutons matching the data on all of $\cN$. We are calculating the Poisson brackets of the data

\subsection{$\theta$ gauge diffeomorphisms and the extension of the phase space}

The conditions (\ref{final_bracket_condition1}) and (\ref{final_bracket_condition2}) define a bracket only between 
gauge invariant smeared data, or more precisely, on smeared data that are invariant under any degeneracy vector of 
the restriction of the symplectic form to $\cC$. If $\varphi$ is not gauge invariant in this sense 
(\ref{final_bracket_condition1}, \ref{final_bracket_condition2}) have no solution $\{\varphi,\cdot\}_\bullet$, and 
if it is invariant then $\{\varphi,\cdot\}_\bullet$ is determined only up to the addition of degeneracy vectors, so
$\{\varphi,\chi\}_\bullet$ is defined only if $\chi$ is also gauge invariant.

The geometric data are largely diffeomorphism invariant. They depend on the diffeomorphism equivalence class of the 
spacetime metric and on a point on $S_0$. Admissible diffeomorphisms, that is, diffeomorphisms generated by admissible 
variations, map $\cN$ and its parts $S_0$, $S_L$ and $S_R$ to themselves, and the effect on the geometric data of such 
a diffeomorphism of the metric field is equivalent to a change of the $\theta$ chart. The diffeomorphism data $s_A$ 
are affected by both the action of the diffeomorphism on $S_0$ and by its action on $S_A$, equivalent to an independent 
change of the $y_A$ chart. A change of the $\theta$ chart, suitably restricted on $\di S_0$, can always be produced by 
a spacetime diffeomorphism that vanishes and has zero gradient at $\di\cN$, and therefore is gauge according to (\ref{boundary_diff}). 
Indeed it will turn out that any generator of diffeomorphisms of the $\theta$ chart is
a degeneracy vector of $\omega_\cN$ restricted to $\cC$. Furthermore, these are the only degeneracy vectors, for once 
this gauge freedom is fixed, by fixing the $\theta$ chart to the $y_L$ chart by setting $s_L = id$, the symplectic 
form $\omega_\cN|_{\cC}$ becomes non-degenerate, as is shown by the existence of a Poisson bracket for the data in precisely this gauge. 
See Subsection \ref{theta_gauge_fixing}.

However the gauge fixing $s_L = id$ is awkwardly asymmetric between the two branches. A nice, natural, and symmetric 
gauge fixing of the $\theta$ chart is obtained by setting $e_{pq} = \dg_{pq}$ on $S_0$, i.e. setting the $\theta^p$ to 
be ``isothermal coordinates'', but this leads to Poisson brackets that are non-local in $\theta$.
We adopt a third way: we do not gauge fix the $\theta$ chart, nor eliminate the $\theta$ dependence from the data. Instead 
an extended phase space is used in which the symplectic form is non-degenerate: $\tau$ is replaced by two independent fields, 
\bearr
\mbox{\textsc{t}}_R & = & \rho_0[d\lam - \tau],\\ 
\mbox{\textsc{t}}_L & = & \rho_0[d\lam + \tau],
\eearr
with the original phase space corresponding to the constraint surface defined by 
\be \label{tau_constraint}
0 = 2\rho_0 d\lam - \sct_L - \sct_R.
\ee
This constraint also generates the transformations of the $\theta$ chart, as will be shown in Subsection \ref{kappa_generates_theta_diff}. 

Introducing a constraint is of course a step backward if our aim is to eliminate all constraints and work with completely free data.
But it leads to a formalism in which the two branches of $\cN$ enter symmetrically and are almost independent, and in which the
$\bullet$ bracket is local in $\theta$ in the sense that only data at equal $\theta$, in fact, on the same generator, have non-zero brackets.
Moreover, the somewhat asymetric completely constraint free canonical formalism corresponding to imposing the constraint and the gauge
fixing $s_L = id$ is easily recovered from the extended phase space $\bullet$ bracket, because the Dirac bracket corresponding to this 
constraint and gauge fixing is easily calculated, as is done in Subsection \ref{theta_gauge_fixing}. 

In the remainder of the paper we will mostly use the fields $\tilde{\tau}_{A\,k} = \sct_{A\,p}{[\di s_A/\di \theta]^{-1}}^p_k$ 
in place of $\sct_A$, because this simplifies the statement of the symplectic form somewhat. $\tilde{\tau}_{A\,k}$ is canonically 
conjugate to $s_A^k$.

\subsection{The Beltrami coefficients $\mu$ and $\bar{\mu}$}

Let us complete the presentation of the initial data by defining an alternative representation of the 
conformal 2-metric $e_{pq}$ which we will use extensively. This representation can be obtained by 
expressing the degenerate line element (\ref{line_element}) on $\cN$ in terms the complex coordinate 
$z = \theta^1 + i\theta^2$:
\be             \label{mu_parametrization}
       ds^2 = h_{pq}d\theta^p d\theta^q
		= \rho (1 - \mu\bar{\mu})^{-1}[dz + \mu\, d\bar{z}][d\bar{z} + \bar{\mu}\,dz].
   \ee
with $\mu$ a complex number valued field of modulus less than 1, called the {\em Beltrami coefficient}. 
$\mu$ encodes the two real degrees of freedom of $e_{pq} = h_{pq}/\rho$.
In terms of $\mu$ and $\bar{\mu}$ the $\theta$ chart components of $e$ are
   \be             \label{mu_parametrization2}
       e_{pq} = \frac{1}{1-\mu\bar{\mu}}\left[\begin{array}{cc} (1+\mu)(1+\bar{\mu}) & -i (\mu - \bar{\mu}) \\
                                         -i (\mu - \bar{\mu}) & (1-\mu)(1-\bar{\mu}) \end{array}\right].
   \ee
Conversely
  \be	\label{mu_formula}
\mu = \frac{\bepsilon\: e\: \bepsilon^T}{2 + \mathrm{tr}\, e}
  \ee
where $\bepsilon$ is the row vector $[1,i]$.

Under an orientation preserving transformation of coordinates $\theta \rightarrow \theta'$ the components of $e$, which is a 
tensor density of weight $-1$, transform according to
\be		\label{e_transformation}
e'_{rs} = \frac{\di \theta^p}{\di \theta'^r}\frac{\di \theta^q}{\di \theta'^s}
det \left[ \frac{\di \theta'}{\di \theta} \right] e_{pq},
\ee
and the corresponding transformation of the Beltrami coefficient is 
\be \label{mu_transformation}
\mu \rightarrow \mu' = \frac{\bar{\bg} + \mu\bar{\ag}}{\ag + \mu\bg},
\ee
with $\ag = \frac{\di z}{\di z'}$ and $\bg = \frac{\di \bar{z}}{\di z'}$.

Although the metric is real in models of the real world, complex metrics will play a
role in the present work. The parametrization (\ref{mu_parametrization}, \ref{mu_parametrization2})
of $e_{pq}$ still works when $e_{pq}$ is complex, with $\mu$ given by (\ref{mu_formula}) and
$\bar{\mu}$ by the same formula with $\bar{\bepsilon} = [1,-i]$ in place of $\bepsilon$, but if $e$ is
not real $\mu$ and $\bar{\mu}$ are not conjugate. That is, $\bar{\mu} = \mu^*$ if and only if $e_{pq}$
is real, where the superscript $*$ indicates complex conjugation.

The conformal metric $e$ is not real when $\bar{\mu} \neq \mu^*$, but it is still equal to its {\em formal
complex conjugate}. The formal complex conjugate of a function (or functional) $f$ of $\mu$ and $\bar{\mu}$
is
\be\label{functional_cc}
\bar{f}(\mu,\bar{\mu}) \equiv [f(\bar{\mu}^*,\mu^*)]^*.
\ee
If $f$ is a polynomial then formal complex conjugation replaces the coefficients by their complex
conjugates and excahnges $\mu$ and $\bar{\mu}$. This amounts to taking the complex
conjugate ``as if'' $\bar{\mu}$ were the conjugate of $\mu$ - hence the name. Note that, unlike the ordinary complex
conjugate $f^*$, the formal complex conjugate at a particular point $(\mu,\bar{\mu})$ is not a
function of the value of $f$ at the same point but rather depends on the value of $f$ at a, generally,
different point $(\bar{\mu}^*,\mu^*)$.

Clearly $\bar{f} = f^*$ when $\bar{\mu} = \mu^*$, so for instance the fact that $\bar{e}_{pq} = e_{pq}$
implies that $e_{pq}$ is real when $\bar{\mu} = \mu^*$. This statement has a partial converse:
If $f$ is a real valued, real analytic function of $R = \Re \mu$ and $I = \Im \mu$ on some range of 
values of these variables then there is a unique analytic extension of this function to complex $R$ and $I$,
or equivalently, independent variables $\mu = R + iI$ and $\bar{\mu} = R - iI$, and this extension is equal 
to its formal complex conjugate. 
The expression (\ref{mu_parametrization2}) provides such an extension, analytic for all $(\mu, \bar\mu) \in \mathbb{C}^2$
save where $\bar{\mu} = 1/\mu$, of the real conformal 2-metrics $e_{pq}$ as a function of $\mu$.

Note that equation (\ref{mu_formula}) and the invariance of $e$ under functional complex conjugation 
implies that the formal complex conjugate of (\ref{mu_transformation}) provides the correct 
transformation law for $\bar{\mu}$ under transformations of the $\theta$ coordinates.

\subsection{The spaces $\cA$, $\cB$ and $\cC$ of variations of the data, and the space $\Phi$ of smeared data}\label{ABCPhi}

In (\ref{auxbracketdef1}) the condition, $\dg A = \omega_\cN[\{A,\cdot\}_\bullet,\dg]$, is imposed for all 
$\dg$ in $\cB$, the set of variations of the initial data corresponding to admissible variations in $L_g^0$. 
The set of variations of the initial data corresponding to admissible variations, $\cA$, is described
in Definition \ref{cA_def} of \ref{a_chart_appendix}. The variations of all data living on $S_0$ are real 
and smooth, $\dg \mu$ and $\dg\bar{\mu}$ are complex conjugates and smooth on $\cN_L$ and $\cN_R$ (and thus 
continuous at $S_0$). Finally, $\dg s_A = 0$ on $\di S_0$ and $\dg\dot{\rho}_A$ vanishes on $S_A$ on both 
branches, so that $\dg \dot{v}_A$ is determined by the variations of $\rho_0$ and $s_A$.

That $\dg$ lies in $L_g^0$ places many further restrictions on the variations of the data. $\dg \in L_g^0$ 
requires that $\dg g_{ab} = 0$ in a spacetime neighborhood of $\di\cN$, so the $a_A$ charts are fixed 
in neighborhoods of $\di\cN_A - S_0$, and the variations of the $a_A$ chart components $e_{ij}$ of the 
conformal 2-metric must also vanish in neighborhoods of $\di\cN_A - S_0$ in $\cN_A$. Since $\theta$ is 
a fixed chart on $S_0$ the variations of $\rho_0$, $\lam$, $\tilde{\tau}_L$, $\tilde{\tau}_L$, $s_L$, 
and $s_R$ must vanish in a neigborhood of $\di S_0$ in $S_0$. 
These conditions are necessary but not sufficient to ensure that $\dg g_{ab} = 0$ in a {\em spacetime} 
neighborhood of $\di\cN$, i.e. also off $\cN$. This places an infinity of further conditions on the variations 
of the data in $\cB$ which we will not work out here.

The set of variations $\cC$ is larger, and more easily characterized, than $\cB$. 
\begin{definition}
$\cC$ is the set of complex variations of the initial data such that:
\begin{itemize}
 \item[1.] The variations of the data $\rho_0$, $\lam$, $\tilde{\tau}_L$, $\tilde{\tau}_L$, $s_L$, 
 and $s_R$ set on $S_0$ are smooth on $S_0$. $\dg\bar{\mu}$ is smooth on $\cN_L$ and $\cN_R$.
 $\dg\mu$ is smooth on $\cN - S_0$ with possible jump discontinuities at $S_0$: On each branch $\dg\mu$ is equal 
 except at $S_0$ to a function that has a smooth extension to all of $\cN_A$. On $S_0$ itself $\dg\mu$ is smooth
 and has smooth limiting values from $\cN_L$ or $\cN_R$, but none of the three need agree.
 \item[2.] On the truncation surface $S_A$ of each branch the variations of both $\dot{\mu}_A$, the Beltrami 
 coefficient corresponding to the complex chart $y^1_A + i y^2_A$, and $\dot{\rho}_A$, the $y_A$ chart 
 area density, vanish.
 \item[3.] $\dg s_A = 0$ on $\di S_0$.
\end{itemize}
\end{definition}
Note that $\dot{\bar{\mu}}_A$ is {\em not} required to be invariant on $S_A$. 

The space $\Phi$ of smeared data for which we will solve (\ref{final_bracket_condition1}) and 
(\ref{final_bracket_condition2}) is defined as follows.
\begin{definition}\label{Phi_def}
A smeared datum $\varphi \in \Phi$ is a sum of integrals of the free null initial data (not including $\dot{v}_A$))
over their domains weighted by weighting functions which satisfy the conditions: 
\begin{itemize}
 \item[1.] The weighting functions of $\mu$ and $\bar{\mu}$ on $\cN_L$, $\cN_R$, and $S_0$ are smooth on each of these domains, 
 but are independent of each other and need not satisfy any boundary conditions.
 \item[2.] The weighting functions of all the data living on $S_0$ are also required to be smooth, and 
 are not subject to boundary conditions except in the case of the weighting functions $f_A^k$ of $\tilde{\tau}_{A\,k}$, 
 which vanish on $\di S_0$.
\end{itemize}
\end{definition}

As we saw in Section \ref{definition_bracket_on_data} this set $\Phi$ includes the linearizations about $g$ of all the observables.

The particular spaces $\cC$ and $\Phi$ adopted were found partly by trial and error. $\cC$ and $\Phi$ should 
be large enough so that (\ref{final_bracket_condition1}, \ref{final_bracket_condition2}) implies (\ref{auxbracketdef1})
so $\cC$ should contain $\cB$, and $\Phi$ should contain the linearized observables.
In addition it was required that the weighting functions in $\Phi$ include at least all $C^\infty$ test functions that are 
compactly supported in the interior of the domain in $\cN$ on which the corresponding datum is set, to ensure that the $\bullet$ 
brackets of all the data are defined at least as distributions. In order that (\ref{final_bracket_condition1}, 
\ref{final_bracket_condition2}) have a solution it is then necessary that the set $\omega_\cN[\cC,\cdot]$ of linear functions on $\cC$
obtained by acting with $\omega_\cN$ on $\cC$, contains the data smeared with these test functions.
Because of the form of $\omega_\cN$ this requires that the variations in $\cC$ satisfy boundary conditions.
On the other hand $\cC$ cannot be too small because it must contain $\{\Phi, \cdot\}_\bullet$. These requierments seem to
oblige us to use our complex space of variations $\cC$. But at present this is not proved.

\section{The symplectic form in terms of free null initial data}\label{omega_free_data}

In \cite{MR13} the symplectic form corresponding to the Hilbert action (without boundary term) 
was evaluated on admissible variations in terms of the free initial data 
and their variations. The result is a sum of contributions, $\omega_L$ and $\omega_R$, from 
the two branches of $\cN$, each of which is in turn a sum of three terms:
\be   \label{omega_free_data1}
\omega_A = \frac{1}{16\pi G} (A_A + B_A + C_A)
\ee
with
\bearr
A_A[\dg_1,\dg_2] & = & \int_{S_0}
\dg_1\lam\, \dg_2\rho_0 + \dg_1\tilde{\tau}_{A\,k}\,\dg_2 s_A^k\: d^2\theta
- (1 \leftrightarrow 2),\label{A}\\
B_A[\dg_1,\dg_2] & = & \frac{1}{4}\int_{S_0}
\dg_1^y \rho_0\, \di_v e_{pq}\, \dg^y_2 e^{pq}\:d^2\theta - (1 \leftrightarrow 2), \label{B}\\
C_A[\dg_1,\dg_2] & = & \frac{1}{2}\int_{S_0} \int_1^{\dot{v}} v^2
\dg^\circ_1 e^{pq}\,\di_v \dg^\circ_2 e_{pq}\: dv\: \rho_0 d^2\theta
- (1 \leftrightarrow 2). \label{C}
\eearr
Here $\dg^y = \dg - {\pounds}_{\xi_\perp}$, where $\xi_\perp
= \dg s_A^k \di_{y_A^k}$ and the partial derivative $\di_{y_A^k}$ is taken at
constant $v$, and
\be	\label{dg_circ_e_def}
\dg^\circ e = \dg^y e - \frac{1}{2}\dg^y \ln \rho_0\, v \di_v e.
\ee

The data in (\ref{A} - \ref{C}) are all represented by their components in the chart 
$(v, \theta^1,\theta^2)$ and expressed as functions of this chart. The variations $\dg$, $\dg^y$, and
$\dg^\circ$ have the following significance: If $f$ is a data field then $\dg f(v, \theta^1,\theta^2) = 
\dg [f(v, \theta^1,\theta^2)]$, that is, the value of $\dg f$ at $(v, \theta^1,\theta^2)$ is the variation 
$\dg$ of $f(v, \theta^1,\theta^2)$. Put another way, the variation $\dg f$ is calculated holding the coordinates
$(v, \theta^1,\theta^2)$ fixed. The variation $\dg^y f(v, \theta^1, \theta^2)$, on the other hand, is obtained 
by transforming $f$ to the chart $(v,y^1,y^2)$, applying the variation $\dg$ to $f$ holding these coordinates 
fixed, and then transforming the variation back to the chart $(v,\theta^1,\theta^2)$. Finally $\dg^\circ f$ 
is obtained by transforming to the chart $(v/{\dot{v}}, y^1,y^2)$,
varying, and transforming back. ($(v/{\dot{v}}, y^1, y^2)$ is the restriction of the $a_A$ chart to $\cN_A$. For this reason $\dg^\circ$ is denoted $\dg^a$ in \cite{MR13}.)

The diferences between these variations arise because the transition functions between the charts
are field dependent, and thus vary along with the data. See \cite{MR13} for a detailed discussion.

\subsection{The constraint $\kappa$ generates $\theta$ diffeomorphisms}\label{kappa_generates_theta_diff}

Now let us verify the claim that the constraint (\ref{tau_constraint}) generates diffeomorphisms 
of the $\theta$ chart. Let $\kappa[\eta]$ be the constraint smeared with a smooth (data independent) 
vector field $\eta$ tangent to $S_0$, and to $\di S_0$ at $\di S_0$:
\begin{equation}\label{theta_diffeo_generator}
\kappa[\eta] \equiv \frac{1}{16\pi G}\int_{S_0} \eta^p [2\rho_0 \di_p \lam
- \tilde{\tau}_{R\,i}\di_p s_R^i - \tilde{\tau}_{L\,m}\di_p s_L^m]\, d^2\theta \label{theta_diff_gen}
\end{equation}
The claim is that $\{\kappa[\eta], f\}_\bullet = \pounds_\eta f$ for all data $f$, where the Lie derivative
acts only on the $\theta$ dependence of the data; or equivalently, that 
\be\label{kappa_action}
\{\kappa[\eta], \varphi\}_\bullet = \pounds_\eta \varphi
\ee
for all smeared data $\varphi$, where $\pounds_\eta$ still acts acts only on the $\theta$ dependence of the data 
and not on the weighting functions against which they are smeared.

To establish this we first show that $\dg\kappa[\eta] = - \omega_{\cN}[\dg,\pounds_\eta]$ for any variation $\dg \in \cC$. 
(Note that $\dg$ acts only on the data and not on the vector field $\eta$.) We start with the first term in the integrand of 
(\ref{theta_diffeo_generator}). Since $\lam$ is a scalar field $\eta^p \rho_0 \di_p \lam = \rho_0 \pounds_\eta \lam$ 
so the variation of this term is
\be
 \dg[\rho_0 \pounds_\eta \lam] = \dg\rho_0 \pounds_\eta \lam + \rho_0 \pounds_\eta \dg\lam 
 = \pounds_\eta [\rho_0 \dg \lam] + \dg\rho_0 \pounds_\eta \lam - \pounds_\eta\rho_0 \dg\lam.
\ee 
% \begin{eqnarray}
%  \dg[\rho_0 \pounds_\eta \lam] & = & \dg\rho_0 \pounds_\eta \lam + \rho_0 \pounds_\eta \dg\lam \nonumber\\ 
%  & = & \pounds_\eta [\rho_0 \dg \lam] + \dg\rho_0 \pounds_\eta \lam - \pounds_\eta\rho_0 \dg\lam.
% \end{eqnarray}

Because $\eta$ is tangent to $\di S_0$ the integral over $S_0$ of $\pounds_\eta [\rho_0 \dg \lam]$, the $\eta$ Lie derivative 
of a scalar density, vanishes. The integral of the variation of the first term in $\kappa[\eta]$ is thus 
\be
\frac{1}{8\pi G}\int_{S_0}  \dg\rho_0 \pounds_\eta \lam - \pounds_\eta\rho_0 \dg\lam\, d^2\theta,
\ee
which is a term in $-\omega_{\cN}[\dg,\pounds_\eta]$.  
Since $s_R^i$ and $s_L^m$ transform as scalar fields like $\lam$, and the $\tilde{\tau}_{A\,k}$ transform as scalar densities like $\rho_0$, 
under diffeomorphisms of the $\theta$ chart, the second and third terms in the integrand can be 
treated in the same way, and we conclude that
\begin{equation}
 \dg\kappa[\eta] = - \frac{1}{16\pi G} (A_R[\dg,\pounds_\eta] +  A_L[\dg,\pounds_\eta]) 
\end{equation}

The remaining, $B$ and $C$, terms of $\omega_{\cN}[\dg,\pounds_\eta]$ must vanish because they can be expressed in terms of fields that 
do not depend on the $\theta$ chart \cite{MR13}. Indeed, they are zero since the $y$ projection of the variation $\pounds_\eta$, 
$\pounds^y_\eta \equiv \pounds_\eta -\pounds_{\xi_\perp}$, vanishes because the vector field $\xi_\perp$ corresponding to the variation 
$\pounds_\eta$ is $\pounds_\eta s_A^k \di_{y_A^k} = \eta$. Thus, as promised,
\begin{equation}
 \dg\kappa[\eta] = - \omega_{\cN}[\dg,\pounds_\eta] 
\end{equation}
for all $\dg \in \cC$. 

Applying this result to $\dg = \{\varphi,\cdot\}_\bullet$ (which lies in $\cC$ by (\ref{final_bracket_condition2})) 
one finds by (\ref{final_bracket_condition1}) that
\begin{equation}
 \{\varphi,\kappa[\eta]\}_\bullet = \omega_{\cN}[\{\varphi,\cdot\}_\bullet,\pounds_\eta] = - \pounds_\eta \varphi.
\end{equation}
The constraint (\ref{tau_constraint}) indeed generates diffeomorphisms of the $\theta$ chart.

In the last equality the fact that $\pounds_\eta$ lies in $\cC$ has been used. Clearly $\pounds_\eta\mu$ and $\pounds_\eta\bar{\mu}$
satisfy the smoothness requierments of variations in $\cC$, and the boundary condition $\pounds_\eta \dot{\mu}_A = 0$ is met because 
$\dot{\mu}_A$ does not depend on the $\theta$ chart.

\subsection{The symplectic form in terms of the Beltrami coefficients $\mu$ and $\bar{\mu}$}

It will be convenient to express 
$\omega_\cN$ in terms of the Beltrami coefficients $\mu$ and $\bar{\mu}$ in place of the 
conformal 2-metric $e_{pq}$. A direct calculation using the expression (\ref{mu_parametrization2}) 
for $e_{pq}$ shows that for any variation $\Dg$
\be     \label{Dg_e_sq}
                \Dg e_{pq} \Dg e^{pq} = -8 \frac{1}{(1 - \mu\bar{\mu})^2}\Dg \mu \Dg \bar{\mu}.
\ee
(This is valid even when $\bar{\mu}$ is not the complex conjugate of
$\mu$ and the two are varied independently.) Substituting $\Dg_1 + a \Dg_2$ for $\Dg$ in 
(\ref{Dg_e_sq}) and taking the part linear in the arbitrary coefficient $a$ one obtains
        \be     \label{Dg_1_Dg_2}
                \Dg_1 e_{pq} \Dg_2 e^{pq} = -4 \frac{1}{(1 - \mu\bar{\mu})^2}
                \{\Dg_1 \mu \Dg_2 \bar{\mu} + \Dg_1 \bar{\mu} \Dg_2 \mu \}.       
        \ee
In particular, when $\Dg_1 = \di_v$ and $\Dg_2 = \dg$
        \be     \label{scndterm}
                \di_v e_{pq} \dg e^{pq} = -4 \frac{1}{(1 - \mu\bar{\mu})^2}
                \{\di_v \mu \dg \bar{\mu} + \di_v \bar{\mu} \dg \mu \}.       
        \ee
(\ref{B}) may thus be rewritten as 
\bearr
B_A[\dg_1,\dg_2] & = &\int_{S_0} \frac{1}{(1 - \mu\bar{\mu})^2} 
\{\di_{v} \mu \dg^y_1 \bar{\mu} + \di_v \bar{\mu} \dg^y_1 \mu \}\: \dg_2^y \rho_0\: d^2\theta
 - (1 \leftrightarrow 2). \label{Bmu}
\eearr
Here $v$ and $y$ are of course the coordinates $v_A$ and $y_A$, associated with $\cN_A$, a fact which will be important further on.

To obtain an expression for the integrand in $C_A[\dg_1, \dg_2]$ one replaces $\dg$ with 
$\dg_1^\circ$ in (\ref{scndterm}), acts with $\dg_2^\circ$ on the resulting expression, and 
antisymmetrizes with respect to interchange of $\dg_1^\circ$ and $\dg_2^\circ$. In this way one 
obtains
 \bearr
\fl\delta^\circ_1 e^{pq} \di_v \delta^\circ_2 e_{pq}
- (1 \leftrightarrow 2)	 = & -4\frac{1}{(1 - \mu\bar{\mu})^2}\{\delta^\circ_1 \mu \di_v \delta^\circ_2 \bar{\mu}
		+ \delta^\circ_1 \bar{\mu} \di_v \delta^\circ_2 \mu \}	\nonumber\\
& -8\frac{1}{(1 - \mu\bar{\mu})^3}[\mu\di_v \bar{\mu} - \bar{\mu}\di_v \mu]
                \delta^\circ_1\mu\delta^\circ_2 \bar{\mu} - (1 \leftrightarrow 2). 
\label{mu_integrand1}
		\eearr
To derive this expression we do not need to assume that the commutators $[\dg_2^\circ, \di_v]$ and 
$[\dg_2^\circ, \dg_1^\circ]$ vanish. The fact that they are themselves variations, that is, derivatives 
along one parameter families of field configurations\footnote{ 
The validity of (\ref{Dg_1_Dg_2}) requires only that $\Dg_1$ and $\Dg_2$ be derivations. That is, that they be linear 
in their arguments and that they satisfy the Leibniz product rule $\Dg_i(fg) = f\Dg_i g + g\Dg_i f$. 
It is easily checked that the commutator of two derivations, $\Dg_1$ and $\Dg_2$, is a derivation: 
\be
[\Dg_1,\Dg_2] (fg) = \Dg_1(f\Dg_2 g + g\Dg_2 f) - (1 \leftrightarrow 2) = f[\Dg_1,\Dg_2]  g + g [\Dg_1,\Dg_2]f.
\ee 
}
suffices. This implies that (\ref{Dg_1_Dg_2}) holds when $\Dg_1$ is a commutator, ensuring that the commutator 
terms cancel in the calculation. 

Equation (\ref{mu_integrand1}) can be cast in a simpler form. Let         
\be   \label{alpha_def}
                \ag(v, \bftheta) = \int_{1}^v  \frac{1}{1 - \mu\bar{\mu}}
                                        [\bar{\mu}\di_{v'} \mu - \mu\di_{v'} \bar{\mu}]\ dv',
        \ee
where the integral is taken along the generator of $\cN_A$ identified by $\bftheta$. 
Then
        \bearr
              \fl  \lefteqn{ \delta^\circ_1 e^{pq} \di_v \delta^\circ_2 e_{pq} 
- (1 \leftrightarrow 2)			}\nonumber \\ 
  =  -4\left\{\frac{e^\ag\delta^\circ_1 \mu}{1 - \mu\bar{\mu}} 
	\di_v \left[\frac{e^{-\ag}\delta^\circ_2 \bar{\mu}}{1 - \mu\bar{\mu}}\right]
      + \frac{e^{-\ag}\delta^\circ_1 \bar{\mu}}{1 - \mu\bar{\mu}} 
	\di_v \left[\frac{e^\ag\delta^\circ_2 \mu}{1 - \mu\bar{\mu}}\right]\right\}
%\nonumber\\ &&        
      - (1 \leftrightarrow 2).
        \eearr
(Note that when the metric is real, so that $\mu$ and $\bar{\mu}$ are complex conjugates, $\ag$ is pure imaginary, 
and consequently $e^\ag$ is a phase). 

This invites us to define
        \be		\label{Box_def}
                \Box = \frac{\sqrt{\rho_0} v e^\ag}{1 - \mu\bar{\mu}}\dg^\circ \mu\ \ \ \
 \ \tilbox = \frac{\sqrt{\rho_0} v e^{-\ag}}{1 - \mu\bar{\mu}}\dg^\circ \bar{\mu}.
        \ee
Then
        \be	\label{C_Box}
C_A[\dg_1,\dg_2] = - 2\int_{S_0} \int_1^{\dot{v}} 
		\Box_1 \di_v \tilbox_2 + \tilbox_1 \di_v \Box_2\: dv d^2\theta 
	    - (1 \leftrightarrow 2),        
        \ee
where the $v$ coordinate and the modified variation $\dg^\circ$ associated with
$\cN_A$ are used. Note that as a consequence of the antisymmetrization in $\dg_1$ and $\dg_2$ moving a factor 
$\sqrt{\rho_0} v$ inside the $v$ derivatives does not affect the result.

$\tilbox$ is {\em almost} the formal complex conjugate of $\Box$ (obtained 
by substituting $\mu$ for $\bar{\mu}$ and {\em vice versa}, and replacing all other quantities
by their complex conjugates). Only the variation $\dg$ itself is not conjugated in 
$\tilbox$. The conjugate variation, $\bar{\dg}$, is defined by $\bar{\dg} \bg = \overline{\dg \bar{\bg}}$
for any $\bg$. An example of a variation that turns out not to be self-conjugate 
is $\dg = \{\mu, \cdot\}_\bullet$, so the difference between $\tilbox$ and $\bar{\Box}$ is important. 

Integrating (\ref{C_Box}) by parts gives an expression for $C_A$ containing no derivatives of $\dg_1 \mu$ or $\dg_2 \mu$:
        \be	\label{C_Box2}
C_A[\dg_1,\dg_2] = 2 \int_{S_0} d^2\theta \left\{\tilbox_1 \Box_2 - \tilbox_2 \Box_1  
	+ 2\int_1^{\dot{v}} 
	\Box_2 \di_v \tilbox_1 - \Box_1 \di_v \tilbox_2\: dv \right\}
        \ee
for admissible variations $\dg_1,\dg_2 \in \cC$. The integration by parts generates a boundary term on $S_0$, but
the corresponding boundary term at the truncation surface $S_A$ is absent because $\dg \dot{\mu}_A = 0$ for variations 
$\dg \in \cC$, and this implies that $\dg^\circ \mu = 0$ on $S_A$: Recall that $\dg^\circ \mu$ can be calculated by 
transforming $\mu$ from the $v\theta$ chart to the chart $(v/\dot{v},y^1,y^2)$, varying with $\dg$ at fixed values 
of these coordinates, and then transforming back to the $v\theta$ chart. The first transformation maps $\mu$,
the Beltrami coefficient for the coordinate $z = \theta^1 + i\theta^2$, to $\dot{\mu}_A$, the corresponding
Beltrami coefficient for the coordinate $z_A = y^1 + i y^2$.  
By (\ref{mu_transformation}),
\be	\label{mu_mu_A_transf}
\mu = \frac{\bar{b} + \dot{\mu}_A \bar{a}}{a + \dot{\mu}_A b},
\ee
where $a = \frac{\di z_A}{\di z}$ and $b = \frac{\di\bar{z}_A}{\di z}$. Since $S_A$ is the surface $v/\dot{v} = 1$
the variation of $\dot{\mu}_A$ at constant $v/\dot{v} = 1$ and $y^1, y^2$ is just $\dg \dot{\mu}_A(y^1,y^2)$ on $S_A$, which vanishes. 
The image $\dg^\circ \mu$ of this variation in the $v\theta$ chart is the variation of $\mu$ due to the variation of $\dot{\mu}_A$
when the transition function between the charts is held fixed. Of course this vanishes when $\dg\dot{\mu}_A$ vanishes. Explicitly,
it follows from (\ref{mu_mu_A_transf}) that 
\be  \label{transformed_dmu}
\dg^\circ \mu = \frac{a\bar{a} - b\bar{b}}{(a + \dot{\mu}_A b)^2}\dg \dot{\mu}_A.
\ee
This expression can also be obtained directly from the definition (\ref{dg_circ_e_def}) of $\dg^\circ e$, without recourse 
to the preceding argument.

One advantage of (\ref{C_Box2}) is that it incorporates the boundary condition $\dg \dot{\mu}_A = 0$. The other advantage is that it  
extends straightforwardly to all of $\cC$. The expression (\ref{C_Box}) for $C_A$ is well defined on 
admissible variations, but it is ambiguous on $\cC$ because $\cC$ includes variations in which $\dg \mu$ has jump discontinuities at $S_0$ 
and these give rise to Dirac delta distributions in the integrand that are supported at the boundary of the domain of integration.
(\ref{C_Box2}) will be adopted as the definition of $C_A$ in (\ref{final_bracket_condition1}).

[ Note that since $\dg \mu$ is defined on $S_0$ for all $\dg \in \cC$ there is actually a natural disambiguation of (\ref{C_Box}), making it 
equivalent to (\ref{C_Box2}) on all of $\cC$: For each branch $\cN_A$ the Dirac delta in $\di_v \Box$ coming from the jump in $\dg\mu$ from $S_0$ to 
$\cN_A-S_0$ is counted as lying entirely within $\cN_A$.] 

\section{The $\bullet$ Poisson brackets of the free null initial data}\label{structure_relations}

Conditions (\ref{final_bracket_condition1}) and 
(\ref{final_bracket_condition2}) determine the Poisson brackets of the initial data
$\tilde{\tau}_{R\,i}$, $\tilde{\tau}_{L\,m}$, $s_R^j$, $s_L^n$, $\lam$, $\rho_0$, $\mu$, and 
$\bar{\mu}$ uniquely as two point distributions on $\cN$ and on $S_0$. The brackets obtained
are as follows:
\bearr
\{\rho_0(\bftheta_1),\lam(\bftheta_2)\}_\bullet 
 =  8\pi G \dg^2(\bftheta_2 - \bftheta_1),	
\label{lam_rho_bracket}\\
\{s_R^i(\bftheta_1),\tilde{\tau}_{R\,j}(\bftheta_2)\}_\bullet 
 =  16\pi G \dg_j^i\dg^2(\bftheta_2 - \bftheta_1),	
\label{omega_s_R_bracket}\\
\{s_L^m(\bftheta_1),\tilde{\tau}_{L\,n}(\bftheta_2)\}_\bullet 
 =  16\pi G \dg_n^m\dg^2(\bftheta_2 - \bftheta_1),	
\label{omega_s_L_bracket}
\eearr
where $\bftheta$ denotes the pair of coordinates $(\theta^1, \theta^2)$.
$s_R$, $s_L$ and $\rho_0$ commute (that is, they have vanishing $\bullet$ brackets) with 
everything except $\tilde{\tau}_R$, $\tilde{\tau}_L$, and $\lam$ respectively. 

The brackets between $\tilde{\tau}_R$, $\tilde{\tau}_L$, and $\lam$ are 
\begin{eqnarray}
\fl\{\lam(\bftheta_1),\lam(\bftheta_2)\}_\bullet = 0 \label{lam_lam_brack}\\
\fl\{\lam(\bftheta),\tau_R[f]\}_\bullet = 8\pi G [\frac{1}{(1 - \mu\bar{\mu})^2}
(\di_{v_R} \bar{\mu} - \di_{v_L} \bar{\mu}) {\pounds}_f \mu]_{\bftheta} 
\label{lam_omega_R_bracket}\\
\fl \{\tau_R[f_1],\tau_R[f_2]\}_\bullet = - 16\pi G
\int_{S_0} \frac{1}{(1 - \mu\bar{\mu})^2}{\pounds}_{f_1} \mu 
(\rho_0 {\pounds}_{f_2}\bar{\mu} - {\pounds}_{f_2}\rho_0\di_{v_R}\bar{\mu})\: d^2\theta - (1 \leftrightarrow 2)
\label{omega_omega_R_bracket} \\
\fl\{\tau_R[f],\tau_L[g]\}_\bullet = 16\pi G \int_{S_0} \frac{1}{(1 - \mu\bar{\mu})^2}
{\pounds}_f \mu(\rho_0 {\pounds}_g\bar{\mu} - {\pounds}_g\rho_0\di_{v_L}\bar{\mu})\: d^2\theta
- (f,R \leftrightarrow g,L), \label{omega_R_omega_L_bracket}
\end{eqnarray}
where $\tau_R[f] = \int_{S_0} \tilde{\tau}_{R\,i} f^i d^2\theta$ with $f^i(\bftheta)$ functions independent of the data that vanish on $\di S_0$, 
and $\tau_L[g]$ is defined similarly in terms of weighting functions $g^m$. $f^i$ and $g^m$ define 
vector fields $f = f^i \di_{y_R^i}$ and $g = g^m \di_{y_L^m}$, and thus Lie derivatives. 
Note that the Lie derivative of the Beltrami coefficient, determined by its transformation law
under diffeomorphisms (\ref{mu_transformation}), is 
\be\label{mu_Lie}
{\pounds}_f \mu = f^z \di_z \mu + f^{\bar{z}} \di_{\bar{z}} \mu
- \mu [\di_z f^z - \di_{\bar{z}}f^{\bar{z}}] - \mu^2 \di_z f^{\bar{z}} + \di_{\bar{z}} f^z.
\ee
In the $\bullet$ brackets given here this Lie derivative of $\mu$ will always be taken at constant $v$, 
both on $S_0$, in which case it just means that the derivative is taken tangent to $S_0$, and off $S_0$
as in (\ref{mu_omegaR_bracket}).

The symmetry between $\cN_L$ and $\cN_R$ allows one to obtain 
$\{\lam(\bftheta),\tau_L[g]\}_\bullet$ and $\{\tau_L[g_1],\tau_L[g_2]\}_\bullet$ from 
(\ref{lam_omega_R_bracket}) and (\ref{omega_omega_R_bracket}) by exchanging $L$ and $R$.

The brackets between $\mu$ and $\bar{\mu}$ are as follows: For $p, q$ 
any pair of $v\theta$ coordinate grid points on $\cN$ 
\be   \label{mu_mu_brack}
0  =  \{\mu(p),\mu(q)\}_\bullet = \{\bar{\mu}(p),\bar{\mu}(q)\}_\bullet.
\ee
When $p$ and $q$ lie on the same branch $\cN_A$
%
%\be	\label{mu_mubar_bracket}
%\{\mu({\mathbf 1}),\bar{\mu}({\mathbf 2})\}_\bullet
% = 4\pi G\, \frac{1}{\rho_0}\dg^2(\bftheta_2 - \bftheta_1)\, H({\mathbf 1},{\mathbf 2}) 
%[\frac{1 - \mu\bar{\mu}}{v_A}]_{\mathbf 1}[\frac{1 - \mu\bar{\mu}}{v_A}]_{\mathbf 2}\:
%e^{\int_{\mathbf 1}^{\mathbf 2} \frac{1}{1 - \mu\bar{\mu}}[\bar{\mu} d\mu - \mu d\bar{\mu}]},
%\ee
%
\be	
\fl\{\mu(p),\bar{\mu}(q)\}_\bullet = 4\pi G\, \frac{1}{\rho_0}\dg^2(\bftheta_p - \bftheta_q)\, H(p,q)
%\nonumber\\&&
[\frac{1 - \mu\bar{\mu}}{v_A}]_p[\frac{1 - \mu\bar{\mu}}{v_A}]_q\:
e^{\int_p^q \frac{1}{1 - \mu\bar{\mu}}(\bar{\mu} d\mu - \mu d\bar{\mu})},
\label{mu_mubar_bracket}
\ee
where $H$ is a step function which is $1$ if $q = p$ or $q$ lies further than $p$ from $S_0$ along the
same generator, and is 0 if $q$ is closer to $S_0$ than $p$.
%\be
%H({\mathbf 1},{\mathbf 2}) = \left\{\begin{array}{cl} 
%1 & \mbox{if ${\mathbf 2} = {\mathbf 1}$ or ${\mathbf 2}$ is further from $S_0$ along the 
%generator than ${\mathbf 1}$} \\
%0 & \mbox{otherwise,} \end{array}\right.
%\ee 
The integral in the exponential is evaluated along the segment of the generator from 
$p$ to $q$.
%If both ${\mathbf 1}$ and ${\mathbf 2}$ lie on $S_0$ then 
%\be		\label{mu_mubar_S0}
%\{\mu({\mathbf 1}),\bar{\mu}({\mathbf 2})\}_\bullet = 4\pi G 
%\, \frac{1}{\rho_0}\dg^2(\theta({\mathbf 2}) - \theta({\mathbf 1})) 
%[(1 - \mu\bar{\mu})^2]_{\mathbf 1}. 
%\ee
Finally, when $p$ and $q$ lie on different branches, and neither lies on $S_0$,
$\{\mu(p),\bar{\mu}(q)\}_\bullet$ vanishes.

There remain the brackets between $\mu$ and the $S_0$ data $\lam$, $\tilde{\tau}_R$, and 
$\tilde{\tau}_L$, and also the brackets of these data with $\bar{\mu}$. For 
$p$ on $\cN_R - S_0$ (i.e. $v_{R\,p} \neq 1$).
\bearr
\{\mu(p),\lam({\bftheta})\}_\bullet 
& = & 4\pi G \frac{1}{\rho_0}\dg^2(\bftheta - \bftheta_p)[v_R \di_{v_R} \mu]_p,
\label{mu_lam_bracket}\\
\{\mu(p), \tau_R[f]\}_\bullet & = & 8\pi G[2{\pounds}_f \mu 
- \frac{{\pounds}_f \rho_0}{\rho_0} v_R\di_{v_R} \mu]_p,
\label{mu_omegaR_bracket}\\
\{\mu(p), \tau_L[g]\}_\bullet & = & 0.   \label{mu_omegaL_bracket}
\eearr
The preceding expressions do not hold for $p$ on $S_0$. In this case. 
\bearr
\{\mu(p),\lam({\bftheta})\}_\bullet & = & 0 \label{mu_lam_0_bracket}\\
\{\mu(p), \tau_R[f]\}_\bullet 
& = & 8\pi G [{\pounds}_f \mu]_p,  \label{mu_omega_R_0_bracket}\\
\{\mu(p), \tau_L[g]\}_\bullet 
& = & 8\pi G [{\pounds}_g \mu]_p.  \label{mu_omega_L_0_bracket}
\eearr

The brackets with $\bar{\mu}(p)$ {\em are} continuous at $p \in S_0$. 
For $p \in \cN_R$ (including $p \in S_0$)
\bearr
\fl\{\bar{\mu}(p),\lam(\bftheta)\}_\bullet 
 = 4\pi G \frac{1}{\rho_0}\dg^2(\bftheta - \bftheta_p)\left\{[v_R \di_{v_R}
\bar{\mu}]_p + [\di_{v_L}\bar{\mu}]_{p_0} \frac{1}{v_R(p)} e^{-2\int_{p_0}^p 
\frac{\mu d \bar{\mu}}{1 - \mu\bar{\mu}}} \right\},
\label{mubar_lam_bracket}\\
\fl\{\bar{\mu}(p),\tau_R[f]\}_\bullet 
=  8\pi G\left\{[2{\pounds}_f \bar{\mu} - \frac{{\pounds}_f \rho_0}
{\rho_0} v_R\di_{v_R} \bar{\mu}]_p - [{\pounds}_f \bar{\mu}]_{p_0}
\frac{1}{v_R(p)} e^{-2\int_{p_0}^p
\frac{\mu d \bar{\mu}}{1 - \mu\bar{\mu}}}\right\},
\label{mubar_omegaR_bracket}\\
\fl\{\bar{\mu}(p), \tau_L[g]\}_\bullet = 8\pi G[{\pounds}_g \bar{\mu}
- \frac{{\pounds}_g \rho_0}{\rho_0}\di_{v_L} \bar{\mu}]_{p_0} 
\frac{1}{v_R(p)} 
e^{-2\int_{p_0}^p \frac{\mu d \bar{\mu}}{1 - \mu\bar{\mu}}}, 
\label{mubar_omegaL_bracket}  
\eearr
where $p_0$ is the base point on $S_0$ of the generator through $p$.
Exchanging $L$ and $R$ in (\ref{mu_lam_bracket} -- \ref{mubar_omegaL_bracket}) gives the 
corresponding brackets for $p$ on $\cN_L$.

We will call equations (\ref{lam_rho_bracket} -- \ref{omega_R_omega_L_bracket}) and (\ref{mu_mu_brack} -- \ref{mubar_omegaL_bracket}), 
expressing the brackets of the data as functionals of the data, the {\em structure relations} of the $\bullet$ bracket.
% These determine the $\bullet$ bracket completely. 

Let us consider now the brackets of the alternative data $\sct_A$. Since $\sct_{A\,p} = \di_p s_A^k \tilde{\tau}_{A\,k}$
\be
\{s_A^k(\bftheta_1),\sct_{B\,l}(\bftheta_2)\}_\bullet 
 =  16\pi G \di_p s_A^k\dg_{AB}\dg^2(\bftheta_2 - \bftheta_1).	\label{sct_s_bracket}
\ee 
The remaining brackets are best stated in terms of 
\be\label{sctf_def}
\sct_A[{\mathfrak f}] = \int_{S_0} \sct_{A\,p} {\mathfrak f}^{\,p} d^2\theta = \tau_A[{\mathfrak f}^{\,p} \di_p s_A^k],
\ee
where $\mathfrak f$ is a vector field with fixed (data independent) components ${\mathfrak f}^{\,p}$ in the $\theta$ chart.
Let $f^k = {\mathfrak f}^{\,p} \di_p [s_A^k]_g$ be the $y_A$ chart components of ${\mathfrak f}$ at a particular solution metric $g$, then at this solution
$\sct_A[{\mathfrak f}] = \tau_A[f]$ and 
\be
\{\phi, \sct_A[{\mathfrak f}]\}_\bullet = \{\phi, \tau_A[f]\}_\bullet
\ee
for all data $\phi$ that commute with $s_A$, that is, for all data save $\sct_A$, or $\tilde{\tau}_A$.

The only bracket that remains to be calculated is $\{\sct_A[{\mathfrak f}_1], \sct_A[{\mathfrak f}_2]\}_\bullet$. By (\ref{sctf_def}) 
\bearr
\fl\{\sct_A[{\mathfrak f}_1], \sct_A[{\mathfrak f}_2]\}_\bullet (g)] & = & \{\tau_A[f_1], \tau_A[f_2]\}_\bullet (g)]\nonumber\\
&& + \tau_A[{\mathfrak f}_1^{\,p} \di_p \{s_A^k, \tau_A[f_2]\}_\bullet (g) + \tau_A[{\mathfrak f}_2^{\,p} \di_p \{\tau_A[f_1], s_A^k\}_\bullet (g).
\eearr
From (\ref{omega_s_R_bracket}, \ref{omega_s_L_bracket})
\be
 \{s_A^k, \tau_B[f_2]\}_\bullet = 16\pi G f_2^k,
\ee
so 
\bearr
\fl\{\sct_A[{\mathfrak f}_1], \sct_A[{\mathfrak f}_2]\}_\bullet & = & - 16\pi G \int_{S_0} \frac{1}{(1 - \mu\bar{\mu})^2}{\pounds}_{{\mathfrak f}_1} \mu 
(\rho_0 {\pounds}_{{\mathfrak f}_2}\bar{\mu} - {\pounds}_{{\mathfrak f}_2}\rho_0\di_{v_R}\bar{\mu})\: d^2\theta - (1 \leftrightarrow 2) \nonumber \\
&& + 16\pi G\: \sct_A[[{\mathfrak f}_1, {\mathfrak f}_2]],
\eearr
where $[{\mathfrak f}_1, {\mathfrak f}_2] = {\mathfrak f}_1^{\,p} \di_p {\mathfrak f}_2^{\,q} - {\mathfrak f}_2^{\,p} \di_p {\mathfrak f}_1^{\,q}$ 
is the Lie bracket of ${\mathfrak f}_1$ and ${\mathfrak f}_2$.

Notice that the geometrical data $\rho_0$, $\lam$, $\sct_A$. $\mu$, $\bar{\mu}$ form a closed Poisson subalgebra. 
The Poisson brackets between these data do not depend on the diffeomorphism data.
With $\tilde{\tau}_A$ in place of $\sct_A$ this is not the case. Since $\mu$ is a function of the $\theta$ chart,
the evaluation of the Lie derivative $\pounds_f \mu$ that appears in brackets of $\tau_A[f]$, such as 
(\ref{omega_omega_R_bracket}), requires the transformation of $f$, given in the basis $\di_{y^k_A}$, to the $\theta$ chart 
coordinate basis $\di_{\theta^p}$ (or the basis $\di_z,\di_{\bar{z}}$ defined by the complex chart $z = \theta^1 + i\theta^2$ that is used in (\ref{mu_Lie})). 
This transformation is determined by $s_A$.

\subsection{Properties of the $\bullet$ brackets}

There are three important properties that the structure relations (\ref{lam_rho_bracket} -- \ref{omega_R_omega_L_bracket}), 
(\ref{mu_mu_brack} -- \ref{mubar_omegaL_bracket}) must have: They must be covariant 
under changes of the $y_L$, $y_R$ and $\theta$ charts labelling the generators of $\cN$, so 
that the bracket they define on observables does not depend on the (arbitrary) choice of these charts; They
must satisfy the Jacobi relations (\ref{Jacobi}), because the $\bullet$ bracket does by 
Proposition \ref{Jacobi_proof}; And finally, the $\bullet$ bracket with the constraint 
(\ref{tau_constraint}) must generate diffeomorphisms of the $\theta$ chart, as demonstrated 
in Subsection \ref{kappa_generates_theta_diff}. 
These properties have been verified directly for the expressions (\ref{lam_rho_bracket} 
-- \ref{omega_R_omega_L_bracket}) and (\ref{mu_mu_brack} -- \ref{mubar_omegaL_bracket}) 
which provides a sensitive check on the rather intricate calculations that lead to these expressions. 

The demonstration of the Jacobi relations from (\ref{lam_rho_bracket} -- \ref{omega_R_omega_L_bracket})
and (\ref{mu_mu_brack} -- \ref{mubar_omegaL_bracket}) is a quite long but
straightforward calculation which will not be reproduced here. 
The demonstration that the constraint (\ref{tau_constraint}) generates 
diffeomorphisms of the $\theta$ chart according to (\ref{kappa_action}) is another
straightforward but considerably shorter calculation. It will not be reproduced here 
either, but the reader interested in verifying the result should remember that 
$\{\kappa[\eta], \cdot\}_\bullet$ acts only on the data.
Thus, for instance $\{\kappa[\eta], \tau_R[f]\}_\bullet = -\tau_R[\eta^p \di_p f]$ even
though $\tau_R[f]$ is invariant under diffeomorphisms of $\theta$ when $f^i(\theta)$
is carried along by the diffeomorphism, precisely because $f$ is {\em not} carried along
by the flow generated by $\kappa[\eta]$.

\subsubsection{Covariance of the structure relations.}

%What about the covariance of the structure relations? 
The equivalence of the structure
relations in different charts requires that the brackets of the data, and the functions
of the data that the structure relations equate these with, transform in the same way under
changes of chart. The bracket acts as a variation (that is, a derivative along some vector 
field tangent to the phase space) on each of its arguments and variations of a phase space 
function transform by the linearization of the transformation of the function itself. For example, 
$\rho_0$ transforms as a scalar density under changes of the $\theta$ chart so, since this 
is a linear transformation law, a variation $\Dg \rho_0$ also transforms as a scalar 
density. The Beltrami coefficient $\mu$, on the other hand, transforms according to the non-linear 
law (\ref{mu_transformation}). Variations of $\mu$ transform according to the linearization of this law:
\be	\label{transf_dg_mu}
\Delta \mu' = \frac{\ag\bar{\ag} - \bg\bar{\bg}}{(\ag + \mu\bg)^2} \Dg \mu,
\ee
where $\ag = \frac{\di z}{\di z'}$ and $\bg = \frac{\di\bar{z}}{\di z'}$.

It follows that a bracket $\{\varphi,\chi\}_\bullet$ between two data transforms in 
the same way as a product $\Dg_1 \varphi \Dg_2 \chi$ of two variations, the linearization of the 
transformation of each of the two arguments acting on the bracket simultaneously. 

Let us verify the covariance of the structure relations, beginning with those, like $\{\mu(p),\mu(q)\} = 0$, 
that set the brackets of data fields to zero. Changes of the $y$ or $\theta$ charts do not mix distinct 
data fields, although of course they do mix the coordinate components of a single multicomponent field, 
such as $\tilde{\tau}_L$, among themselves. Each of the distinct data fields $\mu, \bar{\mu}, \lam, \rho_0, 
\tilde{\tau}_R, s_R,...$ therefore transforms to a function of itself, implying that the transforms of two 
such fields that $\bullet$ commute will also $\bullet$ commute.

Now let us consider non-zero brackets. We begin by verifying covariance under
transformations of the $y$ charts. The only brackets which depend explicitly on the $y$ charts 
are those of $s_A^i(\theta)$, given in (\ref{omega_s_R_bracket}) and (\ref{omega_s_L_bracket}), 
and it follows immediately from the definitions of the data involved that both sides of these 
equations transform in the same way. The only other brackets in which the $y$ charts
enter are those of $\tau_R[f]$ and $\tau_L[g]$. But if we transform the test fields $f^i(\theta)$ 
and $g^m(\theta)$ as components of vectors when the $y^R$ and $y^L$ charts are changed, then 
$\tau_R[f]$ and $\tau_L[g]$ are invariant under such coordinate changes, and the expressions 
for the brackets given on the right side of our structure relations are also invariant. This 
suffices to show that if the structure relations hold for all $f$ and $g$ in one choice of $y_R$ 
and $y_L$ charts, then they hold for all $f$ and $g$ in all choices of these charts.

Consider now transformations of the $\theta$ chart. Equations
(\ref{lam_rho_bracket} -- \ref{lam_lam_brack}) are covariant under such transformations by
inspection. The remaining structure relations involve $\mu$ and $\bar{\mu}$, which transform
according to (\ref{mu_transformation}) and the formal complex conjugate (\ref{functional_cc}) 
of this law, and the variations of which transform according to (\ref{transf_dg_mu}) and its 
formal complex conjugate.
However the explicit forms of these transformation laws are not necessary to conclude that
(\ref{mu_lam_bracket} -- \ref{mu_omega_L_0_bracket}) are covariant. These relations give 
expressions for the brackets of $\mu(p)$ with $\lam(\bftheta)$, $\tau_R[f]$, and $\tau_L[g]$.
$\lam(\bftheta)$ transforms as a scalar field, whereas $\tau_R[f]$, or $\tau_L[g]$ are invariant
under transformations of the $\theta$ chart. Thus $\{\mu(p),\lam(\bftheta)\}_\bullet$ transforms
like a variation of $\mu(p)$ times a scalar function of $\bftheta$, while the brackets of $\mu(p)$
with $\tau_R[f]$ and $\tau_L[g]$ transform simply as variations of $\mu(p)$. This is clearly
also true of the expressions given for these brackets in (\ref{mu_lam_bracket} -- \ref{mu_omega_L_0_bracket}).

To demonstrate the covariance of (\ref{lam_omega_R_bracket} -- \ref{omega_R_omega_L_bracket})
we must show that the right side of (\ref{lam_omega_R_bracket}) transforms as a scalar, and that the integrands of the right
sides of (\ref{omega_omega_R_bracket}) and (\ref{omega_R_omega_L_bracket}) transform as scalar densities.
These are all (sums of) expressions of the form
\be	\label{Dmu_Dmubar_scalars}
\frac{\Dg_1 \mu \Dg_2 \bar{\mu}}{(1 -\mu\bar{\mu})^2},
\ee
with $\Dg_1$ and $\Dg_2$ being variations. Now (\ref{mu_transformation}) shows that
\be	\label{transf_1-mumubar}
1 - \mu'\bar{\mu}' = \frac{\ag\bar{\ag} - \bg\bar{\bg}}{(\ag + \mu\bg)\overline{(\ag + \mu\bg)}} (1 - \mu\bar{\mu}).
\ee
Combining this with (\ref{transf_dg_mu}) one finds that
\be	\label{transf_dg_mu_2}
\frac{\Dg \mu'}{1 - \mu'\bar{\mu}'} = \frac{\overline{\ag + \mu\bg}}{\ag + \mu\bg}\frac{\Dg \mu}{1 - \mu\bar{\mu}}.
\ee
The formal complex conjugate of this equation is also valid:
\be	\label{transf_dg_mubar_2}
\frac{\Dg \bar{\mu}'}{1 - \mu'\bar{\mu}'} 
= \frac{\ag + \mu\bg}{\overline{\ag + \mu\bg}}\frac{\Dg \bar{\mu}}{1 - \mu\bar{\mu}}.
\ee
It follows that expressions of the form (\ref{Dmu_Dmubar_scalars}) indeed are scalars.

To verify the covariance of equations (\ref{mubar_lam_bracket} -- \ref{mubar_omegaL_bracket})
for the brackets of $\bar{\mu}$ with $\lam(\bftheta)$, $\tau_R[f]$, and $\tau_L[g]$ we need to know
how the exponential $\exp(-2\int_{p_0}^p \frac{\mu d \bar{\mu}}{1 - \mu\bar{\mu}})$ transforms.
By (\ref{transf_dg_mubar_2}) and (\ref{mu_transformation})
\bearr
\frac{\mu'\di_v\bar{\mu}'}{1 - \mu'\bar{\mu}'} & = & \frac{\bar{\bg} + \mu\bar{\ag}}{\overline{\ag + \mu\bg}}
\frac{\di_v\bar{\mu}}{1 - \mu\bar{\mu}}	\\
& = & \frac{\bar{\bg}\di_v\bar{\mu}}{\overline{\ag + \mu\bg}} + \frac{\mu \di_v\bar{\mu}}{1 - \mu\bar{\mu}}\\
& = & \di_v\ln(\overline{\ag + \mu\bg}) + \frac{\mu \di_v\bar{\mu}}{1 - \mu\bar{\mu}}. \label{mudmubar_transf}
\eearr
(Note that $\ag$ and $\bg$ are constant along the generators.)
Therefore
\be	\label{exp_int_transf1}
\exp(-2\int_{p_0}^p \frac{\mu' d \bar{\mu}'}{1 - \mu'\bar{\mu}'})
= \frac{[\overline{\ag + \mu\bg}]^2_{p_0}}{[\overline{\ag + \mu\bg}]^2_p}\:
\exp(-2\int_{p_0}^p \frac{\mu d \bar{\mu}}{1 - \mu\bar{\mu}}).
\ee 
Now recall that variations of $\bar{\mu}$ transform according to the formal complex conjugate of
(\ref{transf_dg_mu}). Combining this transformation law with (\ref{exp_int_transf1}) we
find that $\Dg\bar{\mu}(p_0)\exp(-2\int_{p_0}^p \frac{\mu d \bar{\mu}}{1 - \mu\bar{\mu}})$
transforms precisely like $\Dg\bar{\mu}(p)$, the value of the variation at $p$.
From this the covariance of (\ref{mubar_lam_bracket} -- \ref{mubar_omegaL_bracket})
is immediate.

It remains to verify the covariance of equation (\ref{mu_mubar_bracket}) for the bracket
of $\mu(p)$ and $\bar{\mu}(q)$. By
(\ref{transf_dg_mu_2}) and (\ref{transf_dg_mubar_2}) 
\bearr
\fl\frac{1}{[1 - \mu'\bar{\mu}']_p}\{\mu'(p),\bar{\mu}'(q)\}_\bullet
\frac{1}{[1 - \mu'\bar{\mu}']_q} = &\left[\frac{\overline{\ag + \mu\bg}}{\ag + \mu\bg}\right]_p
\left[\frac{\ag + \mu\bg}{\overline{\ag + \mu\bg}}\right]_q
\nonumber \\
\fl & \times \frac{1}{[1 - \mu\bar{\mu}]_p}\{\mu(p),\bar{\mu}(q)\}_\bullet \frac{1}{[1 - \mu\bar{\mu}]_q}.
\eearr
To complete the demonstration of the covariance of (\ref{mu_mubar_bracket}) we only have to show
that $\exp(\int_p^q \frac{1}{1 - \mu\bar{\mu}} [\bar{\mu}d \mu - \mu d\bar{\mu}])$ transforms in the same way,
for then 
\be
\frac{1}{[1 - \mu\bar{\mu}]_p}\{\mu(p),\bar{\mu}(q)\}_\bullet \frac{1}{[1 - \mu\bar{\mu}]_q} 
\exp(-\int_p^q \frac{1}{1 - \mu\bar{\mu}} [\bar{\mu}d \mu - \mu d\bar{\mu}])
\ee
transforms like a scalar, like $\frac{1}{\rho_0}\dg^2(\bftheta_p - \bftheta_q)\, H(p,q)$.
But this follows immediately from (\ref{mudmubar_transf}) and its formal complex conjugate.
The covariance of the structure relations has been demonstrated.

The expression (\ref{mu_mubar_bracket}) for $\{\mu(p),\bar{\mu}(q)\}_\bullet$ has a further 
property. By rewriting the factor $\rho_0 v_A(p) v_A(q)$ in the denominator as $\sqrt{\rho(p)\rho(q)}$
it may be expressed without reference to any particular parametrization of the generators.
This suggests the conjecture that the bracket is given by this expression even when $v$ is not a good 
parameter. For instance, in flat spacetime with $\cN$ swept out by parallel generators 
(neither diverging nor converging). The same cannot be said of all of the brackets. Some become
undefined when the parametrization by $v$ breaks down. Note however that these are brackets of the data
$\lam$, $\tilde{\tau}_L$, and $\tilde{\tau}_R$, and these data themselves become undefined in this event.
The brackets of a set of data, like the original Sachs data \cite{Sachs}, which do not depend on the validity
of the $v$ parametrization may have a natural extension to $\cN$ on which $v$ does not parameterize 
the generators. This seems easy enough to check, but will be left to future investigations for now.
(A possibly more difficult problem is then to demonstrate that the extrapolated brackets are actually correct,
that is, reproduce the Peierls bracket.)
Such an extension of our bracket would be necessary in order to make a direct comparison of the 
present theory with canonical formulations of general relativity linearized about Minkowski space 
(see for instance \cite{Aragone_Gambini}) in terms of data on null hyperplanes. 

\subsubsection{Complex metric modes generated by the $\bullet$ bracket.}\label{complex_modes}

A peculiar feature of the brackets is that they do not preserve the reality of the conformal
2-metric $e_{pq}$. Functionals of the data  that are real on real data, like $\mu(p) + \bar{\mu}(p)$, generate
complex variations of the metric via the $\bullet$ bracket.

The reality of a real metric $e_{pq}$ would be preserved by the flow $\{F,\cdot\}_\cdot$ generated by any formally 
real functional of the data, $F = \bar{F}$ if the bracket itself were real in the sense that it equals its  
formal complex conjugate bracket $\{\varphi,\chi\}_{\bullet c.c.} \equiv \overline{\{\bar{\varphi}, \bar{\chi}\}}_\bullet$.
But this is not the case: $\overline{\{\mu(p),\bar{\mu}(q)\} _\bullet} \neq \{\bar{\mu}(p),\mu(q)\}_\bullet$. 

Taking linear combinations with the complex conjugate bracket one may resolve the $\bullet$ bracket into real 
and imaginary parts, $\{\cdot,\cdot\}_\bullet = \{\cdot,\cdot\}_R + i \{\cdot,\cdot\}_I$, and one finds that
\be\label{real_imaginary_parts}
\fl \{\cdot,\cdot\}_I = \frac{i}{8\pi G}\sum_{A = L, R} \int_{S_A} 
\frac{\dot{\rho}_A}{(1 - \dot{\mu}_A\dot{\bar{\mu}}_A)^2}%\nonumber \\&&
[\{\cdot,\dot{\bar{\mu}}_A\}_\bullet \overline{\{\dot{\bar{\mu}}_A, \bar{\cdot}\}}_{\bullet} 
- \overline{\{\bar{\cdot},\dot{\bar{\mu}}_A\}}_{\bullet}\{\dot{\bar{\mu}}_A,\cdot\}_\bullet] d^2 y_A.
\ee
The variation $\{F,\cdot\}_\bullet$ consists of $\{F,\cdot\}_R$, which preserves the reality of the metric if $F$
is formally real, plus multiples of $\{\dot{\bar{\mu}}_A(q),\cdot\}_\bullet$ and 
$\overline{\{\dot{\bar{\mu}}_A(q), \bar{\cdot}\}}_{\bullet}$ 
with $q$ the endpoint(s), on $S_L$ or $S_R$, of the generator(s) through $p$. 

On observables the bracket is real, as it reproduces the Peierls bracket, which is real.
Indeed observables are unaffected by the modes $\{\dot{\bar{\mu}}_A(q),\cdot\}_\bullet$ and 
$\overline{\{\dot{\bar{\mu}}_A(q), \bar{\cdot}\}}_{\bullet}$. By (\ref{Aflow_DeltaA}), for any observable $B$ of $D[\cN]^\circ$
\be
\{\dot{\bar{\mu}}_A(q), B \}_\bullet = - \Dg_B \dot{\bar{\mu}}_A(q) = 0,
\ee
since $\Dg_B$ vanishes in a neighborhood of $\di\cN$.

In fact, since (\ref{Aflow_DeltaA}) is valid for any linear observable, the preceding equation also applies to any linear observable 
$B[\dg] \equiv \int \theta^{ab} \dg g_{ab}\, \vareg$, and therefore
\be\label{theta_condition}
\int \theta^{ab} \{\dot{\bar{\mu}}_A(q), g_{ab} \}_\bullet \, \vareg = 0
\ee
for any divergenceless, symmetric tensor field $\theta^{ab}$ with compact support contained in $D[\cN]^\circ$. This condition
is satisfied whenever 
\be
\{\dot{\bar{\mu}}_A(q), g_{ab} \}_\bullet = 2 \nabla_{(a}\xi_{b)} = \pounds_\xi g_{ab}
\ee
in $D[\cN]^\circ$, that is, when $\{\dot{\bar{\mu}}_A(q), \cdot \}_\bullet$ produces at most a diffeomorphism there and thus leaves the
geometry unchanged. It is tempting to conclude that all solutions to (\ref{theta_condition}) are of this form, but, frustratingly, the 
author has not been able to prove this.

On the other hand, this result is expected on general grounds. By causality one expects $\{\dot{\bar{\mu}}_A(q), \cdot \}_\bullet$ to act
trivially in $D[\cN]^\circ$ which lies outside the causal domain of influence of $q$. Furthermore, such schock wave solutions, that skim
along a branch $\cN_A$ of $\cN$, without entering $D[\cN]^\circ$, exist in linearized GR: The retarded minus advanced Green's function
of a point on $S_A$ has this property. Indeed, according to Peierls' prescription the Poisson bracket $\{\dot{\bar{\mu}}_A(q), g_{ab} \}$
is proportional to this difference of Green's functions.

It may be shown directly from the field equations that the variation of initial data $\dg \mu = \{\dot{\bar{\mu}}_A(q), \mu \}_\bullet$ 
is precisely of the form corresponding to such a schock wave. To show this it is convenient to start in de Donder gauge.
In this gauge the trace reversed metric perturbation $\bar{\cg}_{ab} = \cg_{ab} - 1/2 g_{ab} \cg$ satisfies $\nabla^a \bar{\cg}_{ab} = 0$,
where $\cg_{ab} = \dg g_{ab}$ and $\cg = g^{ab}\cg_{ab}$. (We follow the notation of \cite{Wald} section 7.5. $\bar{\cg}_{ab}$  
does not denote the formal complex conjugate of $\cg_{ab}$.)
The linearized field equations reduce to
\be
\nabla_c\nabla^c \bar{\cg}_{ab} - 2 R^c{}_{ab}{}^d \bar{\cg}_{cd} = 0.
\ee
A calculation shows that two solutions to this equation (and the gauge condition) match accros $\cN_A$ to form a weak solution if and only 
if the quantities
\be\label{Q1}
\nabla_r \cg_{\ag\bg} + \frac{1}{r} \cg_{\ag\bg}
\ee
and
\be\label{Q2}
\bar{\cg}_{r\bg}
\ee
match across $\cN_A$. Here these quantities have been expressed using the $a_A$ chart $(u, r, y^1, y^2)$ defined in \ref{a_chart_appendix},
the indices $\ag$, $\bg$ being $a_A$ chart coordinate indices. In de Donder gauge certain non-trivial solutions to the past of $\cN_A$ 
can be matched to the solution $\cg_{ab} = 0$ in $D[\cN]^\circ$, to the future of $\cN_A$. In these solutions (\ref{Q1}) and (\ref{Q2}) must 
vanish on $\cN_A$. 

Rather than using de Donder gauge, it will more convenient to put the variation $\{\dot{\bar{\mu}}_A(q), \cdot \}_\bullet$ into a gauge so that it preserves
the $a_A$ chart. That is, we will represent the variation by its effect on the $a_A$ chart components of the metric. On a vacuum solution background $g$ 
de Donder gauge is reached from any other gauge by the transformation $\cg_{ab} \rightarrow \cg_{ab} + 2\nabla_{(a} \xi_{b)}$ with 
$\nabla_a \nabla^a \xi_b = - \nabla^a \bar{\cg}_{ab}$ (see \cite{Wald} section 7.5). This equation can always be solved with $\xi = 0$ on $\cN_A$,
and in this case $\cg_{ij}$, $\cg_{ir}$ and $\cg_{rr}$ are unchanged by the transformation on $\cN_A$. Only the $u\bg$ components of $\cg_{ab}$ are changed.
$\bar{\cg}_{ru} = 1/2 g^{ij} \cg_{ij}$, $\bar{\cg}_{rr} = \cg_{rr}$, and $\bar{\cg}_{ri} = \cg_{ri}$ are also unchanged. (See (\ref{a_line_element} for
the metric components in the $a_A$ chart.)
Thus a variation in $\cC$ corresponds to a de Donder gauge solution which matches the zero solution to the future of $\cN_A$ if
\be\label{match1}
\nabla_r \cg_{ij} + \frac{1}{r} \cg_{ij} = 0
\ee
and
\be\label{match2}
\bar{\cg}_{r\bg} = 0.
\ee
Of the remaining matching conditions $\nabla_r \cg_{ir} + \frac{1}{r} \cg_{ir} = 0$ holds as a consequence of (\ref{match2}) and the fact that the 
generators are geodesics, and the matching of $\nabla_r \cg_{u\bg} + \frac{1}{r} \cg_{u\bg} = 0$, which is affected by the transformation, across
$\cN_A$ partly fixes the solution $2\nabla_{(a}\xi_{b)}$ to the future of $\cN_A$.

The variation of the $a_A$ chart components $\cg_{\ag\bg} = \dg g_{\ag\bg}$ of the metric under $\dg \in \cC$ is $\cg_{u\bg} = \cg_{r\bg} = 0$
$\cg_{ij} = \dot{\rho} r^2 \dg e_{ij}$, so $\cg = 0$ and (\ref{match2}) holds. The matching condition (\ref{match1}) becomes, after some calculation,
\be\label{match1e}
\di_r \dg e_{ij} + \frac{1}{r} \dg e_{ij} - \di_r e_{k(i} e^{kl} \dg e_{j)l} = 0.
\ee
Expressed in terms of $\mu$ and $\bar{\mu}$ this is
\bearr
0 & = & \di_r\left[\frac{r \exp[\int_1^r \frac{\bar{\mu} d \mu - \mu d \bar{\mu}}{1 - \mu\bar{\mu}}]  \dg^\circ \mu}{1 -\mu\bar{\mu}}\right]  \\
0 & = & \di_r\left[\frac{r \exp[- \int_1^r \frac{\bar{\mu} d \mu - \mu d \bar{\mu}}{1 - \mu\bar{\mu}}]  \dg^\circ \bar{\mu}}{1 -\mu\bar{\mu}}\right].
\eearr

But by (\ref{mu_mubar_bracket}) these equations are satisfied by $\dg = \{\dot{\bar{\mu}}_A(q), \cdot \}_\bullet$, so this variation indeed 
corresponds to a shock wave that skims along $\cN_A$ without entering $D[\cN]^\circ$. 

This calculation provides, in addition, a non-trivial, independent verification of the somewhat intricate form of the bracket (\ref{mu_mubar_bracket}).

\subsection{$\theta$ gauge fixing and Dirac bracket on completely free data}\label{theta_gauge_fixing}

In order to avoid gauge fixing the $\theta$ chart on $S_0$ we have extended the phase space by letting
$\tilde{\tau}_R$ and $\tilde{\tau}_L$ be independent fields, and treating the relation (\ref{tau_constraint}), 
which may be expressed as 
\be	
0 = \frac{1}{16\pi G}[ 2\rho_0 \di_p \lam - \tilde{\tau}_{R\,i}\di_p s_R^i - \tilde{\tau}_{L\,m}\di_p s_L^m] \equiv \kappa_p,
\ee
as a constraint.

A completely constraint free theory can be obtained by solving the constraint, fixing gauge, and calculating 
the corresponding Dirac bracket. The Dirac bracket corresponding to the gauge condition $\theta^1 = y_L^1$, 
$\theta^2 = y_L^2$  (i.e. $s_L = \mbox{id}$) is easily obtained from the explicit expressions for the $\bullet$ brackets
on the extended phase space. It is
\be  \label{Dirac_b_kappa}
\fl \{\cdot,\cdot\}_* = \{\cdot,\cdot\}_\bullet - \int_{S_0} \frac{\di\theta^p}{\di y_L^m}
\,[\{\cdot, s_L^m\}_\bullet\{\kappa_p,\cdot\}_\bullet - \{\cdot, \kappa_p\}_\bullet\{s_L^m,\cdot\}_\bullet]\,\:d^2\theta,
\ee
where of course, $\frac{\di\theta^p}{\di y_L^m} = \dg^p_m$ in the chosen gauge. (The same bracket can also
be obtained by solving (\ref{final_bracket_condition1}) and (\ref{final_bracket_condition2}) imposing the constraint
and gauge fixing condition on the variations in $\cC$.)

According to (\ref{omega_s_L_bracket}) and the subsequent discussion, the only datum that has non-zero 
$\bullet$ bracket with $s_L$ is $\tilde{\tau}_L$. The Dirac bracket thus differs from $\bullet$ bracket only 
when one of the arguments depends on $\tilde{\tau}_L$. On the other hand neither $\tilde{\tau}_L$ nor $s_L$ 
are needed to coordinatize the gauge fixed constraint surface, the remaining initial data suffice. Thus the 
$*$ brackets of a complete set of coordinates on the physical phase space, namely our data excluding $s_L$ and 
$\tilde{\tau}_L$, are identical to the corresponding $\bullet$ brackets.

A disadvantage of this gauge fixing is that it introduces an artificial asymmetry between the
two branches of $\cN$. Extending the phase space is a simple way to treat the data on the two branches symmetrically.

\section{Calculation of the brackets}	\label{bracket_calc}

How are the brackets obtained from conditions (\ref{final_bracket_condition1}, \ref{final_bracket_condition2}) or 
(\ref{final_bracket_condition1w}, \ref{final_bracket_condition2})? 
The following is both a derivation indicating how the brackets can be deduced from the conditions, and a proof demonstrating 
that the brackets presented in the last section satisfy these conditions and that they constitute the unique solution. 

More precisely, it is demonstrated that these brackets satisfy (\ref{final_bracket_condition1} 
\be	
\dg \varphi = \omega_\cN[\{\varphi,\cdot\}_\bullet,\dg]\ \ \ \forall\:\dg \in \cC, 
\ee
and (\ref{final_bracket_condition2}), 
\be     
\{\varphi,\cdot\}_\bullet \in \cC.  
\ee
for all $\varphi \in \Phi$, but that $\{\varphi, \cdot\}_\bullet$ is determined uniquely for each $\varphi$ already by the seemingly 
weaker pair of conditions (\ref{final_bracket_condition1w}) 
\be
\dg \varphi = \omega_\cN[\{\varphi,\cdot\}_\bullet,\dg]\ \ \ \forall\:\dg \in \cC \cap \cA,
\ee
and (\ref{final_bracket_condition2}).
The two sets of conditions are thus equivalent, but the fact that (\ref{final_bracket_condition1w}, \ref{final_bracket_condition2}),
determines $\{\varphi, \cdot\}_\bullet$ was used in the demonstration that the $\bullet$ bracket reproduces the Peierls bracket on observables.

We will begin by solving (\ref{final_bracket_condition1w}, \ref{final_bracket_condition2})
for all $\varphi$, but with $\dg$ restricted to satisfy $0 = \dg e = \dg s_L = \dg s_R = \dg \rho_0$. 
This determines all the brackets with $s_L$, $s_R$, and $\rho_0$ as the second argument.
Then it is proved that (\ref{final_bracket_condition1w}, \ref{final_bracket_condition2}) is equivalent
to (\ref{final_bracket_condition1}, \ref{final_bracket_condition2}). That is, it has a unique solution when 
(\ref{final_bracket_condition1}, \ref{final_bracket_condition2}) does.

We then continue solving, but with (\ref{final_bracket_condition1}, \ref{final_bracket_condition2}),
and $\varphi = \langle\mu\rangle$, a weighted average of $\mu$ and $\dg$ restricted by 
$0 = \dg s_L = \dg s_R = \dg \rho_0$. 
Both the case in which $\langle\mu\rangle$ is $\mu$ integrated against a weighting functions on $\cN_L$ and $\cN_R$
and the case in which the weighting function is supported only on $S_0$ will be considered.

This, together with a similar analysis of the case in which $\varphi$ is a smearing of $\bar{\mu}$, 
determines the brackets among $\mu$ and $\bar{\mu}$. The brackets of $\mu$ and $\bar{\mu}$ with the 
remaining data are obtained by once more solving (\ref{final_bracket_condition1}) with 
$\varphi = \langle\mu\rangle$ and $\varphi = \langle\bar{\mu}\rangle$, but now without the restrictions 
on $\dg$. Finally, the brackets among $\lam$, $\tilde{\tau}_R$, and $\tilde{\tau}_L$ are found by solving 
(\ref{final_bracket_condition1}) with $\varphi$ taken to be smearings of these three data over 
$S_0$. In each step the brackets obtained in the previous steps are used.

In this procedure (\ref{final_bracket_condition1}) is solved first, assuming that 
(\ref{final_bracket_condition2}) holds, and then it is shown that the resulting 
bracket indeed satisfies (\ref{final_bracket_condition2}):

\subsection{The brackets of $s_L$, $s_R$, and $\rho_0$}\label{ssrho}

Let us begin then by considering (\ref{final_bracket_condition1}) restricted to variations 
$\dg$ such that $0 = \dg e = \dg s_L = \dg s_R = \dg \rho_0$. For such variations $\xi_\perp = 
\dg s_A^k \di_{y_A^k}$ vanishes, so $\dg^\circ e = \dg^y e = \dg e = 0$. As a consequence
(\ref{final_bracket_condition1}) becomes
\bearr
\dg\varphi &= \omega_\cN[\{\varphi,\cdot\}_\bullet,\dg]
= \frac{1}{16\pi G} A[\{\varphi,\cdot\}_\bullet,\dg] 
\nonumber \\
&= - \frac{1}{16\pi G}\int_{S_0}
2 \{\varphi,\rho_0\}_\bullet\, \dg\lam + \{\varphi,s_R^i\}_\bullet\, \dg \tilde{\tau}_{R\,i}
%\nonumber \\
%&& \ \ \ \ \ \ \ \ \ \ \ \ \ \ \ \ \ \ \ \ \ \ \ \ 
+ \{\varphi,s_L^m\}_\bullet\, \dg \tilde{\tau}_{L\,m}\: d^2\theta, \label{A_phi_delta}
\eearr
where $A = A_L + A_R$. Substituting for $\varphi$ each of the data coordinatizing the extended 
phase space (smeared against weighting functions on $S_0$, $\cN_L$ or $\cN_R$) one 
finds that the only non-zero brackets of $\rho_0$, $s_R$ and $s_L$ are
\bearr
\{\lam(\bftheta_1),\rho_0(\bftheta_2)\}_\bullet = - 8\pi G \dg^2(\bftheta_2 - \bftheta_1), \label{lam_rho_Poi}\\
\{\tilde{\tau}_{B\,i}(\bftheta_1), s_A^j(\bftheta_2)\}_\bullet = - 16\pi G
\dg_{AB}\dg_i^j\dg^2(\bftheta_2 - \bftheta_1).\label{tau_s_Poi}
\eearr
In fact (\ref{A_phi_delta}) determines the brackets $\{\varphi, \rho_0(\bftheta)\}_\bullet$ and $\{\varphi, s_A^k(\bftheta)\}_\bullet$
for all smeared data $\varphi$. All are zero except (\ref{lam_rho_Poi}) and (\ref{tau_s_Poi}). 
Note that because of the smoothness of variations of $\rho_0$ and $s_A^k$ in $\cC$, the requierment that
$\{\varphi,\cdot\} \in \cC$ implies that these brackets, for $\varphi \in \Phi$ are determined as functions, not
just as distributions.

\subsection{Uniqueness proof}\label{uniquenes_proof}

We are now ready to show that (\ref{final_bracket_condition1w}) and (\ref{final_bracket_condition2}) determine 
$\{\varphi,\cdot\}_\bullet$ uniquely. 

The previous subsection shows that (\ref{final_bracket_condition1w}, \ref{final_bracket_condition2}) 
completely determine $\{\varphi, \rho_0\}_\bullet$ and $\{\varphi, s_A^k\}_\bullet$.

Let $X_\varphi$ be the difference between two distinct solutions $\{\varphi,\cdot\}_\bullet$ and $\{\varphi,\cdot\}'_\bullet$ to 
(\ref{final_bracket_condition1w}, \ref{final_bracket_condition2}). Then $X_\varphi \in \cC$ and 
\be \label{varphi_brack_uniqueness}
0 = \omega_\cN[X_\varphi,\dg]
\ee 
for all $dg\in \cC\cap\cA$. But, because the brackets with $\rho_0$ and $s_A$ are unique $X_\varphi \rho_0 = X_\varphi s_A = 0$. 
This together with the restriction $\dg \rho_0 = \dg s_A = 0$ implies that $A_A[X_\varphi,\dg] = B_A[X_\varphi,\dg] = 0$, so 
(\ref{varphi_brack_uniqueness}) reduces to
\be \label{varphi_brack_uniqueness2}
0 = C[X_\varphi,\dg]
\ee
Setting $\dg_1 = X_\varphi$ and $\dg_2 = \dg$ in (\ref{C_Box2}) one obtains 
\be\label{C_Box3}
\fl C[X_{\langle\mu\rangle},\dg] = 4\int_{S_0}
d^2\theta \left\{\tilbox_1 \Box_2 - \tilbox_2 \Box_1  %\right. \nonumber \\&& \left.
+ \sum_{A = L,R} \int_1^{\dot{v}_A}
	 \Box_2 \di_{v_A} \tilbox_1 - \Box_1 \di_{v_A} \tilbox_2\: d{v_A} \right\}.
\ee
(Since $\dg_{1\,R}^\circ = \dg_1 = \dg_{1\,L}^\circ$ and $\dg_{2\,R}^\circ = \dg_2 
= \dg_{2\,L}^\circ$ the surface terms at $S_0$ coming from the two branches are equal.)

Under variations $\dg \in \cC \cap \cA$ both $\dg\mu$ and $\dg\bar{\mu}$ are smooth on $\cN_L$ and on $\cN_R$,
and both vanish on $S_L$ and $S_R$ because admissible variations are real. It is also true that $\dg\bar{\mu} = [\dg\mu]^*$ 
in this case, but stationarity still requires stationarity with respect to independent $\dg\mu$ and $\dg\bar{\mu}$ since 
$\dg\mu + \dg\bar{\mu}$ and $\dg\mu - \dg\bar{\mu}$ are independent.

First let us consider $\dg \mu$ and $\dg\bar{\mu}$ which vanish at $S_0$. Because $X_\varphi \in \cC$ both $X_\varphi \mu$ and 
$X_\varphi \bar{\mu}$ are smooth on $\cN-S_0$, the requierment that (\ref{varphi_brack_uniqueness2}) holds for such $\dg$ implies 
that $\Box_1$ and $\tilbox_1$ are constant save at $S_0$ on each branch. In addition $X_\varphi \in \cC$  
requires that $X_\varphi \dot{\mu}_A = 0$ on each truncating surface $S_A$. 
Thus $\Box_1 = 0$ there, and therefore on all of $\cN - S_0$. It also requires that $X_\varphi \bar{\mu}\}_\bullet$
be smooth throughout each branch. $\tilbox_1$ must therefore have a single constant value on all of $\cN$.
Incorporating these restrictions (\ref{C_Box3}) reduces to
\be
\fl C[X_{\langle\mu\rangle},\dg] = - 4\int_{S_0} \tilbox_1 \Box_2 + \tilbox_2 \Box_1 d^2\theta.
\ee
Demanding that this vanishes also for $\dg \in \cC\cap \cA$ which do not vanish on $S_0$ implies that $\Box_1$ and $\tilbox_1$
vanish on all of $\cN$, establishing that $X_\varphi \mu = X_\varphi \bar{\mu} = 0$. That is $\{\varphi, \mu\}_\bullet$ and 
$\{\varphi, \bar{\mu}\}_\bullet$ are uniquely determined by (\ref{final_bracket_condition2}) and
(\ref{final_bracket_condition1w}) restricted to test variations such that $\dg \rho_0 = \dg s_A = 0$.

What would we have found had we applied the condition (\ref{final_bracket_condition1}) instead of (\ref{final_bracket_condition1w}),
but also restricted to test variations such that $\dg \rho_0 = \dg s_A = 0$? Exactly the same result. This is easily verified explicitly
and it is also a consequence of the fact that, as we shall see, (\ref{final_bracket_condition1}, \ref{final_bracket_condition2})
does have a solution. Thus (\ref{final_bracket_condition1w}) and the apparently stronger condition (\ref{final_bracket_condition1})
are equivalent when the test variations are restricted by $\dg \rho_0 = \dg s_A = 0$. 

To apply (\ref{final_bracket_condition1w}) fully it remains to impose it for test variations under which {\em only} $\rho_0$, $s_L$ and $s_R$
vary. But for such $\dg$ (\ref{final_bracket_condition1w}) is again equivalent to (\ref{final_bracket_condition1}) because the set of
variations of this type in $\cC\cap\cA$ is essentially the same as in $\cC$. In the first set they are required to be real, and in the second not,
which makes no difference to stationarity conditions. 

In sum (\ref{final_bracket_condition1w}, \ref{final_bracket_condition2}) is fully equivalent to (\ref{final_bracket_condition1}, \ref{final_bracket_condition2}).
Since, as we shall see, the second pair of conditions has a unique solution, so does the first pair. 

We can now continue solving for the $\bullet$ bracket, but now using fully the conditions 
(\ref{final_bracket_condition1}, \ref{final_bracket_condition2}). This implies, among other things, that $\{\cdot,\cdot\}_\bullet$
may be assumed to be antisymmetric since we already know that this is a consequence of (\ref{final_bracket_condition1}, \ref{final_bracket_condition2}).
(It is not an evident consequence of (\ref{final_bracket_condition1w}, \ref{final_bracket_condition2}) so it could not be assumed earlier 
without spoiling the uniqueness proof.)

Assuming the antisymmetry of the bracket allows us to conclude from Subsection \ref{ssrho} that
all brackets of $\rho_0$, $s_R$ and $s_L$ with data other than $\lam$, $\tilde{\tau}_R$, and
$\tilde{\tau}_L$, respectively, vanish.

Since the brackets of $\rho_0$, $s_R$, and $s_L$ with all data have been determined,
the expression $\omega_\cN[\{\varphi,\cdot\}_\bullet,\dg]$ may be evaluated for all
$\dg \in \cC$ if $\varphi$ is (a smearing of) $\rho_0$, $s_R$, or $s_L$. It is easily checked
that the result is indeed $\dg \varphi$, as (\ref{final_bracket_condition1}) requiers.
One corollary is that the brackets found so far already ensure that
$\dot{v}_A = (|\det \di_p s_A^k|\dot{\rho}_A/\rho_0)^{\frac{1}{2}}$, which depends only on $\rho_0$, $s_A^k$, 
and the constant field $\dot{\rho}_A$, satisfies
\be    \label{v_bar_eqn}
\dg \dot{v}_A = \omega_\cN[\{\dot{v}_A,\cdot\}_\bullet,\dg]\ \ \ \ \forall \dg \in \cC,
\ee
a result that will be used further on.

\subsection{The brackets among $\mu$ and $\bar{\mu}$}\label{mu_barmu_brackets} 

To obtain the brackets of $\mu$ with $\mu$ and $\bar{\mu}$ we solve 
(\ref{final_bracket_condition1}) with $\varphi = \langle\mu\rangle$, a weighted average 
of $\mu$, and $\dg s_L = \dg s_R = \dg \rho_0 = 0$ but $\dg e$ not necessarily zero. 

Averages of $\mu$ over the branches $\cN_L$ and $\cN_R$ as well as averages over $S_0$ only will be considered.
A subtlety arises when treating the variations of an average $\langle\mu\rangle$ over a branch $\cN_A$
when the weighting function does not vanish at the truncation surface $S_A$: The average $\langle\mu\rangle$ 
varies not only due to the variation
of the field $\mu$ but also due to that of $\dot{v}_A$. This complication can be avoided by restricting
attention to weighting functions that vanish at $\di\cN$, which would be sufficient for our main
purpose of obtaining a bracket that correctly reproduces the Peierls brackets of observables
of the interior of the domain of dependence of $\cN$.
Here arbitrary smooth weighting functions, which need not vanish on any part of $\di\cN$,
will be allowed, because this actually implies only a small additional effort and as a consequence
(\ref{final_bracket_condition1}) and (\ref{final_bracket_condition2}) determine the brackets of
all data on $\cN$ uniquely. (It must be admitted, however, that at present the physical meaning of
the additional brackets obtained in this way, namely those of $\mu$ and $\bar{\mu}$ on $\di\cN$,
is unclear.)

Since $\langle\mu\rangle$ depends only on $\mu$ and $\dot{v}_A 
= (|\det \di_p s_A^k|\dot{\rho}_A/\rho_0)^{\frac{1}{2}}$ for the two branches, it $\bullet$ commutes
with $s_L$, $s_R$, and $\rho_0$. This, together with the conditions $\dg s_L = \dg s_R = \dg \rho_0 = 0$,
implies that $A_A[\{\langle\mu\rangle,\cdot\}_\bullet,\dg]$ and 
$B_A[\{\langle\mu\rangle,\cdot\}_\bullet,\dg]$ vanish, so (\ref{final_bracket_condition1}) 
reduces to 
\be    \label{mu_brack_def}
\dg \langle\mu\rangle = \omega_\cN[\{\langle\mu\rangle,\cdot\}_\bullet,\dg] 
= \frac{1}{16\pi G} C[\{\langle\mu\rangle,\cdot\}_\bullet,\dg],
\ee 
where $C = C_L + C_R$. It also follows that the variation $\Dg_{\langle\mu\rangle} 
\equiv \{\langle\mu\rangle, \cdot \}_\bullet$ generated by $\langle\mu\rangle$ satisfies
$\Dg_{\langle\mu\rangle}^\circ = \Dg_{\langle\mu\rangle}^y = \Dg_{\langle\mu\rangle}$ and
similarly that $\dg^\circ = \dg^y = \dg$. 

Since $\dot{v}_A$ commutes with both $\mu$ and $\bar{\mu}$ the average $\langle\cdot\rangle$ 
can be moved outside the $\bullet$ brackets in $\{\langle\mu\rangle,\mu(p)\}_\bullet$ 
and $\{\langle\mu\rangle,\bar{\mu}(p)\}_\bullet$ for all $p \in \cN$, which implies that 
$C[\{\langle\mu\rangle,\cdot\}_\bullet,\dg] = \langle C[\{\mu,\cdot\}_\bullet,\dg]\rangle$ 
independently of any restrictions on $\dg$ - that is, for all $\dg \in \cC$. 
But $\langle C[\{\mu,\cdot\}_\bullet,\dg]\rangle$ is manifestly independent of any variation of 
$\dot{v}_A$ under $\dg$ or generated by the data. It depends only on the values of $\dot{v}_L$ 
and $\dot{v}_R$ in the data set on which it is being evaluated. If, furthermore, $\dg$ satisfies 
$\dg s_L = \dg s_R = \dg \rho_0 = 0$ then $\dg \dot{v}_A = 0$ and thus $\dg \langle\mu\rangle 
= \langle\dg \mu\rangle$, so (\ref{mu_brack_def}) is equivalent to
\be\label{mu_brack_def2}
\langle\dg \mu\rangle  
= \frac{1}{16\pi G} \langle C[\{\mu,\cdot\}_\bullet,\dg]\rangle.
\ee 
In other words (\ref{mu_brack_def}) may be solved treating $\dot{v}_A$ as fixed.

In solving (\ref{mu_brack_def}) we will, for the sake of clarity, temporarily abandon the
approach of {\em deriving} the solutions from the equations and rather just prove that the 
expressions for $\{\langle\mu\rangle,\mu\}_\bullet$ and 
$\{\langle\mu\rangle,\bar{\mu}\}_\bullet$ determined by (\ref{mu_mu_brack}) and 
(\ref{mu_mubar_bracket}) satisfy (\ref{mu_brack_def}) and also (\ref{final_bracket_condition2}).
The uniqueness proof for $\{\varphi, \mu\}_\bullet$ and $\{\varphi,\bar{\mu}\}_\bullet$ of the last subsection
demonstrates that these are the only solutions.
and that they are the unique expressions that do so.

Let us verify then that the brackets of $\mu$ given in (\ref{mu_mu_brack}) and 
(\ref{mu_mubar_bracket}) satisfy (\ref{mu_brack_def}). Consider first the case in which $\mu$ 
is smeared over the bulk of $\cN$: Let 
$\langle\mu\rangle_h = \sum_{A = L,R} \int_{S_0} \int_1^{\dot{v}_A} h_A \mu\: dv_A\, d^2\theta$, 
where the weighting function $h_A$ is smooth on $\cN_A$ but does not necessarily match the weighting 
function of the other branch at $S_0$, and let $\dg_1 = \{\langle\mu\rangle_h,\cdot\}_\bullet$. According to 
(\ref{mu_mu_brack}) and (\ref{mu_mubar_bracket})
\bearr
\Box_1 = 0	\label{box_mu_h}\\
\tilbox_1(p) = 4 \pi G \int_1^{v(p)} h_A e^{-\ag} \frac{1 - \mu\bar{\mu}}{v \sqrt{\rho_0}} dv
\label{tbox_mu_h}
\eearr 
on each branch $\cN_A$, where the integral is taken along the generator from $S_0$ to $p \in \cN_A$.
It follows that $\tilbox_1 = 0$ on $S_0$ and that
\be
\di_v\tilbox_1 = 4 \pi G\,
h_A e^{-\ag} \frac{1 - \mu\bar{\mu}}{v \sqrt{\rho_0}}.
\ee
The only non-zero term in the integrand of (\ref{C_Box2}) is thus
\be    \label{C_integrand_mu_h}
\Box_2 \di_v\tilbox_1 = 4 \pi G h_A \dg^\circ_2 \mu,
\ee
and consequently 
\be	\label{C_A_mu_h}
C_A[\{\langle\mu\rangle_h,\cdot\}_\bullet,\dg_2]
= 16\pi G  \int_{\cN_A} h_A\dg_2^\circ \mu\: dv\, d^2\theta.
\ee
% \int_{S_0} \int_1^{\bar{v_A}}
If $\dg_2 = \dg$ then $\dg^\circ_2 = \dg$ so, summing the 
contributions of the two branches, one obtains
\be	\label{omega_mu_h}
\omega_\cN[\{\langle\mu\rangle_h,\cdot\}_\bullet,\dg] 
= \frac{1}{16\pi G} C[\{\langle\mu\rangle_h,\cdot\}_\bullet,\dg]
= \dg \langle\mu\rangle_h,
\ee
as expected. 

Notice that (\ref{C_A_mu_h}) is valid for all $\dg_2 \in \cC$. The restrictions 
$\dg_2 s_L = \dg_2 s_R = \dg_2 \rho_0 = 0$ were used only to pass to (\ref{omega_mu_h}).
This was done so that the  expression can be reused further on. For the same reason 
expressions (\ref{box_k} - \ref{C_integrand_mu_k}), (\ref{tilbox_zero} - \ref{mubar_N_avg_C}),
and (\ref{mubar_S0_avg_C}), that determine $C_A[\{\varphi,\cdot\}_\bullet,\dg_2]$ for
$\varphi = \langle\mu\rangle_k$, $\langle\bar{\mu}\rangle_h$, and $\langle\bar{\mu}\rangle_k$
respectively, will be calculated for unrestricted $\dg_2\in \cC$.

Suppose $\mu$ is smeared only over $S_0$. Let 
$\langle\mu\rangle_k = \int_{S_0} k \mu \:d^2\theta$, with $k$ a smooth density on $S_0$,
and let $\dg_1 = \{\langle\mu\rangle_k,\cdot\}_\bullet$. (\ref{mu_mu_brack}) and 
(\ref{mu_mubar_bracket}) then give
\bearr
\Box_1 = 0 \label{box_k}\\
\tilbox_1(p) = 4 \pi G \left[k \frac{1 - \mu\bar{\mu}}{\sqrt{\rho_0}}\right]_{p_0},
\eearr 
where $p_0$ is the base point on $S_0$ of the generator through $p$. As a consequence
% \be	\label{d_box_til_k} 
% \di_v \tilbox_1 = 0
% \ee
$\di_v \tilbox_1 = 0$ and the only non-zero term in the integrand of (\ref{C_Box2}) is
\be   \label{C_integrand_mu_k}
\tilbox_1\Box_2 = 4\pi G k \dg^\circ_2 \mu,
\ee
which leads once more to the result that (\ref{mu_brack_def}) is satisfied:
\be
\omega_\cN[\{\langle\mu\rangle_k,\cdot\}_\bullet,\dg] 
= \dg \langle\mu\rangle_k.
\ee

It is quite easy to check that $\{\langle\mu\rangle,\cdot\}_\bullet$ as defined by the structure
relations (\ref{mu_mu_brack} - \ref{mu_omega_L_0_bracket}) also satisifies (\ref{final_bracket_condition2}), 
that is $\{\langle\mu\rangle,\cdot\}_\bullet \in \cC$. We will return to this issue at the end of this section, 
where it will be shown that all the variations $\{\varphi,\cdot\}_\bullet$ defined by the structure relations 
presented in Section \ref{structure_relations} satisfy (\ref{final_bracket_condition2}).

Now let us turn to the brackets of $\bar{\mu}$ with $\bar{\mu}$ and $\mu$. These are determined, much 
like the brackets of $\mu$, by solving (\ref{final_bracket_condition1}) and
(\ref{final_bracket_condition2}) with $\varphi = \langle\bar{\mu}\rangle$, a smearing of
$\bar{\mu}$, and $\dg$ restricted to variations such that $\dg s_L = \dg s_R = \dg \rho_0 = 0$. 
The uniqueness of these brackets has already been established Subsection \ref{uniquenes_proof}.
% $\{\langle\mu\rangle, \mu\}_\bullet$ and 
% $\{\langle\mu\rangle,\bar{\mu}\}_\bullet$ goes over {\em mutatis mutandis}, showing that 
% $\{\langle\bar{\mu}\rangle, \mu\}_\bullet$ and $\{\langle\bar{\mu}\rangle,\bar{\mu}\}_\bullet$ 
% are uniquely determined. 
And the demonstration that $\dot{v}_A$ can be treated as constant goes over {\em mutatis mutandis} to the present case.
It remains only to show that the brackets (\ref{mu_mu_brack}) and
(\ref{mu_mubar_bracket}) indeed form a solution.

If $\dg_1 = \{\langle\bar{\mu}\rangle_h,\cdot\}_\bullet$, the variation generated by $\bar{\mu}$
smeared over the bulk of $\cN$, then, by (\ref{mu_mu_brack}) and (\ref{mu_mubar_bracket}),
\bearr
\tilbox_1 & = & 0     \label{tilbox_zero}\\
\Box_1 & = & \Box_1^L + \Box_1^R,    \label{box_eq_boxLR}
\eearr 
where $\Box_1^A(p) = 0$ if $p \in \cN - \cN_A$ and 
\be     \label{box_A}
\Box_1^A(p) = - 4 \pi G \int_{v(p)}^{\dot{v}} h_A e^{\ag} 
\frac{1 - \mu\bar{\mu}}{v \sqrt{\rho_0}} dv
\ee
if $p \in \cN_A$, with the integral taken along the generator of $\cN_A$ from $p$ to $S_A$. 
$\Box_1$ is in general discontinuous at $S_0$, since on $S_0$ (and only there) {\em two} 
generators pass through each point, and $\Box_1$ is the sum of the integrals along each of 
these. 

The integrand $\Box_1 \di_v \tilbox_2$ in the expression (\ref{C_Box2}) for $C_A$ is 
nevertheless unambiguously defined, because $\dg_2\bar{\mu}$ (and thus $\tilbox_2$) is 
smooth on $\cN_A$ for any $\dg_2 \in \cC$. In fact, the discontinuity in $\Box_1$ at $S_0$
does not affect the integral over $v$ in (\ref{C_Box2}). Changing the value of the integrand
at $S_0$ only will not change the value of this integral, so one may replace $\Box_1$ by 
$\Box_1^A$ there. Integrating by parts one then obtains
\be    \label{mubar_N_avg_C}
C_A = 2 \int_{S_0} d^2\theta \left\{ \tilbox_2(2\Box_1^A - \Box_1) + 2\int_1^{\dot{v}}  
\tilbox_2 \di_v \Box_1^A\: dv\right\},  
\ee                
where $\tilbox_1 = 0$, by (\ref{tilbox_zero}), and $\Box_1 = \Box_1^A = 0$ on $S_A$,
by (\ref{box_A}), have been used. When $\dg_2 = \dg$ the restrictions on $\dg$ imply that
$\dg_{2\,L}^\circ = \dg_2 = \dg_{2\,R}^\circ$ so the surface terms in the sum 
$C_L + C_R$ cancel and
\be
\tilbox_2 \di_v \Box_1^A = 4 \pi G h_A \dg \bar{\mu}
\ee
on $\cN_A$. As a result
\be \label{mubar_brack_def}
\omega_\cN[\{\langle\bar{\mu}\rangle_h,\cdot\}_\bullet,\dg] 
= \frac{1}{16\pi G} C[\{\langle\bar{\mu}\rangle_h,\cdot\}_\bullet,\dg]
= \dg \langle\bar{\mu}\rangle_h,
\ee
as should be.

When $\dg_1 = \{\langle\bar{\mu}\rangle_k,\cdot\}_\bullet$, the variation generated by 
$\bar{\mu}$ smeared over $S_0$, (\ref{mu_mu_brack}) and (\ref{mu_mubar_bracket}) imply
\bearr
\tilbox_1 = 0\\
\Box_1(p) = \left\{\begin{array}{cl}
- 4 \pi G \left[k \frac{1 - \mu\bar{\mu}}{\sqrt{\rho_0}}\right]_p & \mbox{if $p \in S_0$}\\
0 & \mbox{otherwise}
\end{array}\right.
\eearr 
By (\ref{C_Box2}) 
\be    \label{mubar_S0_avg_C}
C_A = -2\int_{S_0} \Box_1\tilbox_2\: d^2\theta = 8\pi G \dg_{2\,A}^\circ\langle\bar{\mu}\rangle_k.
\ee
Summing the contributions from both branches, recalling $\dg_L^\circ = \dg 
= \dg_R^\circ$, we find that (\ref{mubar_brack_def}) holds also when the average 
over $S_0$, $\langle\bar{\mu}\rangle_k$, replaces $\langle\bar{\mu}\rangle_h$.

\subsection{The brackets of $\mu$ with $\lam$, $\tau_R[f]$, and $\tau_L[g]$}

The brackets of $\mu$ with data other than $\mu$ and $\bar{\mu}$ are obtained by imposing
$\dg \langle\mu\rangle = \omega_\cN[\{\langle\mu\rangle,\cdot\}_\bullet,\dg]$ for 
{\em all} $\dg \in \cC$, using the expressions for the brackets found up to this point to simplify
this condition. 

Let us consider the case in which $\mu$ is smeared over the bulk of $\cN$. 
Since the restrictions on $\dg$ have been lifted $\dg \dot{v}_A$ need not vanish, so 
$\dg \langle\mu\rangle_h$ may not equal $\langle \dg\mu \rangle_h$. Rather
\be
\dg \langle\mu\rangle_h = \langle \dg\mu \rangle_h 
+ \sum_{A = L,R} \int_{S_A} h_A \mu \dg \dot{v}_A\: d^2\theta.
\ee
Nevertheless, just as in our earlier considerations, the variations of $\dot{v}_A$ ultimately
do not affect the calculation of the brackets:
\bearr
\omega_\cN[\{\langle\mu\rangle_h,\cdot\}_\bullet,\dg] 
= \langle\omega_\cN[\{\mu,\cdot\}_\bullet,\dg]\rangle_h %\nonumber\\
+ \sum_{A = L,R} \int_{S_A} h_A\,\mu\,\omega_\cN[\{\dot{v}_A,\cdot\}_\bullet,\dg]\:
d^2\theta,\nonumber\\&&
\eearr
and by (\ref{v_bar_eqn}) the second term is just
$\sum_{A = L,R} \int_{S_A} h \mu\,\dg \dot{v}_A\: d^2\theta$, so 
(\ref{final_bracket_condition1}) is equivalent to
\be  \label{mu_final_bracket_condition1bis}
\langle \dg\mu \rangle_h = \langle\omega_\cN[\{\mu,\cdot\}_\bullet,\dg]\rangle_h
\ \ \ \ \forall \dg\in \cC,
\ee
an equation like (\ref{mu_brack_def2}), in which variations of $\dot{v}_A$ play no role.

Let us therefore solve (\ref{mu_final_bracket_condition1bis}). 
Since we no longer require that $\dg \rho_0$ or $\dg s_A$ vanish
$\langle A_A[\{\mu,\cdot\}_\bullet,\dg]\rangle_h$ and 
$\langle B_A[\{\mu,\cdot\}_\bullet,\dg]\rangle_h$ might not vanish either. By (\ref{Bmu}) 
% \bearr
% \lefteqn{\langle B_A[\{\mu,\cdot\}_\bullet,\dg]\rangle_h}\nonumber\\
% && = \int_{S_0} 
% \frac{\di_{v_A} \mu}{(1 - \mu\bar{\mu})^2} 
% \{\langle\mu\rangle_h,\bar{\mu}\}_\bullet \: 
% (\dg\rho_0 - {\pounds}_{\xi_{\perp\,A}}\rho_0)\: d^2\theta.
% \eearr
\be
\langle B_A[\{\mu,\cdot\}_\bullet,\dg]\rangle_h = \int_{S_0} 
\frac{\di_{v_A} \mu}{(1 - \mu\bar{\mu})^2} 
\{\langle\mu\rangle_h,\bar{\mu}\}_\bullet \: 
(\dg\rho_0 - {\pounds}_{\xi_{\perp\,A}}\rho_0)\: d^2\theta.
\ee
($\dot{v}$ commutes with $\bar{\mu}$ so the average $\langle\cdot\rangle_h$
can be taken inside the bracket.) In the present case this is zero because
\be 
\{\langle\mu\rangle_h,\bar{\mu}(p)\}_\bullet = 4 \pi G 
\left[e^\ag \frac{1 - \mu\bar{\mu}}{v \rho_0}\right]_p 
\int_1^{v(p)} h e^{-\ag} \frac{1 - \mu\bar{\mu}}{v'} dv',
\ee
which vanishes when $p \in S_0$. 

By (\ref{C_A_mu_h})
\be
\langle C_A[\{\mu,\cdot\}_\bullet,\dg]\rangle_h 
 =  16 \pi G \int_{S_0}  \int_1^{\dot{v}_A}  h_A\dg^\circ_A\mu \:dv_A d^2\theta 
\ee
where 
\be   \label{dgcirc_def2}
\dg^\circ_A  =  \dg  - {\pounds}_{\xi_{\perp\,A}} - \frac{1}{2}(\dg \ln \rho_0
- {\pounds}_{\xi_{\perp\,A}} \ln \rho_0)\, v_A \di_{v_A}.
\ee
Adding up the $A$ and $C$ contributions from both branches (and taking into account 
$\xi_\perp = \dg s_A^k \di_{y_A^k}$) one obtains
\be
\langle \omega_\cN[\{\mu,\cdot\}_\bullet,\dg]\rangle_h = \langle\dg\mu\rangle_h 
+ \langle E\rangle_h,
\ee
with
\bearr
\fl E(p)  = & - \left[{\pounds}_{\xi_{\perp\,A}}\mu 
+ \frac{1}{2}(\dg \ln \rho_0 - {\pounds}_{\xi_{\perp\,A}} \ln \rho_0)\, v_A \di_{v_A} \mu 
\right]_p \nonumber \\
\fl &+ \frac{1}{16\pi G} \Big(\{\mu(p),\tau_L[\xi_{\perp\,L}]\}_\bullet
+ \{\mu(p),\tau_R[\xi_{\perp\,R}]\}_\bullet %\nonumber\\&&
+ 2 \int_{S_0} \{\mu(p),\lam\}_\bullet\, \dg\rho_0\: d^2\theta \Big)
\eearr
when $p\in \cN_A$. The first term in $E$, equal to $\dg^\circ_A \mu - \dg \mu$ on $\cN_A$, is ambiguous 
on $S_0$. The average $\langle E\rangle_h$ is a sum of an integral over $\cN_R$ and an 
integral over $\cN_L$, and the corresponding form of the first term of $E$ should be used in 
each. But ultimately $S_0$ makes no contribution to either integral since the integrand is finite 
there and $S_0$ has measure zero in the measure $h\,dv\,d^2\theta$ of the integral.

Equation (\ref{mu_final_bracket_condition1bis}) requiers $\langle E\rangle_h$ to vanish for all 
$h$, $\dg\rho_0$, $\dg s_L$, and $\dg s_R$ (satisfying $\dg s_A$ on $\di S_0$), so 
\bearr
\{\mu(p),\lam({\bftheta})\}_\bullet
& = & 4\pi G \frac{1}{\rho_0}\dg^2(\bftheta - \bftheta_p)
[v_A \di_{v_A} \mu]_p,
\label{mu_lam_bracket2}\\
\{\mu(p), \tau_A[f]\}_\bullet & = & 8\pi G[2{\pounds}_f \mu 
- \frac{{\pounds}_f \rho_0}{\rho_0} v_A\di_{v_A} \mu]_p,
\label{mu_omegaR_bracket2}\\
\{\mu(p), \tau_{-A}[g]\}_\bullet & = & 0,   \label{mu_omegaL_bracket2}
\eearr
hold as distributional equalities on $\cN_A$ with $A = L$ or $R$ ($-A$ indicating in each case the opposite branch). 
The requierment, stemming from (\ref{final_bracket_condition2}), 
that $\{\mu(p), \varphi \}_\bullet$ be smooth in $p$ on $\cN_A-S_0$ then implies that 
(\ref{mu_lam_bracket2} - \ref{mu_omegaL_bracket2}) hold for all $p \in \cN_A-S_0$. 
These expressions agree with the expressions for the same brackets given in 
(\ref{mu_lam_bracket} - \ref{mu_omegaL_bracket}).

Requiering (\ref{final_bracket_condition1}) and (\ref{final_bracket_condition2}) to hold for 
$\varphi = \langle \mu \rangle_k$ 
determines the brackets further. $B_A[\{\langle\mu\rangle_k,\cdot\}_\bullet,\dg]$ does not 
vanish, since by (\ref{mu_mubar_bracket})
\be
\{\langle\mu\rangle_k,\bar{\mu}(p)\}_\bullet = 4\pi G \big[\frac{k}{\rho_0} 
(1 - \mu\bar{\mu})^2\big]_p 
\ee
when $p \in S_0$. Rather
\be
B_A[\{\langle\mu\rangle_k,\cdot\}_\bullet,\dg] = 4\pi G 
\langle (\dg\ln\rho_0 - {\pounds}_{\xi_{\perp\,A}}\ln\rho_0)\di_{v_A}\mu \rangle_k. 
\ee
Equation (\ref{C_Box2}), and (\ref{box_k} - \ref{C_integrand_mu_k}), imply that
\be    \label{C_mu_k}
C_A[\{\langle\mu\rangle_k,\cdot\}_\bullet,\dg] = 8\pi G \langle\dg^\circ_A \mu\rangle_k,
\ee
so 
\be
\omega_\cN[\{\langle\mu\rangle_k,\cdot\}_\bullet,\dg] = \dg\langle\mu\rangle_k 
+ \langle F_L + F_R \rangle_k,
\ee
where 
\bearr
%\lefteqn{F_A(p)}\nonumber\\ 
\fl F_A(p) & = & \frac{1}{16\pi G} \left\{\{\mu(p),\tau_A[\xi_{\perp\,A}]\}_\bullet
+ \int_{S_0} \{\mu(p),\lam\}_\bullet\, \dg\rho_0\: d^2\theta \right\} \nonumber \\
\fl && - \frac{1}{2}\left[{\pounds}_{\xi_{\perp\,A}}\mu
+ \frac{1}{2}(\dg \ln \rho_0 - {\pounds}_{\xi_{\perp\,A}} \ln \rho_0)\, v_A \di_{v_A} \mu 
\right]_p\nonumber\\
\fl && 
+ \frac{1}{4}\left[(\dg\ln\rho_0 - {\pounds}_{\xi_{\perp\,A}}\ln\rho_0)\,\di_{v_A}\mu
\right]_p \\
\fl & = & \frac{1}{16\pi G} \left\{\{\mu(p),\tau_A[\xi_{\perp\,A}]\}_\bullet
+ \int_{S_0} \{\mu(p),\lam\}_\bullet\, \dg\rho_0\: d^2\theta \right\} 
%\nonumber \\&& {}
- \frac{1}{2}{\pounds}_{\xi_{\perp\,A}}\mu (p).
\eearr
Thus (\ref{final_bracket_condition1}) requiers that $\langle F_L + F_R \rangle_k = 0$. 
This leads to a rather simple result for the brackets:
\bearr
\{\mu(p),\lam(\bftheta)\}_\bullet & = & 0 
\label{mu_lam_0_bracket2}\\
\{\mu(p), \tau_R[f]\}_\bullet 
& = & 8\pi G {\pounds}_f \mu(p),  \label{mu_omega_R_0_bracket2}\\
\{\mu(p), \tau_L[g]\}_\bullet 
& = & 8\pi G {\pounds}_g \mu(p),  \label{mu_omega_L_0_bracket2}
\eearr
in agreement with (\ref{mu_lam_0_bracket} - \ref{mu_omega_L_0_bracket}). Once more the
smoothness requierment included in (\ref{final_bracket_condition2}) implies that 
(\ref{mu_lam_0_bracket2} - \ref{mu_omega_L_0_bracket2}) hold not only almost everywhere
but for all $p \in S_0$.

\subsection{The brackets of $\bar{\mu}$ with $\lam$, $\tau_R[f]$, and $\tau_L[g]$}

The brackets of $\bar{\mu}$ with $\lam$, $\tau_R[f]$, and $\tau_L[g]$ are obtained in a 
similar way. The calculations are a little more intricate, but on the other hand the two 
cases in which the average $\langle\bar{\mu}\rangle$ is taken over the bulk of $\cN$ and 
in which it is restricted to $S_0$ are treated simultaneously. In the following 
$\langle\cdot\rangle$ will represent both $\langle\cdot\rangle_h$ and $\langle\cdot\rangle_k$.

As in the calculation of the corresponding brackets of $\mu$, (\ref{final_bracket_condition1}) 
is equivalent to
\be  \label{mubar_final_bracket_condition1bis}
\langle \dg\bar{\mu} \rangle = \langle\omega_\cN[\{\bar{\mu},\cdot\}_\bullet,\dg]\rangle.
\ee
(\ref{Bmu}), (\ref{mu_mu_brack}), and (\ref{mu_mubar_bracket}) imply that
\bearr
\fl\sum_{A = L,R} \langle B_A[\{\bar{\mu},\cdot\}_\bullet,\dg]\rangle%\nonumber\\&&
& =  \sum_{A = L,R} \int_{S_0} \frac{\di_{v_A} \bar{\mu}}{(1 - \mu\bar{\mu})^2}
\{\langle\bar{\mu}\rangle,\mu\}_\bullet \: 
(\dg\rho_0 - {\pounds}_{\xi_{\perp\,A}}\rho_0)\: d^2\theta\\
\fl & = 16\pi G\langle I\rangle,
\eearr
with
\bearr
%\lefteqn{
\fl I(p) = - \frac{1}{4} \left[e^\ag \frac{1 - \mu\bar{\mu}}{v}\right]_p
%}\nonumber\\&& \times
\left[(\dg\ln\rho_0 - {\pounds}_{\xi_{\perp\,L}}\ln \rho_0)
\frac{\di_{v_L} \bar{\mu}}{1 - \mu\bar{\mu}}\right. % \nonumber\\ && \ \
+ \left. (\dg\ln\rho_0 - {\pounds}_{\xi_{\perp\,R}}\ln \rho_0)
\frac{\di_{v_R} \bar{\mu}}{1 - \mu\bar{\mu}}\right]_{p_0},\nonumber\\
\eearr
where $p_0$ is the base point on $S_0$ of the generator through $p \in \cN$.

When $\langle\cdot\rangle$ is an average over the bulk of $\cN$ (\ref{box_eq_boxLR} -- \ref{mubar_N_avg_C})
imply that $\langle C[\{\bar{\mu},\cdot\}_\bullet,\dg]\rangle = 16 \pi G \langle J \rangle$ with
\be
J(p) = \dg^\circ_A\bar{\mu}(p) - \frac{1}{2}\left[e^\ag \frac{1 - \mu\bar{\mu}}{v_A}\right]_p
\left[\frac{\dg^\circ_A\bar{\mu} - \dg^\circ_{-A}\bar{\mu}}{1 - \mu\bar{\mu}}\right]_{p_0}.
\ee
(Once more $A$ labels the branch of $p$ while $-A$ labels the opposite branch: $-L = R$, $-R = L$).
This expression is also valid when $\langle\cdot\rangle$ is an average over $S_0$. In that 
case (\ref{mubar_S0_avg_C}) shows that $\langle C[\{\bar{\mu},\cdot\}_\bullet,\dg]\rangle_k =
8\pi G\langle \dg^\circ_L\bar{\mu} + \dg^\circ_R\bar{\mu}\rangle_k$, which coincides 
with $16 \pi G \langle J \rangle_k$.

Thus
\bearr
%\lefteqn{
\fl \langle\omega_\cN[\{\bar{\mu},\cdot\}_\bullet,\dg]\rangle  = \langle I + J \nonumber\\
\fl \ \ \ \ \ \ \ \ \ + \frac{1}{16\pi G} \left\{\{\bar{\mu}(p),\tau_L[\xi_{\perp\,L}]\}_\bullet
+ \{\bar{\mu}(p),\tau_R[\xi_{\perp\,R}]\}_\bullet
+ 2 \int_{S_0} \{\bar{\mu}(p),\lam\}_\bullet\, \dg\rho_0\: d^2\theta \right\}\rangle.
%\nonumber\\ && {}
\eearr
Equation (\ref{final_bracket_condition1}) therefore requires that
\bearr
\fl 0 & = &\langle - \left[{\pounds}_{\xi_{\perp\,A}}\bar{\mu}
+ \frac{1}{2}(\dg \ln \rho_0 - {\pounds}_{\xi_{\perp\,A}} \ln \rho_0)\, v_A \di_{v_A} \bar{\mu} 
\right]_p\nonumber\\
\fl && {} + \frac{1}{2}\left[{\pounds}_{\xi_{\perp\,A}}\bar{\mu} 
- {\pounds}_{\xi_{\perp\,-A}}\bar{\mu}
- (\dg \ln \rho_0 - {\pounds}_{\xi_{\perp\,-A}} \ln \rho_0)\, \di_{v_{-A}}\bar{\mu} 
\right]_{p_0} \frac{1}{v_A(p)}\: e^{-2\int_{p_0}^p \frac{\mu d\bar{\mu}}{1 - \mu\bar{\mu}}}
\nonumber\\
\fl && {} + \frac{1}{16\pi G} \left\{\{\bar{\mu}(p),\tau_L[\xi_{\perp\,L}]\}_\bullet
+ \{\bar{\mu}(p),\tau_R[\xi_{\perp\,R}]\}_\bullet
+ 2 \int_{S_0} \{\bar{\mu}(p),\lam\}_\bullet\, \dg\rho_0\: d^2\theta \right\}\rangle.
%\nonumber\\ && {}
\label{mubar_S0_brack_conditions}
\eearr
(Here the identity    
$e^{\ag(p)}\: [1 - \mu\bar{\mu}]_p/[1 - \mu\bar{\mu}]_{p_0}
 = e^{-2\int_{p_0}^p \frac{\mu d\bar{\mu}}{1 - \mu\bar{\mu}}}$
has been used.) This implies the brackets
\bearr
\fl \{\bar{\mu}(p),\lam(\bftheta)\}_\bullet 
& = & 4\pi G \frac{1}{\rho_0}\dg^2(\bftheta - \bftheta_p)\left\{[v_A \di_{v_A} 
\bar{\mu}]_p 
+ [\di_{v_{-A}}\bar{\mu}]_{p_0}\frac{1}{v_A(p)} e^{-2\int_{p_0}^p 
\frac{\mu d \bar{\mu}}{1 - \mu\bar{\mu}}} \right\},
%\nonumber\\ && {}
\label{mubar_lam_bracket2}\\
\fl \{\bar{\mu}(p),\tau_A[f]\}_\bullet & = & 
8\pi G\left\{[2{\pounds}_f \bar{\mu} - \frac{{\pounds}_f \rho_0}
{\rho_0} v_A\di_{v_A} \bar{\mu}]_p 
- [{\pounds}_f \bar{\mu}]_{p_0}
\frac{1}{v_A(p)} e^{-2\int_{p_0}^p 
\frac{\mu d \bar{\mu}}{1 - \mu\bar{\mu}}}\right\},
%\nonumber\\ && {}
\label{mubar_omegaR_bracket2}\\
\fl \{\bar{\mu}(p), \tau_{-A}[g]\}_\bullet & = & 
8\pi G [{\pounds}_g \bar{\mu} 
- \frac{{\pounds}_g \rho_0}{\rho_0}\di_{v_{-A}} \bar{\mu}]_{p_0} 
\frac{1}{v_A(p)} 
e^{-2\int_{p_0}^p 
\frac{\mu d \bar{\mu}}{1 - \mu\bar{\mu}}},
\label{mubar_omegaL_bracket2}  
\eearr
equivalent to (\ref{mubar_lam_bracket}-\ref{mubar_omegaL_bracket}). The smoothness
of variations of $\bar{\mu}$ in $\cC$ implies that these equations hold for all $p \in \cN$.

\subsection{The brackets between $\lam$, $\tilde{\tau}_{R\,i}$, and $\tilde{\tau}_{L\,m}$}

It remains to find the brackets between $\lam$, $\tilde{\tau}_{R\,i}$, and 
$\tilde{\tau}_{L\,m}$. These will be obtained by imposing (\ref{final_bracket_condition1}) with 
$\varphi$ taken to be smearings of each of the three data in turn. The expressions for the brackets obtained
so far imply that
\be
\omega_\cN[\{\langle \lam \rangle_k, \cdot\}_\bullet , \dg] = \dg \langle \lam \rangle_k 
+ \frac{1}{16 \pi G} \langle K_R + K_L \rangle_k,
\ee
where
%\begin{widetext} 
\bearr
\fl K_A(p) & = & \int_{S_0} \{\lam(p), \lam\}_\bullet\, \dg \rho_0\: d^2\theta 
+ \{\lam(p), \tau_A[\xi_{\perp\,A}]\}_\bullet 
\nonumber\\
\fl && + \,8\pi G \left[\frac{1}{(1 - \mu\bar{\mu})^2}(\di_{v_A} \mu\, \dg^{y_A}\bar{\mu} +
\di_{v_A}\bar{\mu}\, \dg^{y_A}\mu)\right]_p \nonumber\\
\fl && + \int_{S_0} \frac{1}{(1 - \mu\bar{\mu})^2}\{\lam(p), \bar{\mu}\}_\bullet \di_{v_A} \mu\, 
\dg^{y_A} \rho_0\: d^2\theta 
%\nonumber\\&&
+ C_A[\{\lam(p), \cdot\}_\bullet, \dg].
\eearr
The individual terms can be simplified further. By (\ref{mubar_lam_bracket2})
\be
%\lefteqn{
\fl\int_{S_0} \frac{1}{(1 - \mu\bar{\mu})^2}\{\lam(p), \bar{\mu}\}_\bullet \di_{v_A} \mu 
\,\dg^{y_A} \rho_0\: d^2\theta
%}\nonumber\\&& \ \ \ \ \ \ \ \ \ \ \ \ \
= - 4\pi G \left[\frac{\di_{v_R} \bar{\mu} + \di_{v_L} \bar{\mu}}{(1 - \mu\bar{\mu})^2}
\di_{v_A} \mu\, \dg^{y_A} \ln\rho_0\right]_p.
\ee
$C_A[\{\lam(p), \cdot\}_\bullet, \dg]$ can be evaluated as follows:
Let $\Dg_\lam = \{\lam(p), \cdot\}_\bullet$, then by (\ref{dgcirc_def2})
\be 
\Dg^\circ_\lam = \{\lam(p), \cdot\}_\bullet + 4\pi G \frac{1}{\rho_0}\dg^2(\bftheta - \bftheta(p))
v_A \di_{v_A}.
\ee 
Thus by (\ref{mu_lam_bracket2}), (\ref{mu_lam_0_bracket2}), and 
(\ref{mu_mubar_bracket})
\bearr     %\label{dg_lam_circ_mu}
\Dg^\circ_\lam \mu(q) & = & 4\pi G \frac{1}{\rho_0} \dg^2(\bftheta(q) - \bftheta(p))
(\di_{v_A}\mu)_p H(q,p)\\
& = & -\left[\frac{\di_{v_A}\mu}{(1 -\mu\bar{\mu})^2}\right]_p \{\bar{\mu}(p), \mu(q)\}_\bullet.
\label{dg_lam_circ_mu}
\eearr
(Since $p \in S_0$, $\dg^2(\bftheta(q) - \bftheta(p)) H(q,p)$ vanishes except when $q = p$.)
Similarly, by (\ref{mubar_lam_bracket2}),
\bearr
\Dg^\circ_\lam \bar{\mu}(q) & = & - 4\pi G \frac{1}{\rho_0}\dg^2(\bftheta(q) - \bftheta(p))
\frac{1}{v_A(q)} e^{-2\int_p^q \frac{\mu d \bar{\mu}}{1 - \mu\bar{\mu}}} 
(\di_{v_{-A}}\bar{\mu})_p\\
& = & -\left[\frac{\di_{v_{-A}}\bar{\mu}}{(1 -\mu\bar{\mu})^2}\right]_p 
\{\mu(p), \bar{\mu}(q)\}_\bullet,
\eearr
so, by (\ref{C_mu_k}) and (\ref{mubar_S0_avg_C}),
\bearr
C_A[\{\lam(p), \cdot\}_\bullet, \dg] 
& = & - \left[\frac{\di_{v_{-A}}\bar{\mu}}{(1 - \mu\bar{\mu})^2}\right]_p
C_A[\{\mu(p), \cdot\}_\bullet, \dg]\nonumber\\ 
&& - \left[\frac{\di_{v_A}\mu}{(1 - \mu\bar{\mu})^2}\right]_p
C_A[\{\bar{\mu}(p), \cdot\}_\bullet, \dg]\\
& = & {}- 8\pi G\left[\frac{1}{(1 - \mu\bar{\mu})^2}(\di_{v_A}\mu\, \dg_A^\circ \bar{\mu}
		+ \di_{v_{-A}}\bar{\mu}\, \dg_A^\circ \mu)\right]_p.
\eearr
Summing all the terms one finds that
\bearr
\fl \langle K_R + K_L \rangle_k
& = & 2 \int_{S_0} \{\langle \lam \rangle_k, \lam\}_\bullet\, \dg \rho_0\: d^2\theta 
+ \{\langle \lam \rangle_k, \tau_R[\xi_{\perp\,R}]\}_\bullet 
+ \{\langle \lam \rangle_k, \tau_L[\xi_{\perp\,L}]\}_\bullet 
\nonumber\\
\fl && {}- 8\pi G \langle \frac{1}{(1 - \mu\bar{\mu})^2} (\di_{v_R}\bar{\mu} - \di_{v_L}\bar{\mu})  
({\pounds}_{\xi_{\perp\,R}}\mu - {\pounds}_{\xi_{\perp\,L}}\mu ) \rangle_k. 	\label{KR_KL}
\eearr

By (\ref{final_bracket_condition1}) $\langle K_R + K_L \rangle_k$ must vanish. Thus the 
coefficients of $\xi_{\perp\,R} = \dg s_R^i \di_{y^i}$ and 
$\xi_{\perp\,L} = \dg s_L^m \di_{y^m}$ in this expression must vanish, which leads immediately 
to the formula (\ref{lam_omega_R_bracket}) for the bracket of $\lam$ and $\tau_R[f]$, and an 
analogous formula for the bracket of $\lam$ with $\tau_L[g]$. Only the first term on the right 
hand side of (\ref{KR_KL}) depends on $\dg\rho_0$, so the bracket of $\lam$ with $\lam$ 
vanishes, in agreement with (\ref{lam_lam_brack}).

Now let us solve (\ref{final_bracket_condition1}) with $\varphi = \tau_A[f]$.
For definiteness we take $A = R$, the calculation for $A = L$ being entirely analogous. 
\be
\omega_\cN[\{\tau_R[f], \cdot\}_\bullet, \dg] = \dg \tau_R[f] 
+ \frac{1}{16 \pi G} M[f],
\ee
where 
\bearr
M[f]
& = & 2\int_{S_0} \{\tau_R[f], \lam\}_\bullet\, \dg \rho_0\: d^2\theta 
+ \{\tau_R[f], \tau_R[\xi_{\perp\,R}]\}_\bullet + \{\tau_R[f], \tau_L[\xi_{\perp\,L}]\}_\bullet 
\nonumber\\&&
+ 4\pi G \int_{S_0}{\pounds}_f \rho_0\,\di_{v_R} e_{pq}\,\dg^{y_R} e^{pq}\:d^2\theta \nonumber\\
&& {}- \frac{1}{4}\int_{S_0} (\{\tau_R[f], e^{pq}\}_\bullet + 16\pi G{\pounds}_f e^{pq}) 
\di_{v_R}e_{pq}\,\dg^{y_R}\rho_0\: d^2\theta 
\nonumber\\&&
 - \frac{1}{4}\int_{S_0} \{\tau_R[f], e^{pq}\}_\bullet  
\di_{v_{L}}e_{pq}\,\dg^{y_{L}}\rho_0\: d^2\theta \nonumber\\
&& {}+ C[\{\tau_R[f], \cdot\}_\bullet, \dg].	\label{M2}
\eearr
Here the relations $\{\tau_R[f], s^i_R\}_\bullet 
= - 16\pi G f^i$ and $\{\tau_R[f], s^m_L\}_\bullet = 0$ have been used. These imply that if 
$\Dg_{\tau_R} \equiv \{\tau_R[f], \cdot\}_\bullet$ then 
$\Dg^{y_R}_{\tau_R} = \{\tau_R[f], \cdot\}_\bullet + 16\pi G {\pounds}_f$ and
$\Dg^{y_{L}}_{\tau_R} = \{\tau_R[f], \cdot\}_\bullet$.

Simplifying the terms of (\ref{M2}) we find
\bearr
\fl M[f] & = & {} - 16\pi G\int_{S_0} \frac{1}{(1 - \mu\bar{\mu})^2}[\di_{v_R}\bar{\mu} 
- \di_{v_{L}}\bar{\mu}]{\pounds}_f\mu\, \dg \rho_0\: d^2\theta
\nonumber\\
\fl && 
+ \{\tau_R[f], \tau_R[\xi_{\perp\,R}]\}_\bullet + \{\tau_R[f], \tau_L[\xi_{\perp\,L}]\}_\bullet 
\nonumber\\
\fl && {} - 16\pi G \int_{S_0}\frac{1}{(1 - \mu\bar{\mu})^2}\,{\pounds}_f \rho_0[\di_{v_R}\mu
(\dg\bar{\mu} - {\pounds}_{\xi_{\perp\,R}}\bar{\mu}) + \di_{v_R}\bar{\mu}
(\dg\mu - {\pounds}_{\xi_{\perp\,R}}\mu)]\:d^2\theta \nonumber\\
\fl && {} - 8\pi G \int_{S_0}\frac{1}{(1 - \mu\bar{\mu})^2}[{\pounds}_f \mu\,\di_{v_R}\bar{\mu} + 
({\pounds}_f \bar{\mu} - {\pounds}_f \ln \rho_0\,\di_{v_R}\bar{\mu})\di_{v_R}\mu
\nonumber\\ \fl && \ \ \ \ \ \ \ \ \ \ \ \ \ \ \ \ \ \ \ \ \ \ \ \ \ \ \ \ \  
-2({\pounds}_f \mu\,\di_{v_R}\bar{\mu}
+ {\pounds}_f \bar{\mu}\,\di_{v_R}\mu)] \dg^{y_R}\rho_0\: d^2\theta \nonumber\\
\fl && {} - 8\pi G \int_{S_0}\frac{1}{(1 - \mu\bar{\mu})^2}[{\pounds}_f \mu\,\di_{v_{L}}\bar{\mu} +
({\pounds}_f \bar{\mu} - {\pounds}_f \ln \rho_0\,\di_{v_R}\bar{\mu})\di_{v_{L}}\mu] 
\dg^{y_{L}}\rho_0\: d^2\theta 
\nonumber\\
\fl && 
+ C[\{\tau_R[f], \cdot\}_\bullet, \dg].
\eearr

To evaluate $C[\{\tau_R[f], \cdot\}_\bullet, \dg]$ note that the $R$ branch modified 
variation corresponding to $\Dg_{\tau_R} \equiv \{\tau_R[f], \cdot\}_\bullet$ is
\be   \label{dg_tau_RR}
\Dg^\circ_{\tau_R\,R} = \{\tau_R[f], \cdot\}_\bullet 
+ 16\pi G{\pounds}_f - 8\pi G{\pounds}_f \ln\rho_0\, v_R \di_{v_R},
\ee
while the $L$ branch modified variation is simply
\be   \label{dg_tau_RL}
\Dg^\circ_{\tau_R\,L} = \{\tau_R[f], \cdot\}_\bullet.
\ee

Thus 
\be    \label{tau_boxes}
\begin{array}{ll}
\Box_{\tau_R\,R}(p) = 0 & \ \ \ \forall p \in \cN_R - S_0\\
\Box_{\tau_R\,L}(p) = 0 & \ \ \ \forall p \in \cN_L - S_0\\
\Box_{\tau_R\,R}(p) = 8\pi G \big[\frac{\sqrt{\rho_0}}{1 - \mu\bar{\mu}}
({\pounds}_f\mu - {\pounds}_f \ln\rho_0\, \di_{v_R}\mu)\big]_p 
& \ \ \ \forall p \in S_0\\
\Box_{\tau_R\,L}(p) 
= - 8\pi G \big[\frac{\sqrt{\rho_0}}{1 - \mu\bar{\mu}}{\pounds}_f\mu\big]_p 
& \ \ \ \forall p \in S_0\\
\tilbox_{\tau_R\,R}(p) 
= 8\pi G \big[\frac{\sqrt{\rho_0}}{1 - \mu\bar{\mu}}{\pounds}_f\bar{\mu}\big]_{p_0}
& \ \ \ \forall p \in \cN_R\\
\tilbox_{\tau_R\,L}(p) 
= - 8\pi G \big[\frac{\sqrt{\rho_0}}{1 - \mu\bar{\mu}}({\pounds}_f\bar{\mu}
- {\pounds}_f\ln\rho_0\, \di_{v_R}\bar{\mu})\big]_{p_0}
& \ \ \ \forall p \in \cN_L
\end{array}
\ee
Note that $\Box_{\tau_R}$ and $\di_v \tilbox_{\tau_R}$ both vanish everywhere on 
$\cN - S_0$. Thus, substituting into (\ref{C_Box2}), one obtains
\bearr
\fl C[\{\tau_R[f], \cdot\}_\bullet, \dg] = - 16\pi G \int_{S_0} & 
\frac{\rho_0}{(1 - \mu\bar{\mu})^2}[ 
%}\nonumber\\ &&  
({\pounds}_f\mu - {\pounds}_f\ln\rho_0\, \di_{v_R}\mu)\,\dg_R^\circ \bar{\mu}
   - {\pounds}_f\bar{\mu}\,\dg_R^\circ\mu \nonumber\\
\fl &%\ \ \ \ \ \ \ \ \ \ \ \ \ \ \ \ \ \ \ \ \ \ \ \ \ \ \ \ 
{} - {\pounds}_f\mu\,\dg_L^\circ\bar{\mu}
   + ({\pounds}_f\bar{\mu} - {\pounds}_f\ln\rho_0\, \di_{v_R}\bar{\mu})\,\dg_L^\circ \mu]
   \:d^2\theta.
\eearr

After a rather large number of cancellations the sum of all contributions to $M[f]$ becomes
\bearr
\fl M[f] = & \{\tau_R[f], \tau_R[\xi_{\perp\,R}]\}_\bullet
+ \{\tau_R&[f], \tau_L[\xi_{\perp\,L}]\}_\bullet \nonumber\\
\fl & {} + 16\pi G\int_{S_0} \frac{\rho_0}{(1 - \mu\bar{\mu})^2}\big[&
{\pounds}_f\mu ({\pounds}_{\xi_{\perp\,R}}\bar{\mu} - {\pounds}_{\xi_{\perp\,R}}\ln\rho_0\, 
\di_{v_R}\bar{\mu}) \nonumber\\ 
\fl &&- {\pounds}_{\xi_{\perp\,R}}\mu ({\pounds}_f\bar{\mu}
- {\pounds}_f\ln\rho_0\, \di_{v_R}\bar{\mu})\nonumber\\
\fl && 
{}- {\pounds}_f\mu ({\pounds}_{\xi_{\perp\,L}}\bar{\mu} - {\pounds}_{\xi_{\perp\,L}}\ln\rho_0\, 
\di_{v_L}\bar{\mu})\nonumber\\
\fl && {} + {\pounds}_{\xi_{\perp\,L}}\mu ({\pounds}_f\bar{\mu} 
- {\pounds}_f\ln\rho_0\, \di_{v_R}\bar{\mu})\big]\:d^2\theta. \label{M_final}
\eearr

(\ref{final_bracket_condition1}) requires that $M[f] = 0$, so the brackets of $\tau_R$ with
$\tau_R$ and $\tau_L$ can be read off directly from (\ref{M_final}), and are found to coincide 
with (\ref{omega_omega_R_bracket}) and (\ref{omega_R_omega_L_bracket}). Repeating the 
calculation for $\varphi = \tau_L[g]$ yields analogous expressions for brackets of the twist variables 
with $L$ and $R$ interchanged, including an expression for $\{\tau_L[g], \tau_R[f]\}_\bullet$ 
equivalent to the one already obtained. 
%(given the antisymmetry of $\{\cdot,\cdot\}_\bullet$).

\subsection{Verification of $\{\varphi,\cdot\}_\bullet \in \cC$}

We have now imposed (\ref{final_bracket_condition1}) for all weighted averages $\varphi$ of the 
initial data over $\cN$ or $S_0$, and shown that the brackets presented in
(\ref{lam_rho_bracket} -- \ref{omega_R_omega_L_bracket}), (\ref{mu_mu_brack} -- \ref{mubar_omegaL_bracket})
form the unique solution to these 
conditions, {\em provided} (\ref{final_bracket_condition2}), $\{\varphi,\cdot\}_\bullet \in \cC$,
holds for all these $\varphi$. Now we must verify that (\ref{final_bracket_condition2}) in fact 
holds with the brackets that have been found. 

This amounts to demonstrating that $\{\varphi,\dot{\mu}_A \}_\bullet = 0$ and $\{\varphi, s_A \}_\bullet = 0$, for 
$A = L,R$, that $\{\varphi,\bar{\mu} \}_\bullet = 0$ is smooth on each branch of $\cN$ and continuous at $S_0$, 
that $\{\varphi, \mu \}_\bullet$ is smooth on $\cN_L - S_0$, $\cN_R - S_0$ and $S_0$ with a possible step discontinuities 
between these three manifolds, and that the variations under $\{\varphi, \cdot \}_\bullet$ of the remaining data 
are smooth on $S_0$. 
The smoothness and continuity conditions, as well as $\{\varphi, s_A \}_\bullet = 0$ follow immediately from
the smoothness conditions satisfied by the weighting functions in $\Phi$ and the explicit expressions for the brackets.

As to the condition $\{\varphi,\dot{\mu}_A \}_\bullet = 0$, by (\ref{transformed_dmu}) this condition 
is equivalent to requiering that
\be
\Dg^\circ_{\varphi\,A} \mu(q) = 0\ \ \ \forall q \in S_A 
\ee
on each of the two branches, where $\Dg^\circ_{\varphi\,A}$ is the modified variation 
corresponding to $\Dg_\varphi \equiv \{\varphi,\cdot\}_\bullet$. 

For $\varphi$ a smearing of $\mu$, $\rho_0$, $s_L$, or $s_R$ the variation
$\Dg^\circ_{\varphi\,A} \mu(q)$ vanishes because these data $\bullet$ commute with $\mu$, 
$\rho_0$, $s_L$, and $s_R$. The case $\varphi = \langle\bar{\mu}\rangle$ is also simple. Because
$\bar{\mu}$ $\bullet$ commutes with $\rho_0$, $s_L$, and $s_R$ the modified variation
generated by $\langle\bar{\mu}\rangle$ is just
$\Dg^\circ_{\langle\bar{\mu}\rangle}\mu(q) = \{\langle\bar{\mu}\rangle, \mu(q)\}_\bullet$. 
But when $q \in S_A$ this bracket vanishes. By (\ref{mu_mubar_bracket}) it receives contributions
only from $\bar{\mu}$ on $S_A$ which have zero weight in the average $\langle\bar{\mu}\rangle$.
(See (\ref{box_A}).)

There remain two non-trivial cases to consider: $\varphi = \langle\lam\rangle_k$ and 
$\varphi = \tau_A[f]$. By (\ref{dgcirc_def2}) 
\be   \label{dg_lam_circ}
\Dg^\circ_{\lam\, A} = \{\lam(p), \cdot\}_\bullet 
+ 4\pi G \frac{1}{\rho_0}\dg^2(\theta - \theta(p))v_A \di_{v_A}.
\ee 
$\Dg^\circ_{\lam\,A} \mu(q)$ was calculated in (\ref{dg_lam_circ_mu}), and is zero for 
$q \in S_A$. The bracket (\ref{mu_lam_bracket2}) ensures that the first and second terms in 
(\ref{dg_lam_circ}) cancel when applied to $\mu(q)$. Similarly, the form of $\Dg^\circ_{\tau_R}$ 
on each branch is given in (\ref{dg_tau_RR}) and (\ref{dg_tau_RL}), with analogous forms 
holding for $\Dg^\circ_{\tau_L}$, and their action on $\mu(q)$ is found to vanish for $q \in S_A$. 
(See the first two lines of (\ref{tau_boxes}).) 

Equation (\ref{final_bracket_condition2}) thus holds. This completes the demonstration that the brackets 
we have proposed constitute the unique solution to (\ref{final_bracket_condition1}) and (\ref{final_bracket_condition2}).
Unlike in \cite{MR07} no auxiliary assumptions were made about the 
brackets. In particular, it was assumed in \cite{MR07} that the brackets respect causality, that is, that 
functions of the gravitational field on causally unconnected domains commute. Here this 
emerges as a consequence of the form of the symplectic form.

\section{Discussion}

One of the first thing one would want, presented with such an intricate and intricately derived result as is the Poisson
bracket on free null initial data obtained here is check it on some examples to see if it is correct. This has been done
in a sense in the context of cylindrical gravitational waves. Starting from standard canonical comutation relations for 
spacelike initial data, Korotkin and Samtleben \cite{KorotkinSamtleben} obtain the Poisson brackets for an alternative complete 
set of data which they call the ``monodromy matrix''. In \cite{FR17} A. Fuchs and the author obtain the same Poisson brackets 
for these data, starting from (the cylindrically symmetric version of) the Poisson brackets on free null initial data found here.

One would also want to see what the underlying structure is that is hiding in all this apparent complexity. The monodromy matrix
of \cite{KorotkinSamtleben} can be regarded as null initial data, and has simpler Poisson brackets, associated with a Yangian 
algebra, so in the cylindrically symmetric case there are the beginnings of some understanding of the structure.
In \cite{FR17} we also point out that the monodromy matrix captures all the degrees of freedom of the conformal 2-metric {\em except}
except the shock wave modes that do not propagate into the interior of the domain of dependence of $\cN$. (Although we do not give a proof 
of this claim there). As a result the bracket on these data preserve reality in a straightforward way.

It would also be very interesting to know the extent to which the form of the $\bullet$ bracket depends on choices made 
in calculating them, and to what extent they are inevitable. The author believes that the brackets of the so called geometrical data
are essentially fixed by general relativity, because on generic spacetimes, for sufficiently small $\cN$ there exist many observables 
in the sense of our definition, and these fix the brackets. Another strong indication that there is little room to alter the bracket,
at least the brackets among $\mu$ and $\bar{\mu}$, is the calculation in Subsection \ref{complex_modes}, which shows that the perturbation of
$\mu$ and $\bar{\mu}$ by a source localized at a point on $\cN_A$ must be of the form produced by the brackets found here. 
To settle this question definitively one would have to find necessary and sufficient conditions for the matching of the $\bullet$ bracket 
to the Peierls bracket on observables, instead of only sufficient conditions. The chief obstacle is of course that we don't know much 
about the set of observables. This is probably a hard way to improve the results presented here, but characterizing the set of observables 
is very interesting in its own right.

A way forward that might be easier is to try to define the Poisson bracket as much as possible in the standrad way as the inverse of
the symplectic form. This requires a complete characterization of the gauge and non-gauge diffeomorphisms, which in turn requires a good
understanding of the boundary term $\omega_\cN[\pounds_\xi, \dg]$ (\ref{boundary_diff}). This is also interesting for other purposes:
developing Hamiltonians to evolve the data to other double null sheets; defining quasi-local energy, momentum, and angular momentum;
understanding what, if any, is the physical significance of the diffeomorphism data $s_L$ and $s_R$. 

A smaller project is to find out whether, or to what extent the condition (\ref{auxbracketdef}) and the hypothesys that $\{\cdot,\cdot\}_\bullet$
is a Poisson bracket, an antisymmetric biderivation satisfying the Jacobi relations, determines the $\bullet$ bracket. Here $\{\cdot,\cdot\}_\bullet$
has been treated as {\em a priori} simply a derivation in each of its arguments, and conditions stronger than (\ref{auxbracketdef}) have been imposed 
which ensure among other things that it is a Poisson bracket. But this may be introducing arbitrariness in the calculation.
In fact the bracket was originally found more or less in this way, from (\ref{auxbracketdef}) and the Jacobi relations, but by {\em ad hoc} 
procedures which would have to be formalized.

One would also want to understand the appearance of the imaginary part in the bracket. As we have argued, the imaginary part of the bracket
is not a problem in a theory of the interior of the domain of dependence of $\cN$, but it is certainly mysterious.
It does seem to be necessary though, within the present formalism, because if one simply drops the imaginary part and keeps only the real part of the bracket
in the expansion (\ref{real_imaginary_parts}) then one is left with the pre-Poisson bracket of \cite{MR07}, which does not satisfy the Jacobi relations.

The Poisson bracket on free null initial data was calculated chiefly as a starting point for a canonical quantum theory of General Relativity
without constraints. A promising aspect of the formalism is that a polarization of the phase space immediately suggests itself: $\mu$, $\rho_0$,
$s_L$ and $s_R$ commute with the other half of the initial data $\bar{\rho}$, $\lam$, $\tilde{\tau}_L$, and $\tilde{\tau}_R$. This could be the 
basis of quantization. In \cite{FR17} A. Fuchs and the author have advanced in another direction, namely, importing the detailed (although incomplete)
results about the quantization of cylindrical gravitational waves of \cite{KorotkinSamtleben} to null canonical gravity, with 
and without cylindrical symmetry.

\section{Acknowledgements}

%%\centerline{------}

I thank R. Gambini and C. Rovelli for key discussions, and the institutions at which this work 
was carried out, CPT in Luminy, AEI in Potsdam, UNAM in Morelia, PI in Waterloo, and the Universidad de la 
Rep\'ublica, for their support. I have benefitted from the questions, comments, and encouragement of 
J. Zapata, L. Smolin, R. Epp, P. Aichelburg, A. Perez, L. Freidel, J. Lewandowski, A. Rendall, H. Friedrich, 
C. Kozameh, M. Bradley, I. Bengtsson, J. Peraza, and M. Paternain among others. 

This work has been supported by PAPIIT-Universidad Nacional Aut\'onoma de M\'exico through 
grant IN109415.

\appendix

\section{The $a_A^\alpha$ charts and admissible variations}\label{a_chart_appendix}

In this appendix the definitions of admissible variations, and of the $a_A$ charts, given in \cite{MR13}, 
are recalled for the convenience of the reader.

\begin{definition}\label{admissible_def} 
A variation of the spacetime metric is {\em admissible} if 

\begin{itemize}
 \item[1] it is a real, smooth solution to the linearized field equations,
 \item[2] it preserves the double null sheet $\cN$. That is, the varied generators from $S_0$ sweep out the same
 null branches $\cN_A$ in spacetime as do the original unvaried generators,
 \item[3] it leaves each generator that is contained in the boundary $\di\cN$ unchanged,
 \item[4] it leaves the area density $\dot{\rho}_A$ on the truncation surfaces $S_A$ unchanged in the fixed
charts $y_A$,
\end{itemize}
and finally,
\begin{itemize}
 \item[5] within a spacetime neighborhood of each $S_A$ it leaves invariant a special 
 metric dependent chart formed by the $a_A$ coordinates defined below.
\end{itemize}
\end{definition}

In \cite{MR13} it is shown that one may limit the variations $\dg$ appearing in the condition (\ref{auxbracketdef}),
\be                    
        \dg A = \omega_\cN[\{A,\cdot\}_\bullet,\dg]\ \ \ \forall \dg\in L_g^0,
\ee
to admissible ones without weakening this condition, and that one may always chose among the variations
$\{A,\cdot\}$ that satisfy this condition an admissible one. The reason is, in sum, that because of diffeomorphism
gauge invariance, because $\dg$ is already limited to variations that vanish in a neighborhood of $\di\cN$, and because
it is assumed that the area parameter $v$ is non-stationary along the geodesics, one may add diffeomorphism
generators to both $\dg$ and $\{A,\cdot\}$ without altering either side of (\ref{auxbracketdef}), or the condition 
that $\dg$ vanish near $\di\cN$, in such a way that the variations thus modified are admissible.

One may thus restrict the variations in (\ref{auxbracketdef}) to admissible ones. Since admissible variations of the metric
on $\cN$ may be expressed in terms of variations of our free null initial data on $\cN$ this turns into a condition on
the $\bullet$ bracket of these data.

The coordinates $a_A^\ag = (u_A, r_A, y_A^1, y_A^2)$ form a chart on a spacetime neighborhood 
of $\cN_A$. On $\cN_A$ itself $u_A = 0$ and the $a_A$ chart reduces to a coordinate system 
$r_A, y_A^1, y_A^2)$ defined in much the same way as the $v_A\theta$ chart, but with the truncating 
2-surface $S_A$ playing the role of $S_0$: $y^1$ 
and $y^2$ are constant on the generators of $\cN_A$ and coincide on $S_A$ with the
fixed $y$ chart already introduced to define the diffeomorphism datum $s$.
% \footnote{
% %
% To lighten notation the branch index, $A$, will be supressed when there
% is little risk of confusion.}
% %
$r$ is an area parameter along the generators like $v$, but normalized to 1 on
$S_A$, so $r = \sqrt{\rho_y/\dot{\rho}} = v/\dot{v}$, where $\rho_y$ is the
area density on cross sections of $\cN_A$ in the $y$ chart, and $\dot{\rho}$ is
the area density on $S_A$ in this chart. 

To obtain a chart on a four dimensional neighborhood $r$, $y^1$, and $y^2$ are extended off
$\cN_A$ by holding them constant on the null geodesics normal to the equal $r$
cross sections of $\cN_A$ and transverse to $\cN_A$.
Finally $u$ is a parameter along these geodesics set to 0 on $\cN_A$ and chosen
such that $\di_u \cdot \di_r = -1$. For definiteness $u$ will be taken to be an affine parameter,
but this will not affect our considerations.

% On $\cN_A$ the transformation between the $a$ and $b$ charts is quite simple:
% \be	\label{btoa_transformation}
% r = v/\dot{v}(\theta)\ \ \ 
% y^i = s^i(\theta)\ \ \
% u = 0.
% \ee
In the $a$ chart the spacetime line element at $\cN_A$ takes the simple form
\be             \label{a_line_element}
ds^2 = -2du dr + h_{ij}dy^i dy^j = -2du dr + r^2 \dot{\rho} e_{ij}dy^i dy^j,
\ee
where  $e_{ij}$ are the $y$ chart components of the conformal 2-metric.

Essentially all smooth variations of the free null initial data on $\cN$ correspond to admissible variations of the spacetime metric. 
\begin{definition}\label{cA_def}
The set $\cA$ of admissible variations of the data on $\cN$ consists of all variations of the data such that
\begin{itemize}
 \item[1] $\dg e_{pq}$ is smooth on each branch of $\cN$, including in each case $S_0$.
 \item[2] $\dg \rho_0$, $\dg\lam$, $\dg s_L$, $\dg s_R$, and $\dg \tau$ are smooth on $S_0$. As a consequence $\dg\tilde{\tau}_A$ and $\dg\sct_A$ for $A = L,R$ are also smooth.
 \item[3] $\dg s_L$ and $\dg s_R$ vanish on $\di S_0$.
 \item[4] $\dg \dot{\rho}_A = 0$ on $S_A$ for $A = L,R$.
\end{itemize}
The last condition only fixes $\dg \dot{v}$ as a function of the variations of the other data.
\end{definition}

It is clear that the variations of the initial data corresponding to an admissible variation of the metric satisfies these conditions.
The following is an outline of a proof that establishes that, conversely, there always exists an admissible variation of the metric matching 
variations of the initial data satisfying conditions 1 - 4. The non-trivial part is to establish the existence of a smooth solution, $\dg g_{ab}$, 
to the linearized field equation, $\dg G_{ab} = 0$, matching the given variations of the initial data, which preserves $\cN$ as a double null sheet. ($\cN$ already 
is a double null sheet with respect to the unperturbed metric.) Once this is done the arguments of \cite{MR13} show that a diffeomorphism generator 
$\pounds_\xi$ may always be added to $\dg$ so that it also leaves the $a$ charts invariant in a neighborhood of each truncations surface $S_A$.
With that the result is established, since conditions 3 and 4 of the definition of admissible variations are equivalent to the conditions 3 and 4 above.

%%%% note first of all that $\cN$ has a globally hyperbolic neighborhood $\cM$ (Prop. B. 16 \cite{MR07}) which we will take to be spacetime in the following. Note also

As preliminaries to establishing the existence of the solution recall that spacetime $\cM$ is assumed to be globally hyperbolic and note 
that spacetime can be covered with open sets of compact closure (for instance the images of balls of $\Real^4$ under the exponential map) 
so the compact set $\cN$ has an open neighborhood $U$ of compact closure in $\cM$. Finally $J^-[D[\cN]]$, the causal past of the domain of 
dependence of $\cN$, is cut into two disjoint pieces of by $\cN$, and so is therefore $U$. Call $U_-$ the part to the past of $\cN$ and 
$U_+$ the part to the future.

Now to the outline proof: As a first step we use the algorithm of Sachs \cite{Sachs}, or Dautcourt \cite{Dautcourt}, for calculating the solution metric and all its 
derivatives to obtain a metric perturbation $\dg g_{ab}$, and all its spacetime derivatives, on $\cN$, such that the linearized field equation and all its 
derivatives holds on $\cN$ and the double nul sheet character of $\cN$ is maintained. 

What remains is to extend this to a smooth solution on spacetime. To do this we use a trick used in \cite{Rendall}. We make an arbitrary smooth extension 
$\dg' g_{ab}$ of the variation of the metric on $\cN$ to spacetime that matches $\dg g_{ab}$ and all its spacetime derivatives on $\cN$ and has support contained 
in $U$. That this can be done is guaranteed by the Whitney extension theorem \cite{Whitney} and the smoothness of the variations of the data, of the unperturbed spacetime 
metric, and the compactness of $S_0$. See \cite{Rendall} for a discussion.

Consider now $\dg' G_{ab}$, the variation of the Einstein tensor corresponding to $\dg' g_{ab}$. Imposing the de Donder gauge we may solve the linearized field equations 
with $\dg' G_{ab}$ as a source term by integrating it against the advanced Green's function, recovering $\dg' g_{ab}$ as the result. But, what if instead we set the source term to 
zero in the future of $\cN$, and use only the restriction of $\dg' G_{ab}$ to $U_-$ as a source? $\dg' G_{ab}$ is smooth, has support only in $U$ and vanishes on $\cN$.
In fact all its derivatives vanish on $\cN$, so replacing $\dg' G_{ab}$ by zero in $U_+$ leaves a smooth function (within $J^-[D[\cN]]$). Integrating it against the 
advanced Green's function from a point $p \in \cN$ reproduces $\dg g_{ab}$ on $\cN$ because setting the source to zero to the future of $\cN$ does not affect this integral.
On the other hand, if the point $p$ lies in the interior of $D[\cN]$ the result is changed, to a solution of the linearized field equation without source term to the future of 
$\cN$. In other words, we obtain a solution of this equation matching the variations of the initial data on $\cN$ and preserving the double null sheet character of $\cN$. 
Furthermore, since the source is smooth within the support of the advanced Green's function, which is contained in $J^-[D[\cN]]$, and the linearized Einstein equation is 
normally hyperbolic \cite{Bar_Ginoux_Pfaffle} in de Donder gauge, the resulting solution is smooth. This completes the argument.

\end{document}